\documentclass{aastex7}
\usepackage[T1]{fontenc}
\usepackage{amsmath}	% Advanced maths commands
\usepackage{amssymb}	% Extra maths symbols
\usepackage{bm}		% Bold maths symbols, including upright Greek
\usepackage{threeparttable}

\begin{document}

\title{Global resistive MHD accretion flows around spinning AGNs: impact of resistivity on MAD state}

%%%%%%%%%%%%%%%%%%%%%%%%%%%%%%%% Authors %%%%%%%%%%%%%%%%%%%%%%%%%%%%%%%%%%%%%%%%%%%%%%%%%%%%%%%%%%

\author[0000-0002-3672-6271]{Ramiz Aktar}
\affiliation{Department of Physics and Institute of Astronomy, National Tsing Hua University, 30013 Hsinchu, Taiwan}
\email[show]{ramizaktar@gmail.com}  

\author[0000-0002-1473-9880]{Kuo-Chuan Pan}
\affiliation{Department of Physics and Institute of Astronomy, National Tsing Hua University, 30013 Hsinchu, Taiwan}
\affiliation{Center for Theory and Computation, National Tsing Hua University, Hsinchu 30013, Taiwan}
\affiliation{Physics Division, National Center for Theoretical Sciences, National Taiwan University, Taipei 10617, Taiwan}
\email{kuochuan.pan@gapp.nthu.edu.tw}

\author{Toru Okuda}
\affiliation{Hakodate Campus, Hokkaido University of Education, Hachiman-Cho 1-2, Hakodate
040-8567, Japan}
\email{bbnbh669@ybb.ne.jp}

%%%%%%%%%%%%%%%%%%%%%%%%%%%%%%%%%%%%%%%%%%%%%%%%%%%%%%%%%%%%%%%%%%%%%%%%%%%%%%%%%%%%%%%%%%%%%%%%%%


%%%%%%%%%%%%%%%%%%%%%%%%%%%%%%%% Abstract %%%%%%%%%%%%%%%%%%%%%%%%%%%%%%%%%%%%%%%%%%%%%%%%%%%%%%%%%

\begin{abstract}

In this study, we investigate the effect of resistivity on the dynamics of global magnetohydrodynamic accretion flows (Res-MHD) around a spinning supermassive black hole. We perform a comparative study of 2D and 3D resistive models around black holes. We examine accretion flow dynamics considering globally uniform resistivity values, ranging from $\sim 0$ to 0.1. During the simulation time of $t \lesssim 1000~t_g$, we find that the mass accretion rate is comparable for both the 2D and 3D models. However, as the flow becomes increasingly turbulent, non-axisymmetric effects begin to dominate, resulting in significant differences in the mass accretion rates between the 3D and 2D. All the resistive models in a highly magnetized flow belong to the Magnetically Arrested Disk (MAD) state. We propose an efficient and physically motivated approach to examine the magnetic state by estimating the spatial average plasma beta parameter across the computational domain. We find that when the average plasma beta is close to or below unity (\( \beta_{\text{ave}} \lesssim 1 \)), the accretion flow enters the MAD state. Additionally, we find that high-resistivity flow reduces magnetorotational instability (MRI) turbulence in the accretion flow, while the turbulence structures remain qualitatively similar in low-resistivity flows. Moreover, we observe indications of plasmoid formations in low-resistivity flow compared to high-resistivity flow. Furthermore, we do not find a clear relationship between the variability of the accretion rate, magnetic flux, and resistivity. Lastly, our findings suggest that low-resistivity models produce higher power jets than those with higher resistivity.

\end{abstract}

%%%%%%%%%%%%%%%%%%%%%%%%%%%%%%%%%%%%%%%%%%%%%%%%%%%%%%%%%%%%%%%%%%%%%%%%%%%%%%%%%%%%%%%%%%%%%%%%%%


%\keywords{\uat{Galaxies}{573} --- \uat{Cosmology}{343} --- \uat{High Energy astrophysics}{739} --- \uat{Interstellar medium}{847} --- \uat{Stellar astronomy}{1583} --- \uat{Solar physics}{1476}}

%%%%%%%%%%%%%%%%%%%%%%%%%%%%%%%%  Introduction %%%%%%%%%%%%%%%%%%%%%%%%%%%%%%%%%%%%%%%%%%%%%%%%%%%%

\section{Introduction} \label{sec:intro}

Accretion onto black holes is one of the most powerful phenomena in the universe. Active galactic nuclei (AGNs) act as the central engines of galaxies. For instance, Sagittarius A* (Sgr A*) is a low-luminosity AGN located at the center of our own Milky Way. AGNs are commonly associated with jets and outflows. In the case of M87, highly relativistic jets have been observed \citep{Junor-etal-99, Cui-etal-23}. However, Sgr A* shows episodic X-ray and near-infrared flares that are detected roughly on a daily basis \citep{Baganoff-etal-2001, Genzel-etal-2003, Meyer-etal-2008, Degenaar-etal-2013, Neilsen-etal-2013, Neilsen-etal-2015, Ponti-etal-2015, Fazio-etal-2018, Boyce-etal-2019}, and there is no clear evidence of associated jets. Recently, \citet{GRAVITY-Collaboration-18} reported the detection of near-infrared flaring events very close to the black hole's event horizon (approximately 10 gravitational radii). They found that the speed of these flares was about 30\% of the speed of light. Furthermore, the flares exhibited continuous rotation of the polarization angle, and the polarization signature was consistent with orbital motion in a strong poloidal magnetic field.

In recent years, a significant magnetic state of accretion flow has been identified, known as the ``magnetically arrested disk'' (MAD) \citep{Narayan-etal-03}. During the accretion process in a highly magnetized flow, magnetic flux accumulates near the black hole’s event horizon. The resulting magnetic pressure balances the ram pressure in the disk, thereby impeding further mass accretion \citep{Narayan-etal-03, Igumenshchev-08, Tchekhovskoy-etal11}. Over the years, several simulation studies have examined the MAD state in accretion flows \citep{McKinney-etal-12, Narayan-etal12, Dihingia-etal21, Mizuno-etal-21, Chatterjee-Narayan-22, Fromm-etal-22, Janiuk-James-22, Dhang-etal23, Jiang-etal-23, Aktar-etal24a, Aktar-etal24b, Zhang-etal-24}. The MAD state is commonly used to describe AGNs with jets. In this regard, the Event Horizon Telescope (EHT) Collaboration has recently achieved a remarkable milestone by capturing groundbreaking images of black holes in M87 and the Galactic Center (Sgr A*) \citep{Event-Horizon-etal-2019, Event-Horizon-etal-2022a}. Furthermore, it has been confirmed that an ordered magnetic field favors the MAD state, as evidenced by comparisons between radio images and post-processed general relativistic magnetohydrodynamics (GRMHD) simulations \citep{Event-Horizon-etal-2021, Event-Horizon-etal-2024}. Similarly, the GRAVITY Collaboration has also confirmed the importance of the MAD state based on their high angular resolution near-infrared (NIR) observations of flares in Sgr A* \citep{GRAVITY-Collaboration-18, GRAVITY-Collaboration-20}.

Resistivity plays a crucial role in the dynamics of accretion by facilitating magnetic reconnection, an essential process for energy dissipation, particle acceleration, and angular momentum transport. By regulating the formation of current sheets and the efficiency of reconnection, resistivity shapes the behavior of accretion flows and their observational signatures, such as high-energy flares and variations in jets. Generally, global accretion flows and jets are effectively modeled using ideal magnetohydrodynamics (MHD) simulations. In ideal MHD flows, magnetic field lines are ``frozen'' into the plasma, meaning they move along with it \citep{Alfven-1942}. However, in ideal MHD, dissipation occurs within the accretion flows, and the frozen-in condition breaks down due to numerical resistivity, which is not a physical or reliable method for examining reconnection mechanisms and the formation of plasmoids. Several simulation studies have explored magnetic reconnection in accretion disks triggered by numerical resistivity, considering ideal MHD \citep{Ball-etal-2016, Ball-etal-2018, Nathanail-etal-20, Ripperda-etal-2022, Jiang-etal-23}. These models are primarily based on the numerical methods and resolution used, rather than on an underlying physical model. In this regard, \citet{Ripperda-etal-2019} developed a resistive model in a three-dimensional GRMHD code to examine highly magnetized flows. They also performed axisymmetric simulations to model reconnection and plasmoid formation across a range of resistive accretion flows \citep{Ripperda-etal-2020}. Recently, \citet{Nathanail-etal-2024} investigated how resistivity affects variability in the accretion flow around black holes. Motivated by this work, our study examines the impact of resistivity on highly magnetized accretion flows. For our analysis, we consider resistive MHD (Res-MHD) flows in both two-dimensional (2D) and three-dimensional (3D) models. We conduct a comparative study between these 2D and 3D models in a highly magnetized flow and also examine the impact of resistivity on the MAD state, considering a globally uniform resistivity in the flow.

We organize the paper as follows. In Section \ref{simulation-scheme}, we present the description of the numerical model and governing equations. In  Section \ref{results}, we discuss the simulation results of our model in detail. Finally,  we draw the concluding remarks and summary in Section \ref{conclusion}.

%%%%%%%%%%%%%%%%%%%%%%%%%%%%%%%%%%%%%%%%%%%%%%%%%%%%%%%%%%%%%%%%%%%%%%%%%%%%%%%%%%%%%%%%%%%%%%%%%%%

%%%%%%%%%%%%%%%%%%%%%%%%%%%%%%%%%%%%%%%%%%%%%%%%%%
\section{Simulation Setup}
\label{simulation-scheme}
%%%%%%%%%%%%%%%%%%%%%%%%%%%%%%%%%%%%%%%%%%%%%%%%%%

We conduct simulations of resistive magnetohydrodynamics (Res-MHD) using the publicly available numerical simulation code PLUTO\footnote{\url{http://plutocode.ph.unito.it}} \citep{Migone-etal07}. In our analysis, we adopt a unit system in which \( G = M_{\rm BH} = c = 1 \), where \( G \) is the gravitational constant, \( M_{\rm BH} \) is the mass of the black hole, and \( c \) is the speed of light. As a result, we measure distance, velocity, and time using the following definitions: \( r_g = \frac{G M_{\rm BH}}{c^2} \), \( c \), and \( t_g = \frac{G M_{\rm BH}}{c^3} \), respectively. In this work, we do not consider any explicit radiative processes within the flow.

%%%%%%%%%%%%%%%%%%%%%%%%%%%%%%%%%%%%%%%%%%%%%%%%%%
\subsection{Governing Equations for Resistive-MHD}
%%%%%%%%%%%%%%%%%%%%%%%%%%%%%%%%%%%%%%%%%%%%%%%%%%

Here, we present resistive MHD governing equations and are as follows \citep{Migone-etal07} \\
\begin{align}
& \label{governing_eq_1} \frac{\partial \rho}{\partial t} + \nabla \cdot (\rho \bm{v}) =0,\\
& \label{governing_eq_2} \frac{\partial (\rho \bm{v})}{\partial t} + \nabla \cdot (\rho \bm{v} \bm{v} - \bm{B}\bm{B}) + \nabla P_t = -\rho \nabla \Phi, \\
& \frac{\partial E }{\partial t} + \nabla \cdot [(E+P_t)\bm{v} - (\bm{v}.\bm{B})\bm{B} + \eta \bm{J} \times \bm{B} ]  = -\rho \bm{v}.\nabla \Phi,\\
& \frac{\partial \bm{B}}{\partial t} + \nabla \times (-\bm{v} \times \bm{B}+\eta \bm{J}) =0,
\end{align}
where $\rho$, $\bm{v}$, and $\bm{B}$ are the mass density, fluid velocity, and magnetic field, respectively. $P_t$ is the total pressure with contribution from gas pressure $(P_{\rm gas})$ and magnetic pressure ($B^2/2$) i.e., $P_t = P_{\rm gas} + B^2/2$. $E$ is the total energy density given by
\begin{align}
E = \frac{P_{\rm gas}}{\gamma -1} + \frac{1}{2} \rho v^2 + \frac{1}{2} B^2,
\end{align}
where, $\gamma$ is the adiabatic index. The resistive tensor is defined as $\eta$. Here, the resistive tensor is assumed to be diagonal as $\eta \equiv \rm{diag}(\eta_{x1}, \eta_{x2}, \eta_{x3})$ \citep{Migone-etal07}. The electric current density is defined as $\bm{J} = \nabla \times \bm{B}$. Here, $\Phi$ is the gravitational potential of the black hole.

%%%%%%%%%%%%%%%%%%%%%%%%%%%%%%%%%%%%%%%%%%%%%%%%%%
\subsection{Gravitational potential of the black hole}
%%%%%%%%%%%%%%%%%%%%%%%%%%%%%%%%%%%%%%%%%%%%%%%%%%
\label{Kerr_pot}

In this study, we model the gravitational effects surrounding a spinning black hole using the effective Kerr potential \citep{Dihingia-etal18b}. We implement this effective Kerr potential in the PLUTO code, following the methodology of our earlier works \citep{Aktar-etal24a, Aktar-etal24b}. The effective Kerr potential is given by \citep{Dihingia-etal18b, Aktar-etal24a, Aktar-etal24b}
 \begin{align} \label{DDMC18_pot}
 \Phi^{\rm eff}\left(r,z,a_k, \lambda\right)= \frac{1}{2} \ln \left(\frac{A (2 R - \Sigma) r^2 - 4 a_k^2 r ^4}{\Sigma \lambda \left(\Sigma \lambda  R^2 + 4 a_k r ^2 R - 2 \lambda  R^3\right)- A \Sigma r ^2 }\right),
 \end{align}
 where $R=\sqrt{r^2+z^2}$ = spherical radial distance, $\Delta=a_k^2+R^2-2 R$, $\Sigma=\frac{a_k^2 z^2}{R^2}+R^2$ and $A=\left(a_k^2+R^2\right)^2-\frac{a_k^2 r^2 \Delta}{R^2}$. Here, $\lambda$ and $a_k$ are specific flow angular momentum and black hole spin, respectively. The dimensionless black hole spin is defined as \(a_k = \frac{J}{M_{\text{BH}}}\), where \(J\) represents the angular momentum of the black hole. The Keplerian angular momentum can be obtained from equation (\ref{DDMC18_pot}) in the equatorial plane $(z \rightarrow 0)$ as $\lambda_K = \sqrt{r^3 \frac{\partial \Phi^{\rm eff}}{\partial r}|_{\lambda \rightarrow 0}}$. The angular frequency is obtained as $\Omega = \lambda/r^2$.
 
 It is important to note that \citet{Dihingia-etal18b} conducted a comprehensive examination of various comparative studies between General Relativity (GR) and an effective Kerr potential based on analytical methods. They demonstrated a strong agreement in the transonic properties when using the effective Kerr potential in the semi-relativistic regime, in comparison to GR around the Kerr black hole. Motivated by these findings, our simulation approach employs the effective Kerr potential to accurately capture the essential spacetime features near black holes. A significant advantage of this method is its ability to achieve much higher spatial resolution compared to GRMHD simulations while using similar computational resources \citep{Aktar-etal24a, Aktar-etal24b}. This capability makes our approach particularly well-suited for exploring the complex dynamics of 3D accretion flows, including dissipation and radiation.
 
% Consequently, our numerical approach successfully captures the essential features of space-time around black holes by implementing the effective Kerr potential. Additionally, our simulation model is capable of achieving higher spatial resolutions than GRMHD simulations while utilizing the same computing resources \citep{Aktar-etal24a, Aktar-etal24b}. More specifically, our simulation approach using an effective Kerr potential is highly beneficial for exploring multi-dimensional (3D) flow dynamics. However, it is important to note that our simulation method is less accurate compared to GRMHD simulations. Nonetheless, we believe that the overall qualitative features will remain consistent with those observed in GRMHD simulations.

%%%%%%%%%%%%%%%%%%%%%%%%%%%%%%%%%%%%%%%%%%%%%%%%%%
\subsection{Set up of initial equilibrium Torus and atmospheric condition}
\label{torus_set_up}
%%%%%%%%%%%%%%%%%%%%%%%%%%%%%%%%%%%%%%%%%%%%%%%%%%

The initial equilibrium torus is constructed using the Newtonian analog of relativistic tori as described by \cite{Abramowicz-etal78}. In this study, we adopt the same formalism to establish the initial equilibrium torus outlined by \citet{Aktar-etal24a}. The density distribution of the torus is derived by considering a constant angular momentum flow \citep{Matsumoto-etal96, Hawley-00, Kuwabara-etal05, Aktar-etal24a, Aktar-etal24b}:
\begin{align}
\Phi^{\rm eff}\left(r,z, a_k, \lambda\right) + \frac{\gamma}{\gamma -1}\frac{P_{\rm gas}}{\rho}={\cal C},
\end{align}
where `$\mathcal{C}$' is the integration constant and is evaluated under the condition of zero gas pressure \((P_{\rm gas} \rightarrow 0)\) at the location \(r = r_{\rm min}\) in the equatorial plane. Here, \(r_{\rm min}\) denotes the inner edge location of the torus. Consequently, the density distribution can be expressed using the adiabatic equation of state, \(P_{\rm gas} = K \rho^{\gamma}\), as follows:
\begin{align}
\rho=\left[\frac{\gamma-1}{K\gamma}\left({\cal C}-\Phi^{\rm eff}\left(r,z, a_k, \lambda\right)\right)\right]^{\frac{1}{\gamma -1}},   
\end{align}
where \(K\) is determined based on the maximum density condition \((\rho_{\rm max})\) at the location \(r = r_{\rm max}\) in the equatorial plane. It is given by:
\begin{align}
K=\frac{\gamma - 1}{\gamma}\left[\mathcal{C}-\Phi^{\rm eff}\left(r_{\rm max},0, a_k, \lambda\right)\right]\frac{1}{\rho_{\rm max}^{\gamma-1}}.   
\end{align}
Here, the term \(\Phi^{\rm eff}\) refers to the effective Kerr potential as mentioned in Equation \ref{DDMC18_pot}.

In this work, we consider a different density distribution outside the torus compared to our previous studies \citep{Aktar-etal24a, Aktar-etal24b}. We assume that the atmosphere evolves freely and that the initial density and pressure in the space surrounding the torus are very low. The density and pressure distributions outside the torus are defined as \(\rho_{\rm atm} = \rho_{\rm floor} r^{-3/2}\) and \(P_{\rm atm} = P_{\rm floor} r^{-5/2}\), with the floor values chosen to be very small (see section \ref{floor-condition}). These atmospheric conditions are generally considered in various GRMHD simulation code \citep{Gammie-etal03, Porth-etal-16, White-etal-2016, Liska-etal-2018}.

%%%%%%%%%%%%%%%%%%%%%%%%%%%%%%%%%%%%%%%%%%%%%%%%%%%%%%%%%%%%%%%%%%%%%%%%%
%                        Table 1
%%%%%%%%%%%%%%%%%%%%%%%%%%%%%%%%%%%%%%%%%%%%%%%%%%%%%%%%%%%%%%%%%%%%%%%%%
\begin{table*}
\centering

%\begin{minipage}{180mm}
%\ContinuedFloat
\caption{Code unit system used in this paper }
\label{Table-1}
 \begin{tabular}{@{}c c   } 
 \hline
 Units &    code units   \\ 
 \hline  
 Density  &  $\rho_0 = 1 \times 10^{-10}$\\
 Length   &    $ r_g = GM_{\rm BH}/c^2$\\
 Velocity &  $c$ \\
 Time     &  $t_g = GM_{\rm BH}/c^3$\\
 Magnetic field  & $B = \frac{B_0}{c \sqrt{4 \pi \rho_0}}$\\
 \hline

%$**$ $r_g = \frac{GM_{\rm BH}}{c^2}$% = Schwarzschild radius.

 \end{tabular}
 %\end{minipage}
 
\end{table*}

%====================================================================

%%%%%%%%%%%%%%%%%%%%%%%%%%%%%%%%%%%%%%%%%%%%%%%%%%%%%%%%%%%%%%%%%%%%%%%%%
%                        Table 2
%%%%%%%%%%%%%%%%%%%%%%%%%%%%%%%%%%%%%%%%%%%%%%%%%%%%%%%%%%%%%%%%%%%%%%%%%

\begin{table*}[htbp]
\centering
\begin{threeparttable}
\caption{Simulation models.}
\label{Table-2}

\begin{tabular}{@{}c c c c c c c c c c c@{}}
\hline
Model\tnote{$a$} & $r_{\rm min}$\tnote{b} & $r_{\rm max}$\tnote{c} & $\rho_{\rm max}$\tnote{d} & $\beta_0$\tnote{e} & $a_k$\tnote{f} & $\lambda$\tnote{g} & $\eta$\tnote{h} & $(n_r \times n_\phi \times n_z)$\tnote{i} & $[\sigma_{\dot{M}_{\rm acc}}, \sigma_{\dot{\phi}_{\rm acc}}]$ \tnote{j} \\
      & ($r_g$) & ($r_g$) &  & & & & & & \\
\hline
2D Model \\
${\rm 2D1}$  & 20 & 40 & $10^{-10}$ & 10 & 0.95 & 6.20 & 0.1 & $(860 \times 0 \times 860)$ & [1.3128, 1.4115] \\
${\rm 2D2}$  & '' & '' & '' & '' & '' & '' & 0.01 & '' & [1.6000, 1.2389] \\
${\rm 2D3}$  & '' & '' & '' & '' & '' & '' & $10^{-3}$ & '' & [1.2747, 0.9528] \\
${\rm 2D4}$  & '' & '' & '' & '' & '' & '' & $10^{-4}$ & '' & [1.6453, 1.0756] \\
${\rm 2D5}$  & '' & '' & '' & '' & '' & '' & $10^{-5}$ & '' & [1.4824, 1.0017] \\
${\rm 2D6}$  & '' & '' & '' & '' & '' & '' & $\sim 0.0~(10^{-20})$ & '' & [1.3349, 0.9645] \\ \\

3D Model \\
${\rm 3D1}$  & '' & '' & '' & '' & '' & '' & 0.1 & $(860 \times 92 \times 860)$ & [1.8909, 1.3872] \\
${\rm 3D2}$  & '' & '' & '' & '' & '' & '' & 0.01 & '' & [1.0969, 1.9827] \\
${\rm 3D3}$  & '' & '' & '' & '' & '' & '' & $10^{-3}$ & '' & [1.2565, 1.5483] \\
${\rm 3D4}$  & '' & '' & '' & '' & '' & '' & $10^{-4}$ & '' & [1.3949, 1.2941] \\
${\rm 3D5}$  & '' & '' & '' & '' & '' & '' & $10^{-5}$ & '' & [1.3023, 1.3132] \\
${\rm 3D6}$  & '' & '' & '' & '' & '' & '' & $\sim 0.0~(10^{-20})$ & '' & [1.5944, 1.3324] \\
\hline
\end{tabular}

\begin{tablenotes}
\footnotesize
\item[] 
$(a):$ Model names. 
$(b):$ inner edge of the initial equilibrium torus.
$(c):$ location of the maximum pressure of the initial equilibrium torus.
$(d):$ maximum density at $r_{\rm max}$.
$(e):$ initial plasma beta parameter.
$(f):$ dimensionless spin parameter of the black hole.
$(g):$ specific flow angular momentum.
$(h):$ resistivity parameter of the flow.
$(i):$ resolution of the simulation.
$(j):$ Variability based on mass accretion rate ($\sigma_{\dot{M}_{\rm acc}}$) and normalized magnetic flux ($\sigma_{\dot{\phi}_{\rm acc}}$)
\end{tablenotes}

\end{threeparttable}
    
\end{table*}

%====================================================================

%%%%%%%%%%%%%%%%%%%%%%%%%%%%%%%%%%%%%%%%%%%%%%%%%%
\subsection{Initial magnetic field configuration in the torus}
\label{Mag_config}
%%%%%%%%%%%%%%%%%%%%%%%%%%%%%%%%%%%%%%%%%%%%%%%%%%

In this paper, we initialize a poloidal magnetic field by applying a purely toroidal component of the vector potential \citep{Hawley-etal02}. The expression for the toroidal component of the vector potential is given by \citep{Hawley-etal02, Aktar-etal24a}:
\begin{align}
A_\phi = B_0 [\rho(r,\phi,z) - \rho_{\rm min}],
\end{align}
where \(B_0\) represents the normalized initial magnetic field strength and \(\rho_{\rm min}\) is the minimum density within the torus. The initial magnetic field strength is parameterized by the ratio of gas pressure to magnetic pressure, known as the plasma beta parameter, defined as \(\beta_0 = \frac{2 P_{\rm gas}}{B_0^2}\). Additionally, we define the magnetization parameter as \(\sigma_{\rm M} = \frac{B^2}{\rho}\) \citep{Dihingia-etal21, Dihingia-etal22, Dhang-etal23, Curd-Narayan23, Aktar-etal24a, Aktar-etal24b}. The magnetization parameter effectively represents the ratio of magnetic energy to rest mass energy in the flow.

%%%%%%%%%%%%%%%%%%%%%%%%%%%%%%%%%%%%%%%%%%%%%%%%%%
\subsection{Initial and boundary conditions}
%%%%%%%%%%%%%%%%%%%%%%%%%%%%%%%%%%%%%%%%%%%%%%%%%%

To simulate resistive magnetohydrodynamic (Res-MHD) accretion flows, we utilize the PLUTO simulation code \citep{Migone-etal07}. In our simulations, we apply the HLLC Riemann solver, use second-order linear interpolation for spatial discretization, and implement the second-order Runge-Kutta method for time integration. To retain the divergence-free condition given by $\nabla \cdot \bm{B} = 0$, we utilize the hyperbolic divergence cleaning method \citep{Migone-etal07}. The inner boundary conditions are set to be absorbing condition around the black hole \citep{Okuda-etal19, Okuda-etal22, Okuda-etal23, Aktar-etal24a, Aktar-etal24b}, with the inner boundary located at \(R_{\rm in} = 2.5 r_g\), which acts as the event horizon in our model. At the inner boundary, \( r \leq R_{\rm in} \), the radial velocity is set to the free-fall value, while the remaining variables are obtained by interpolation from the outer region near the boundary. All other boundaries are configured with outflow conditions \citep{Aktar-etal24a, Aktar-etal24b}. For our simulations, we consider the resolution in the axisymmetric approximation to be ($n_r \times n_\phi \times n_z$) = ($860 \times 0 \times 860$) for the 2D model, and for 3D models, the resolution is set to ($n_r \times n_\phi \times n_z$) = ($860 \times 92 \times 860$). We maintain the same resolution in the $r$ and $z$ directions for comparative studies, as summarized in Table \ref{Table-2}.

There is widespread agreement in the literature regarding the dimensions of the torus for MAD model in GRMHD simulations \citep{Wong-etal-21, Fromm-etal-22}. For the MAD state, the inner edge (\(r_{\rm min}\)) and the location of maximum pressure (\(r_{\rm max}\)) are typically found at \(r_{\rm min} \approx 20 r_g\) and \(r_{\rm max} \approx 40 r_g\), respectively. Accordingly, we establish the initial equilibrium torus with the inner edge positioned at \(r_{\rm min} = 20 r_g\) and the maximum pressure location at \(r_{\rm max} = 40 r_g\). We set the minimum density of the torus to \(\rho_{\rm min} = 0.1 \rho_{\rm max}\), where \(\rho_{\rm max}\) represents the maximum density of the torus. For our simulation, we designate \(\rho_{\rm max} = \rho_0\), with \(\rho_0\) being the reference density. In this work, we fix the reference density at \(\rho_0 = 10^{-10}\), and the constant adiabatic index is set to \(\gamma = \frac{4}{3}\). A summary of the code unit definitions is provided in Table \ref{Table-1} \citep{Aktar-etal24a}.

%%%%%%%%%%%%%%%%%%%%%%%%%%%%%%%%%%%%%%%%%%%%%%%%%%
\subsection{Floor values and ceiling conditions}
%%%%%%%%%%%%%%%%%%%%%%%%%%%%%%%%%%%%%%%%%%%%%%%%%%
\label{floor-condition}

To avoid encountering negative density and pressure in supersonic and highly magnetized flows, particularly near the event horizon, we establish very low minimum values for both density and pressure. Specifically, we set the minimum density at \(\rho_{\text{floor}} = 10^{-6} \rho_0\) and the minimum pressure at \(P_{\text{floor}} = 10^{-8} P_0\), where $\rho_0$ and $P_0$ are the reference density and pressure, respectively. Additionally, we impose a ceiling condition on magnetization, limiting it to \(\sigma_{\text{M}} < 100\) in our simulation model, which is a typical consideration in GRMHD simulations \citep{Tchekhovskoy-etal11, Porth-etal-16, Dihingia-etal21, Narayan-etal-22, Chatterjee-Narayan-22, Dhang-etal23, Jiang-etal-23, Dhruv-etal-2025}.

%%%%%%%%%%%%%%%%%%%%%%%%%%%%%%%%%%%%%%%%%%%%%%%%%%
\subsection{Identification of magnetic state of the flow}
\label{magnetic_state}
%%%%%%%%%%%%%%%%%%%%%%%%%%%%%%%%%%%%%%%%%%%%%%%%%%

To investigate the magnetic state of our model, we examine the magnetic flux accumulated at the horizon. We define a dimensionless normalized magnetic flux threading the black hole horizon, denoted as \(\dot{\phi}_{\rm acc}\), which is known as the MAD parameter. The MAD parameter is defined in Gaussian units \citep{Tchekhovskoy-etal11, Narayan-etal12, Dihingia-etal21, Dhang-etal23, Aktar-etal24a} as follows:

For the 2D model:
\begin{align}
\dot{\phi}_{\rm acc} =  \frac{\pi \sqrt{4 \pi}}{\sqrt{\dot{M}_{\rm acc}}}  \int |B_r|_{R = R_{\rm in}} \, r\, dz,
\end{align}
and for the 3D model:
\begin{align}
\dot{\phi}_{\rm acc} = \frac{\sqrt{4 \pi}}{2 \sqrt{\dot{M}_{\rm acc}}} \int \int |B_r|_{R = R_{\rm in}} \, r\, dz \, d\phi,
\end{align}
where \(R_{\rm in}\) represents the inner boundary or the horizon for our model, and \(B_r\) is the radial component of the magnetic field. In our model, we define the normalized magnetic flux in Gaussian units by incorporating the factor of \(\sqrt{4 \pi}\) \citep{Narayan-etal-22}. 

The mass accretion rate, \(\dot{M}_{\rm acc}\), is given by:

For the 2D model:
\begin{align}
\dot{M}_{\rm acc} = -2\pi \int \rho (r,z) \, r \, v_r \, dz,
\end{align}
and for the 3D model:
\begin{align}
\dot{M}_{\rm acc} = - \int \int \rho (r,z, \phi) \, r \, v_r \, dz \, d\phi,
\end{align}
where the negative sign indicates the inward direction of the accretion flow.

The accretion state transitions to the MAD state when the normalized magnetic flux \(\dot{\phi}_{\rm acc}\) reaches a critical value. It is generally accepted that the MAD state is achieved when the critical value of \(\dot{\phi}_{\rm acc} \gtrsim 50\) in Gaussian units \citep{Tchekhovskoy-etal11, Narayan-etal12, Dihingia-etal21, Jiang-etal-23, Aktar-etal24b}.

%%%%%%%%%%%%%%%%%%%%%%%%%%%%%%%%%%%%%%%%%%%%%%%%%%%
%%                        Figure 1
%%%%%%%%%%%%%%%%%%%%%%%%%%%%%%%%%%%%%%%%%%%%%%%%%%%
\begin{figure*}
	\begin{center}
        \includegraphics[width=0.48\textwidth]{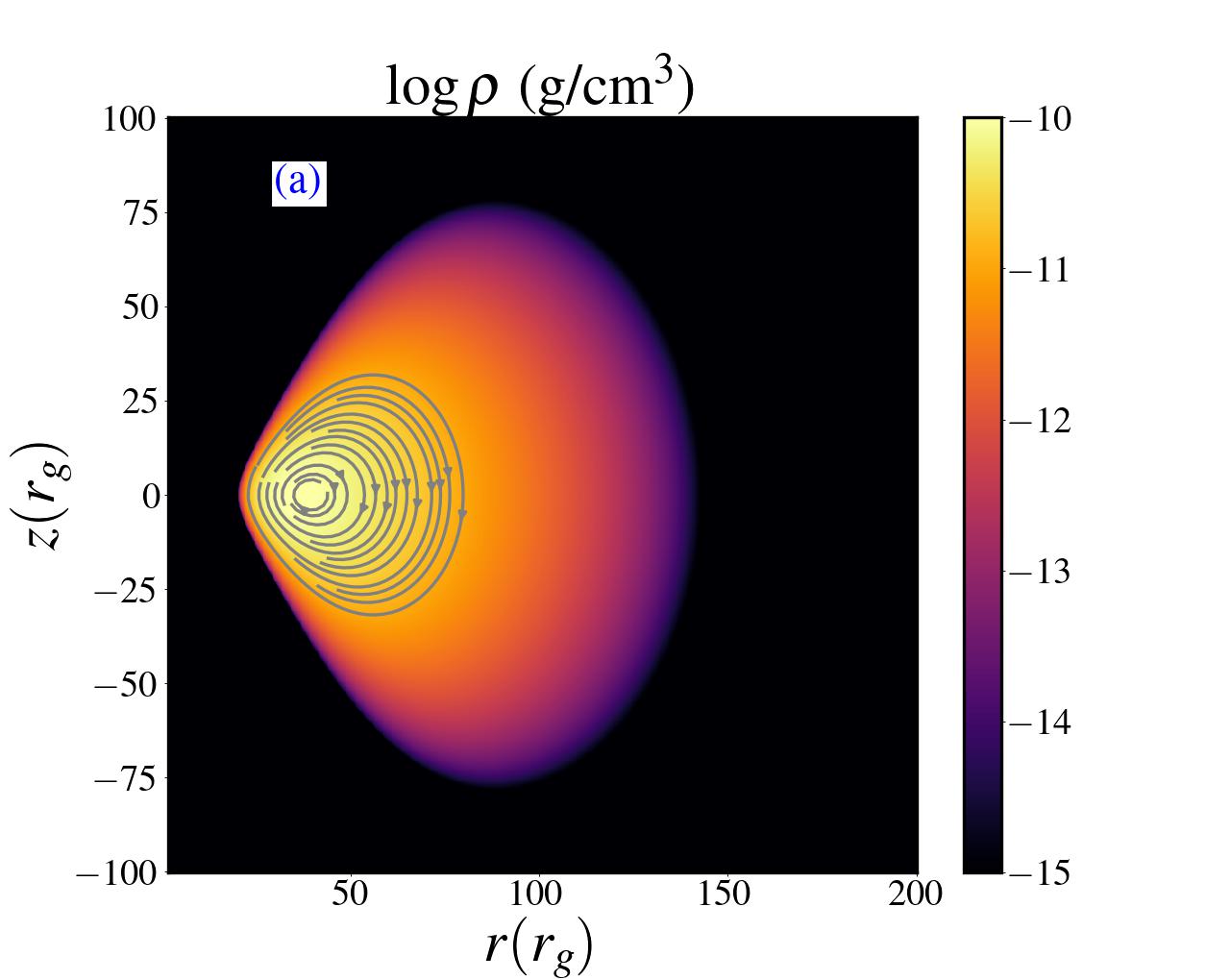} 
        \includegraphics[width=0.48\textwidth]{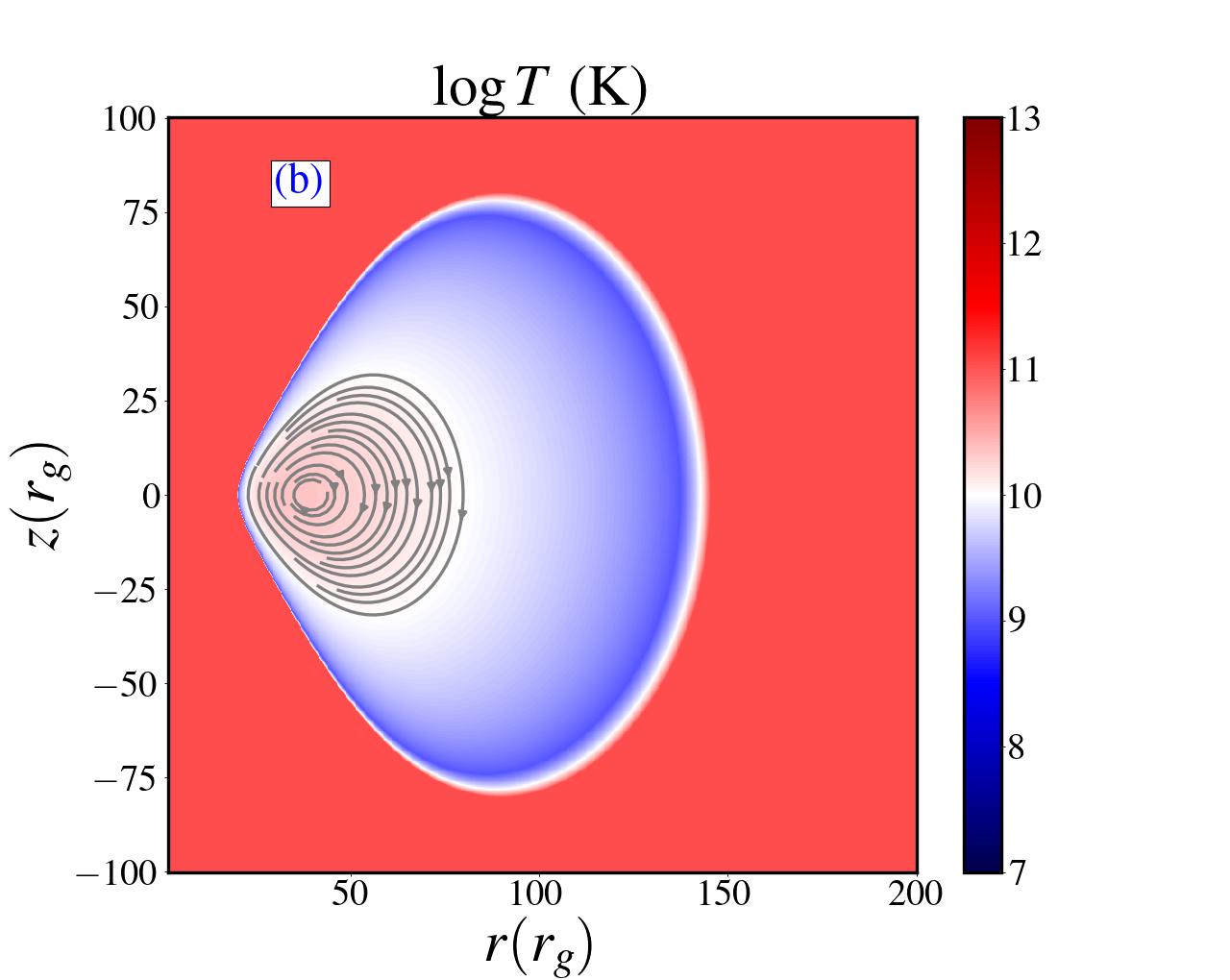} 
	\end{center}
	\caption{Distribution of $(a)$: density ($\log \rho $) and $(b)$: temperature ($\log T $) of the initial equilibrium torus at $t = 0~t_g$. Here, the grey lines represent magnetic field lines.}
	\label{Figure_1}
\end{figure*}
%%%%%%%%%%%%%%%%%%%%%%%%%%%%%%%%%%%%%%%%%%%%%%%%%%%%

%%%%%%%%%%%%%%%%%%%%%%%%%%%%%%%%%%%%%%%%%%%%%%%%%%%
%%                        Figure 2
%%%%%%%%%%%%%%%%%%%%%%%%%%%%%%%%%%%%%%%%%%%%%%%%%%%
\begin{figure}
	\begin{center}
        \includegraphics[width=0.49\textwidth]{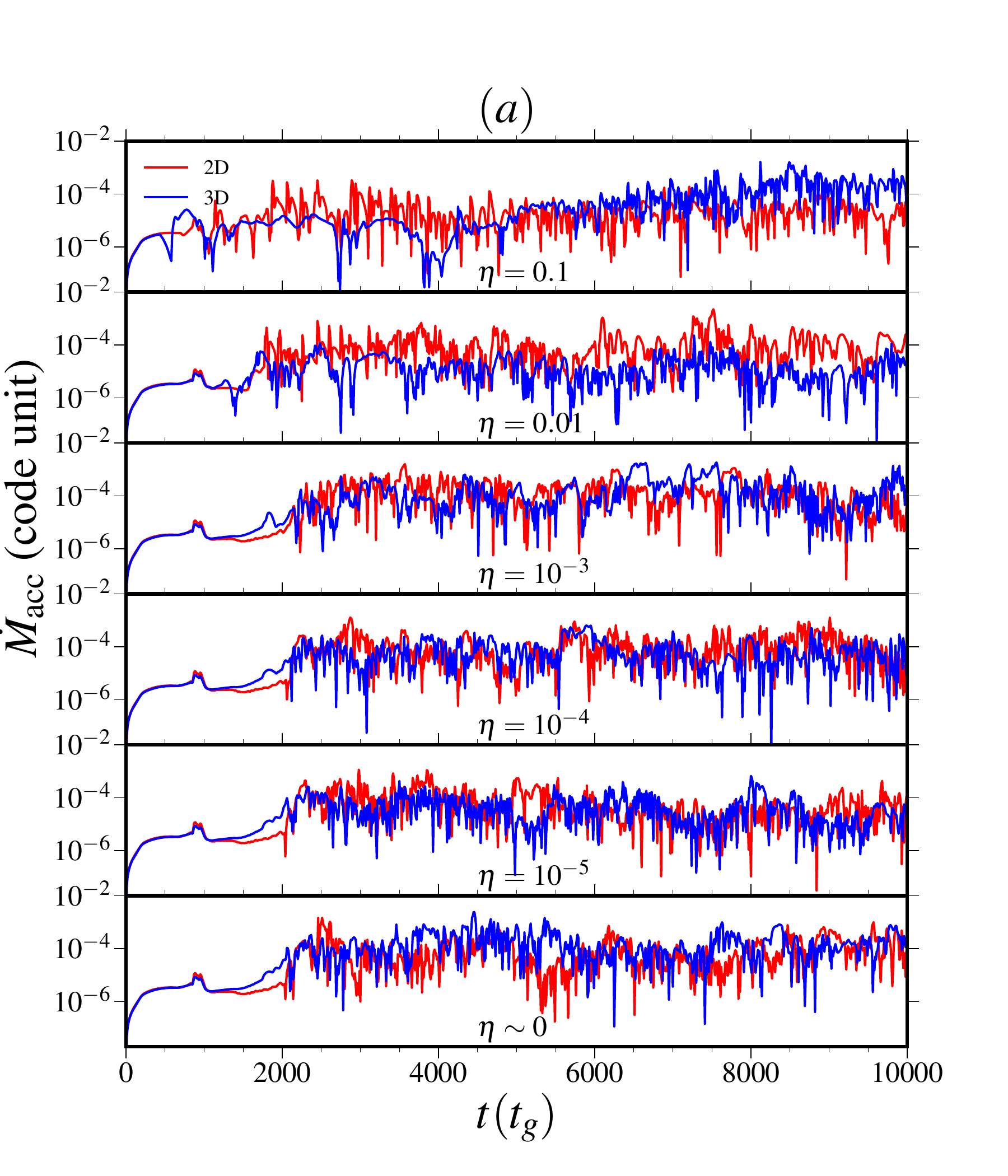} 
	\includegraphics[width=0.49\textwidth]{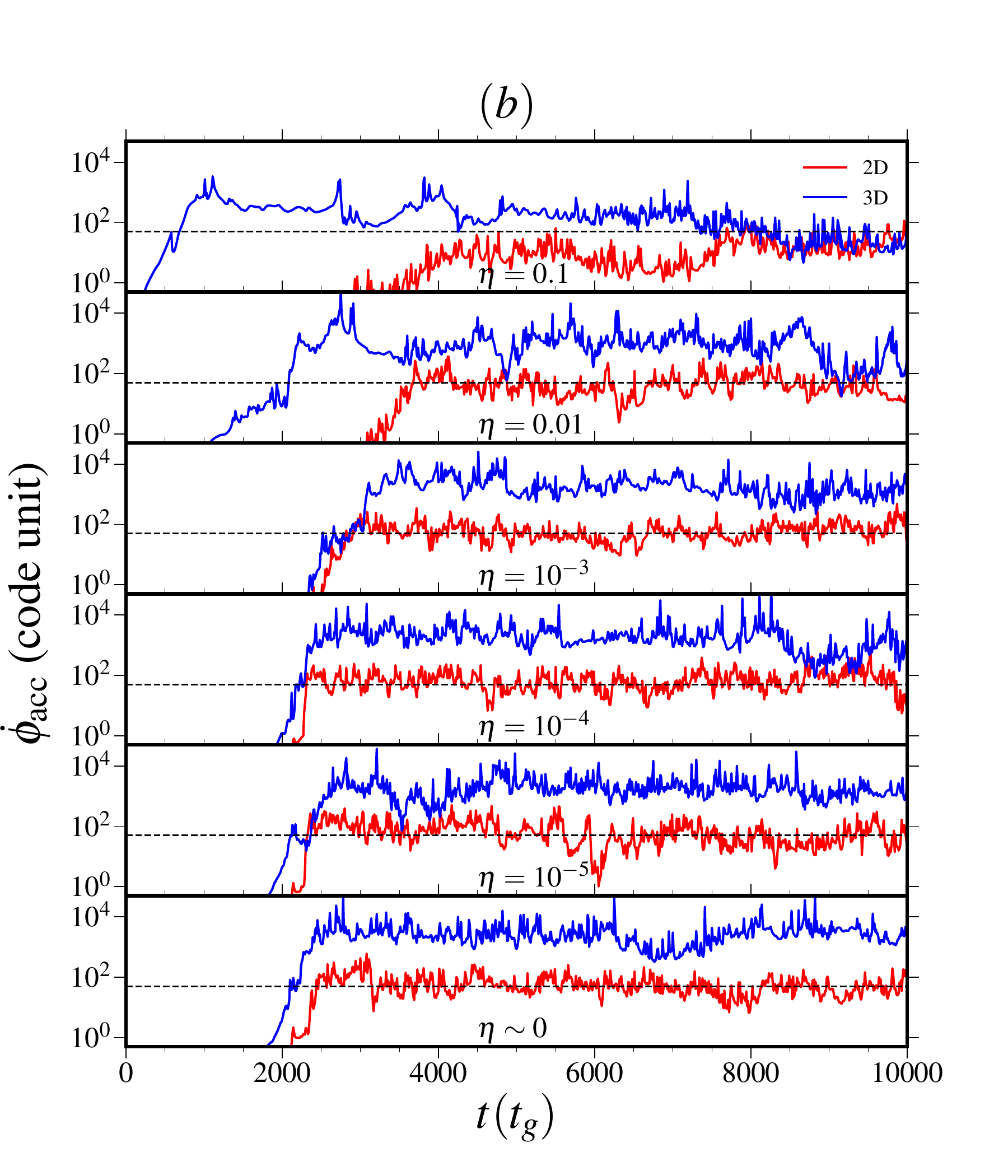} 
        \vskip -3.0mm
        \includegraphics[width=0.49\textwidth]{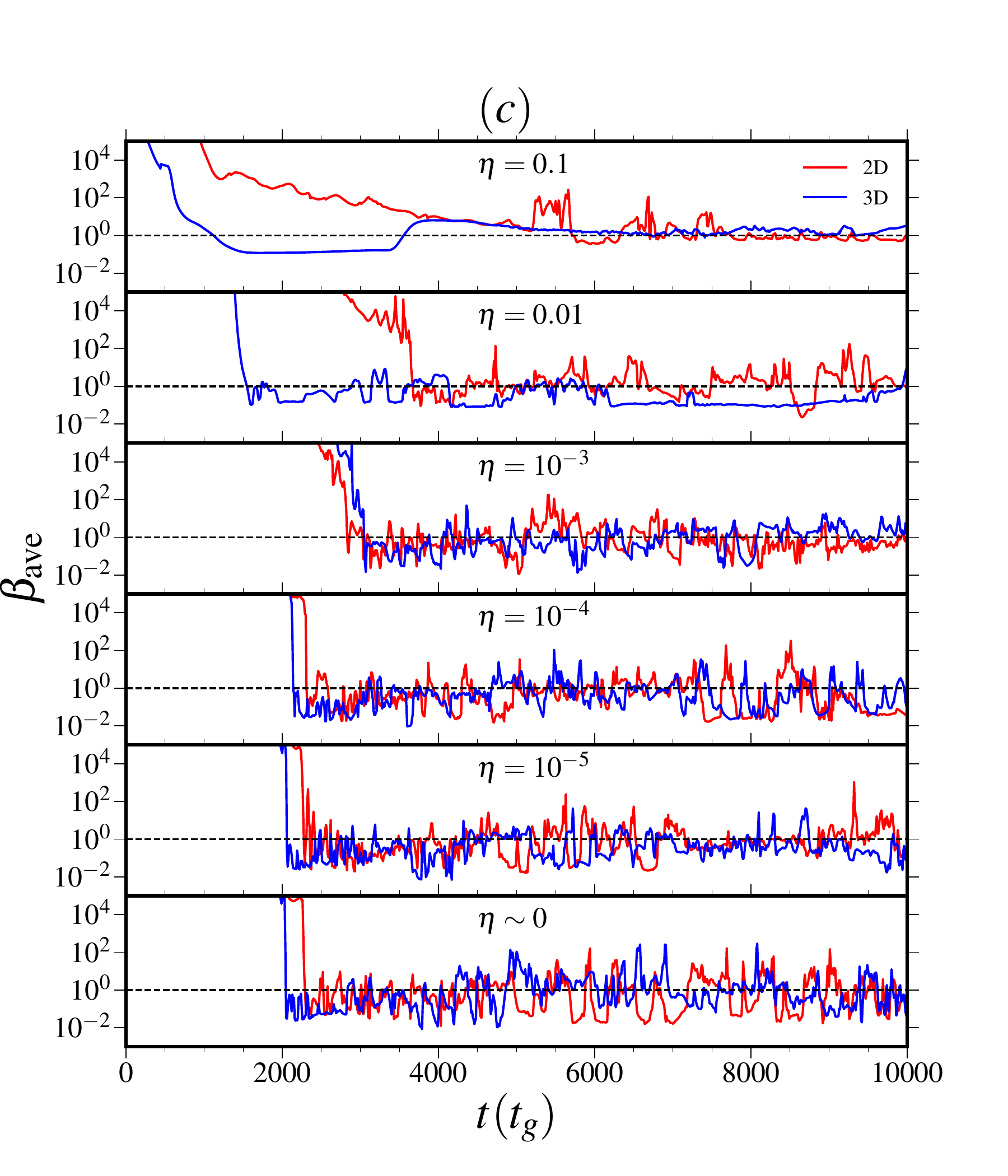} 
        \includegraphics[width=0.49\textwidth]{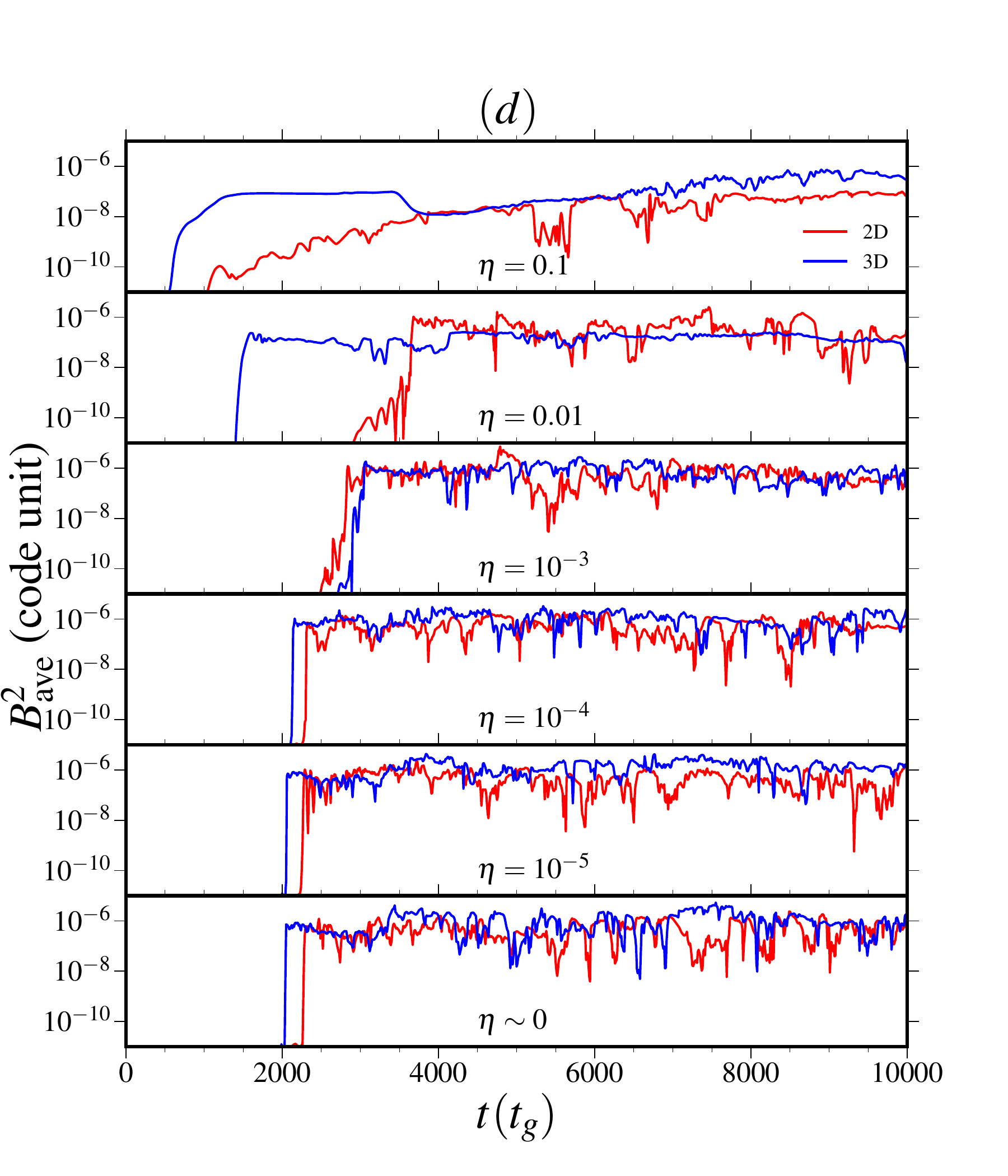} 
	\end{center}
	\caption{Comparison of temporal evolution of $(a)$: mass accretion rate ($\dot{M}_{\rm acc}$), $(b)$: normalized magnetic ﬂux ($\dot{\phi}_{\rm acc}$) accumulated at the 
    black hole horizon, ($c$): spatial average plasma-$\beta$ ($\beta_{\rm ave}$) and ($d$): spatial average magnetic energy ($B^2_{\rm ave}$)  with the simulation time for different resistivity. Here, we consider the resistivity as $\eta = 0.1, 0.01, 10^{-3}, 10^{-4}, 10^{-5}$, and $\sim 0$. See the text for details.}
	\label{Figure_2}
\end{figure}
%%%%%%%%%%%%%%%%%%%%%%%%%%%%%%%%%%%%%%%%%%%%%%%%%%%%

%%%%%%%%%%%%%%%%%%%%%%%%%%%%%%%%%%%%%%%%%%%%%%%%%%%
%%                        Figure 3
%%%%%%%%%%%%%%%%%%%%%%%%%%%%%%%%%%%%%%%%%%%%%%%%%%%
\begin{figure}
	\begin{center}
        \includegraphics[width=0.60\textwidth]{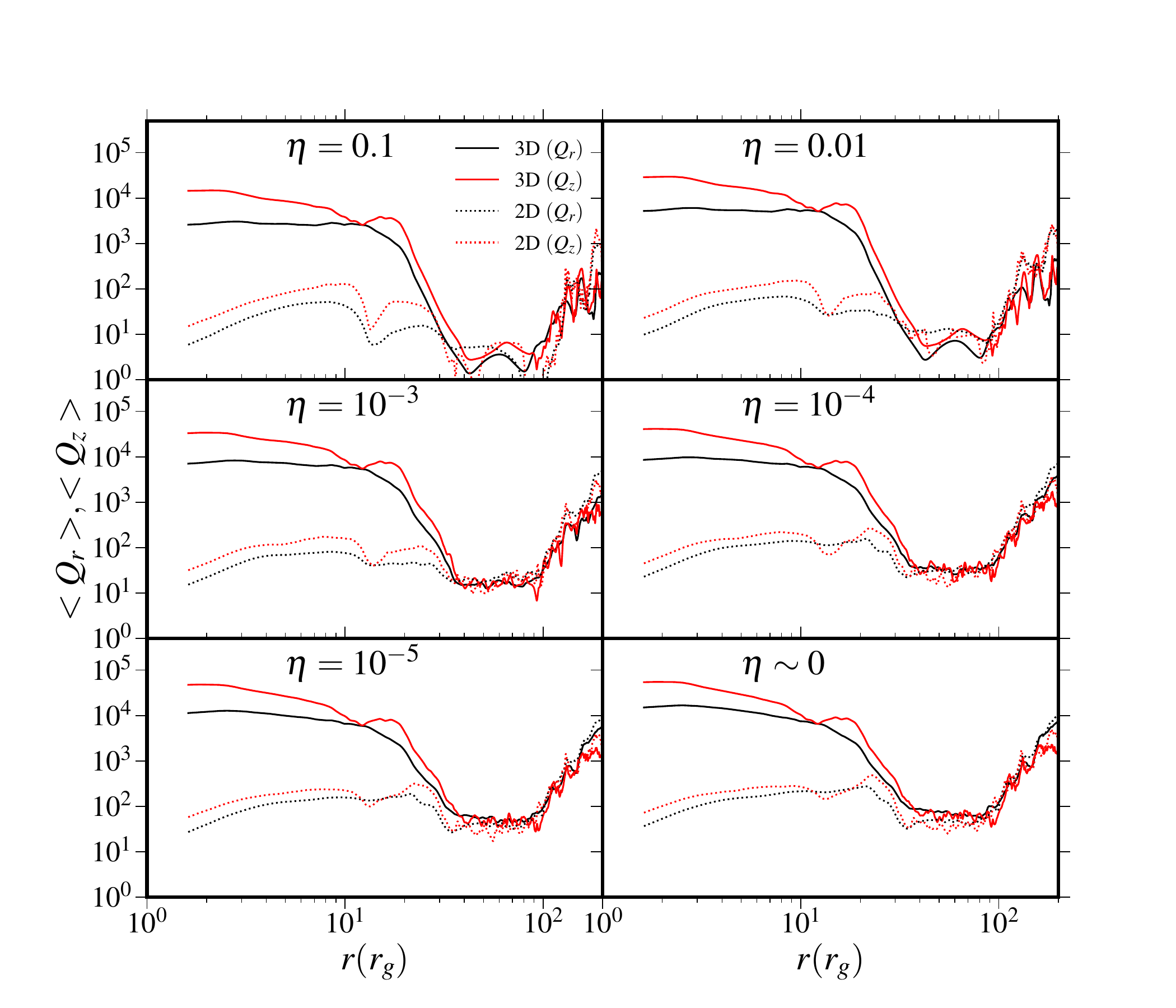} 
	\end{center}
	\caption{{Radial variation of space-averaged MRI quality factor $Q_r$ (black) and $Q_z$ (red) for various resistivity $\eta = 0.1, 0.01, 10^{-3}, 10^{-4}, 10^{-5}$, and $\sim 0$. Solid and dotted curves are for 3D and 2D models, respectively.}}
	\label{Figure_3}
\end{figure}
%%%%%%%%%%%%%%%%%%%%%%%%%%%%%%%%%%%%%%%%%%%%%%%%%%%%

%%%%%%%%%%%%%%%%%%%%%%%%%%%%%%%%%%%%%%%%%%%%%%%%%%%
%%                        Figure 4
%%%%%%%%%%%%%%%%%%%%%%%%%%%%%%%%%%%%%%%%%%%%%%%%%%%
\begin{figure}
	\begin{center}
        \includegraphics[width=0.49\textwidth]{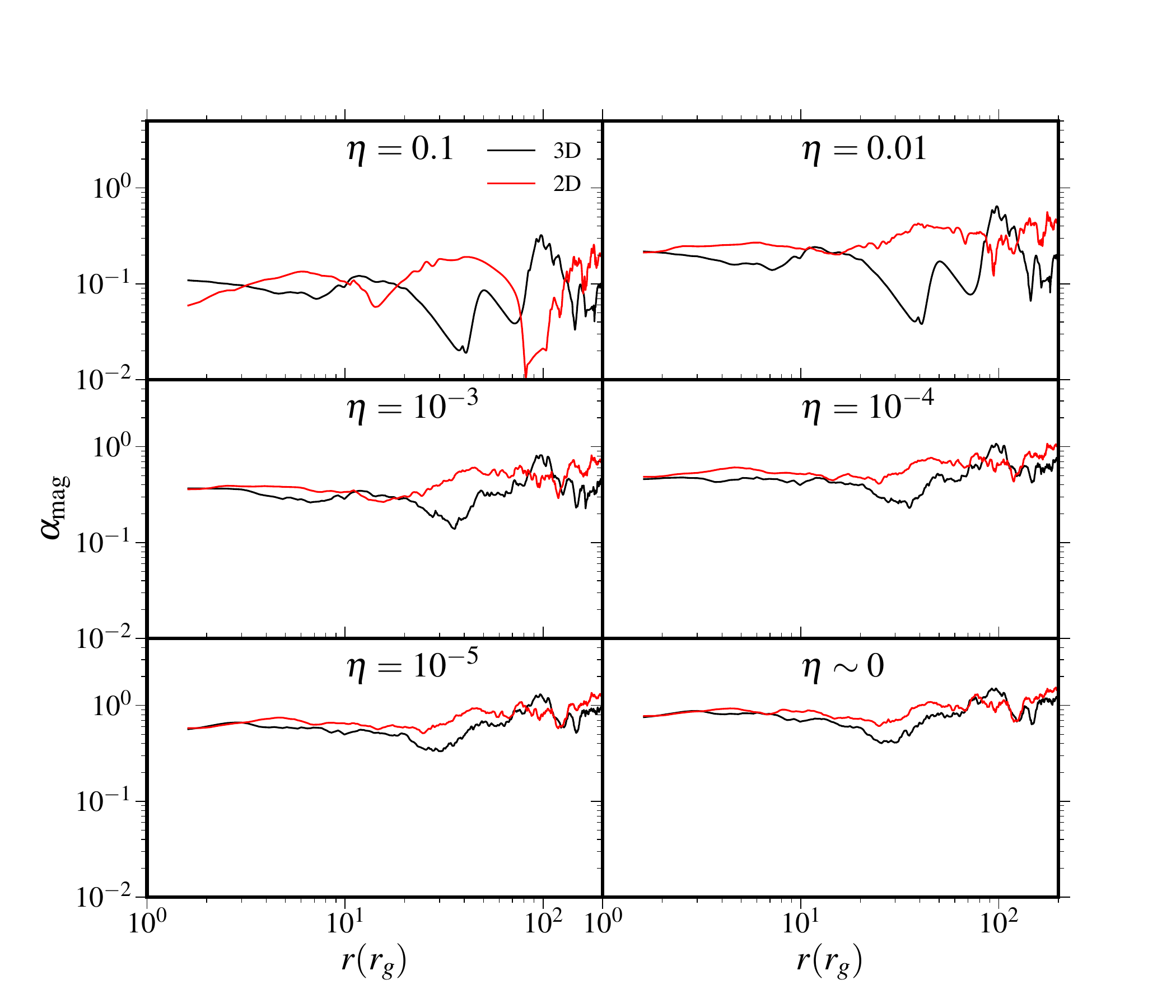} 
        \includegraphics[width=0.49\textwidth]{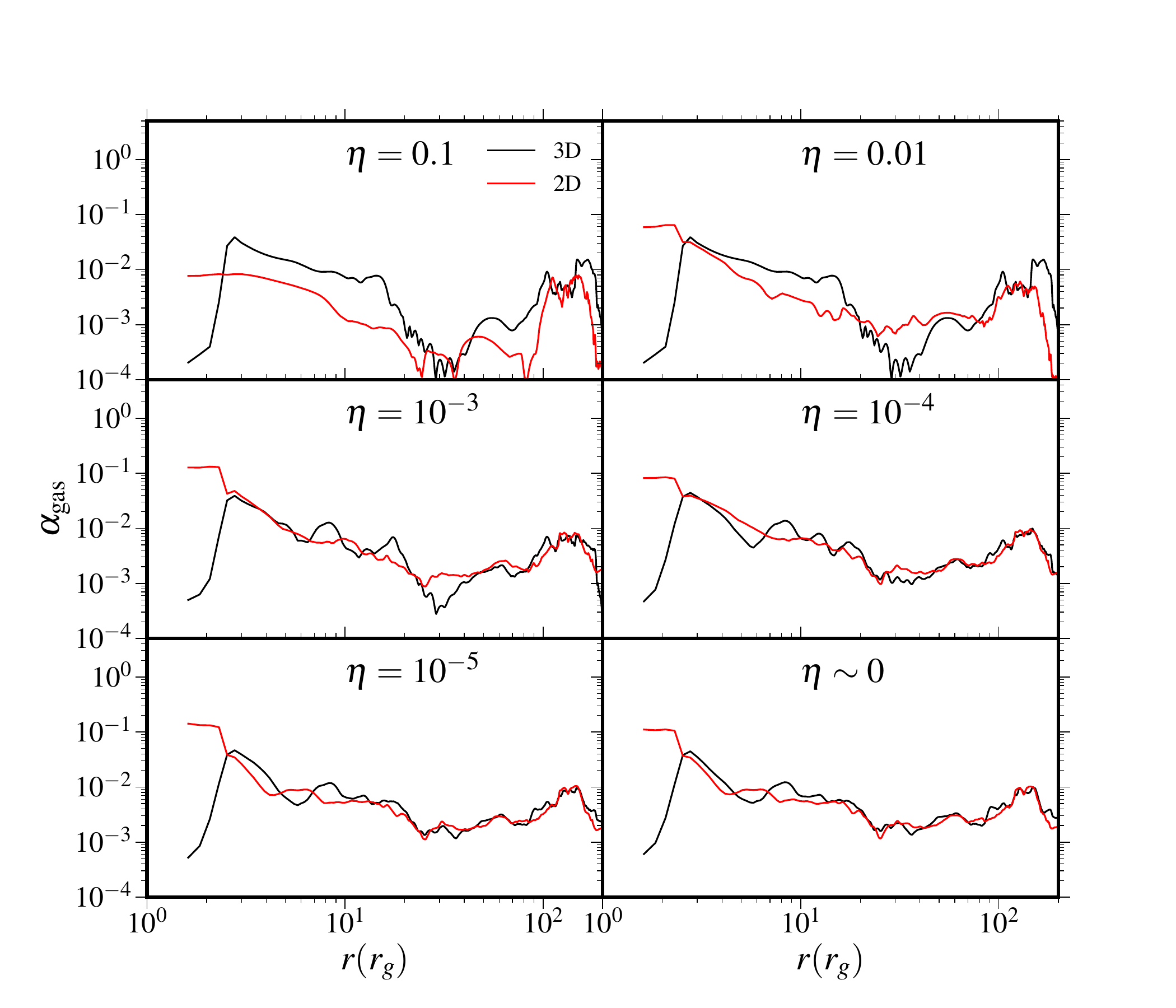} 
	\end{center}
	\caption{Radial variation of Left: Maxwell stress $(\alpha_{\rm mag})$ and Right: Reynold stress $(\alpha_{\rm gas})$ for various resistivity $\eta = 0.1, 0.01, 10^{-3}, 10^{-4}, 10^{-5}$, and $\sim 0$. Black and red curves are for 3D and 2D models, respectively.}
	\label{Figure_4}
\end{figure}
%%%%%%%%%%%%%%%%%%%%%%%%%%%%%%%%%%%%%%%%%%%%%%%%%%%%

%%%%%%%%%%%%%%%%%%%%%%%%%%%%%%%%%%%%%%%%%%%%%%%%%%%
%%                        Figure 5
%%%%%%%%%%%%%%%%%%%%%%%%%%%%%%%%%%%%%%%%%%%%%%%%%%%
\begin{figure*}
	\begin{center}
        \hskip -2.5mm
        \includegraphics[width=0.17\textwidth]{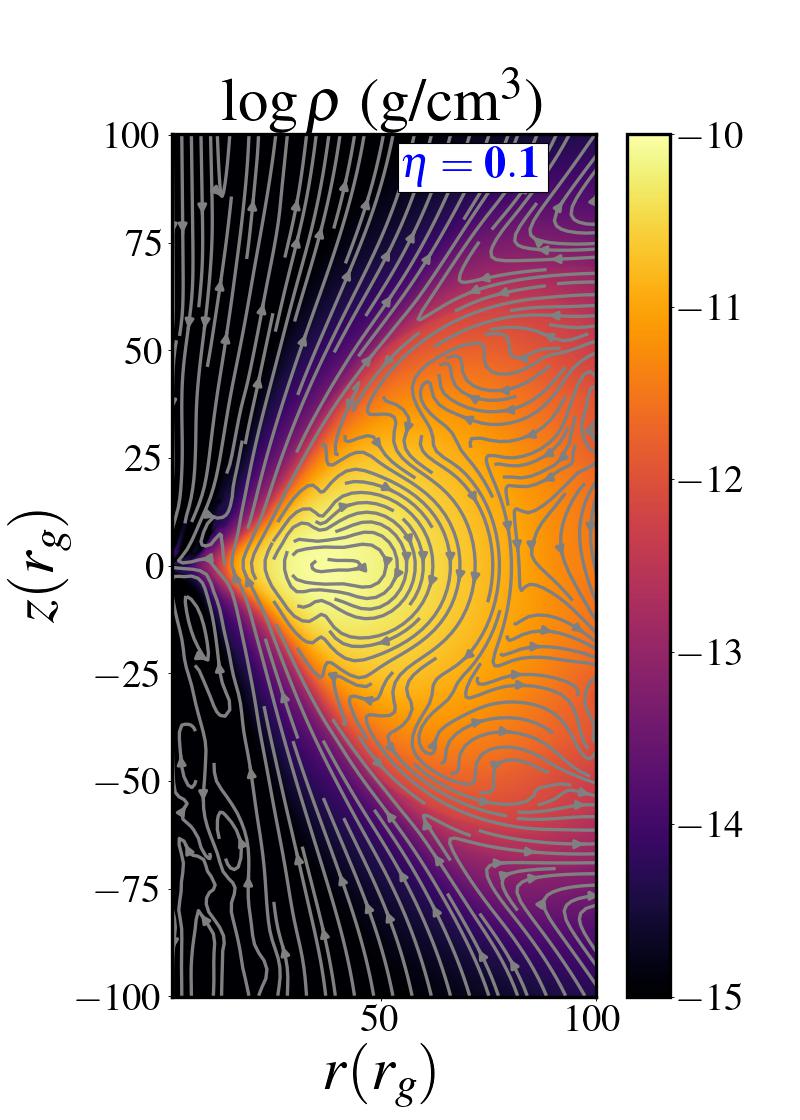} 
        \hskip -2.5mm
        \includegraphics[width=0.17\textwidth]{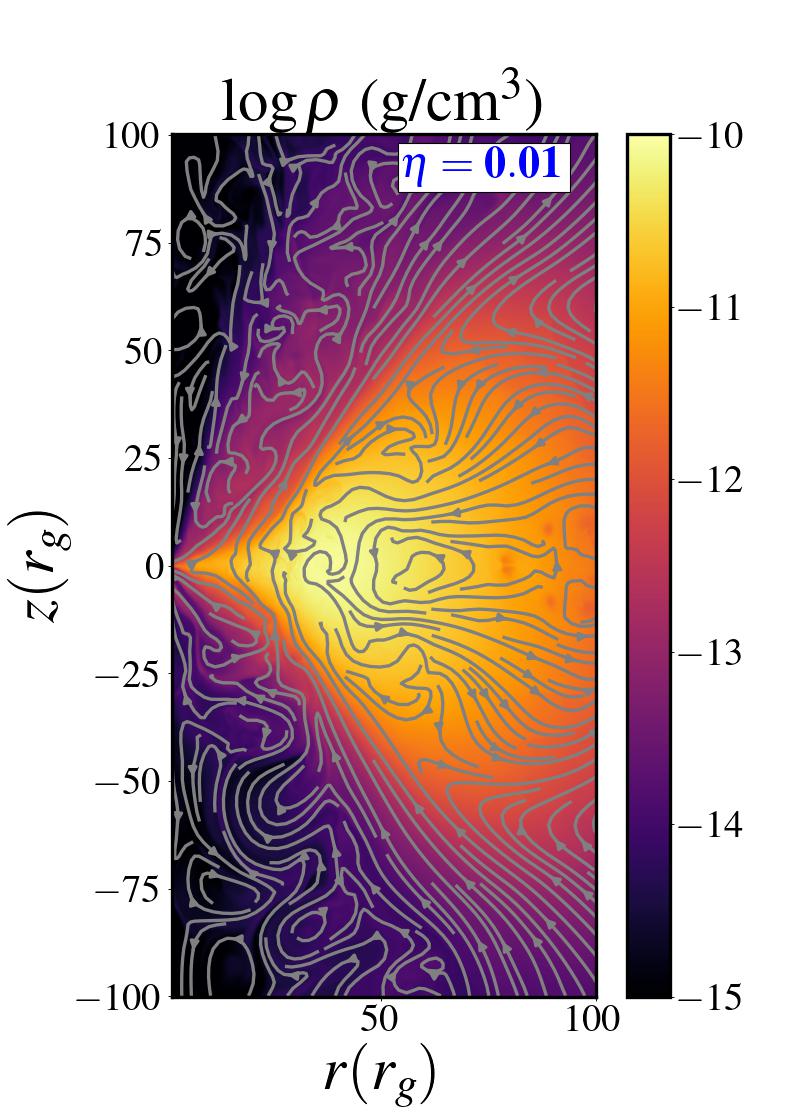} 
        \hskip -2.5mm
	\includegraphics[width=0.17\textwidth]{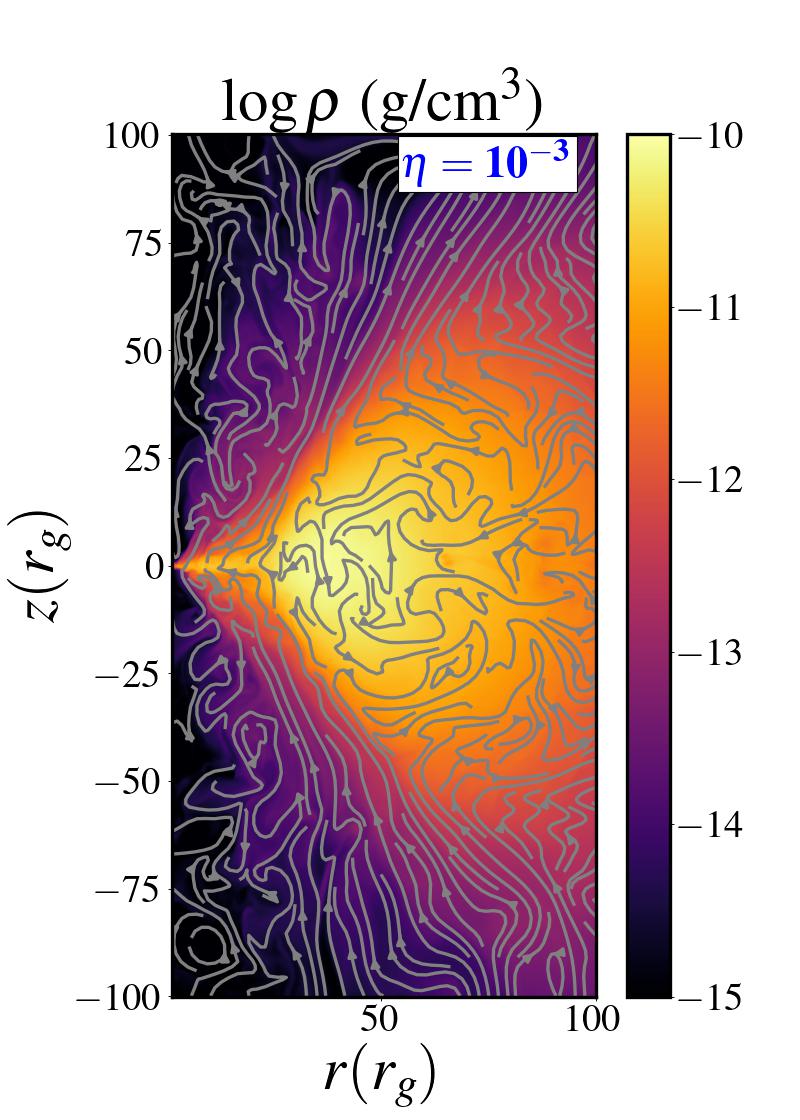} 
        \hskip -2.5mm
        \includegraphics[width=0.17\textwidth]{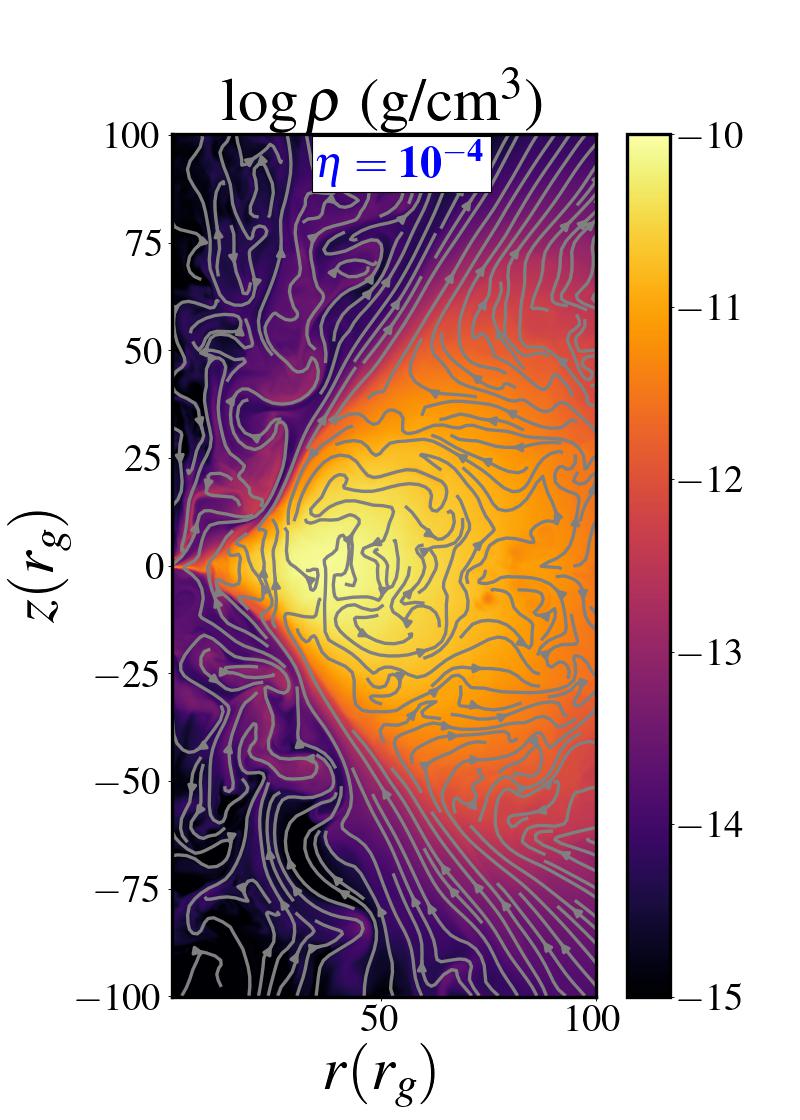} 
        \hskip -2.5mm
        \includegraphics[width=0.17\textwidth]{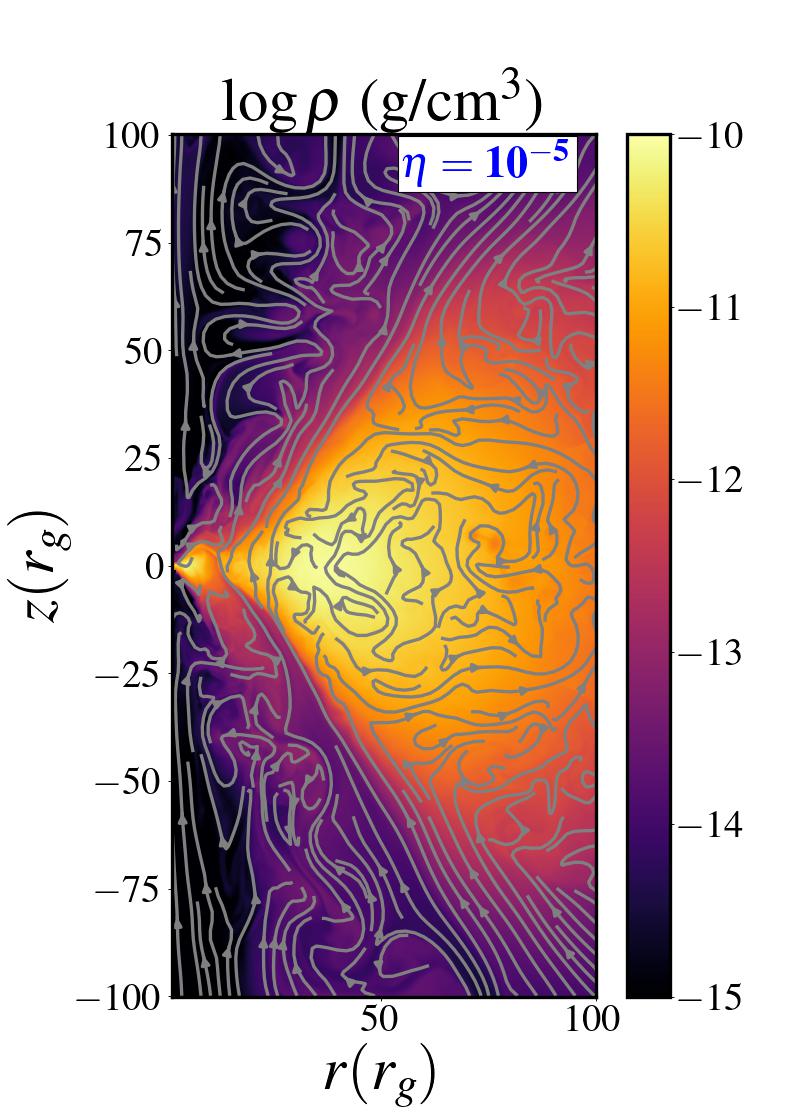} 
        \hskip -2.5mm
        \includegraphics[width=0.17\textwidth]{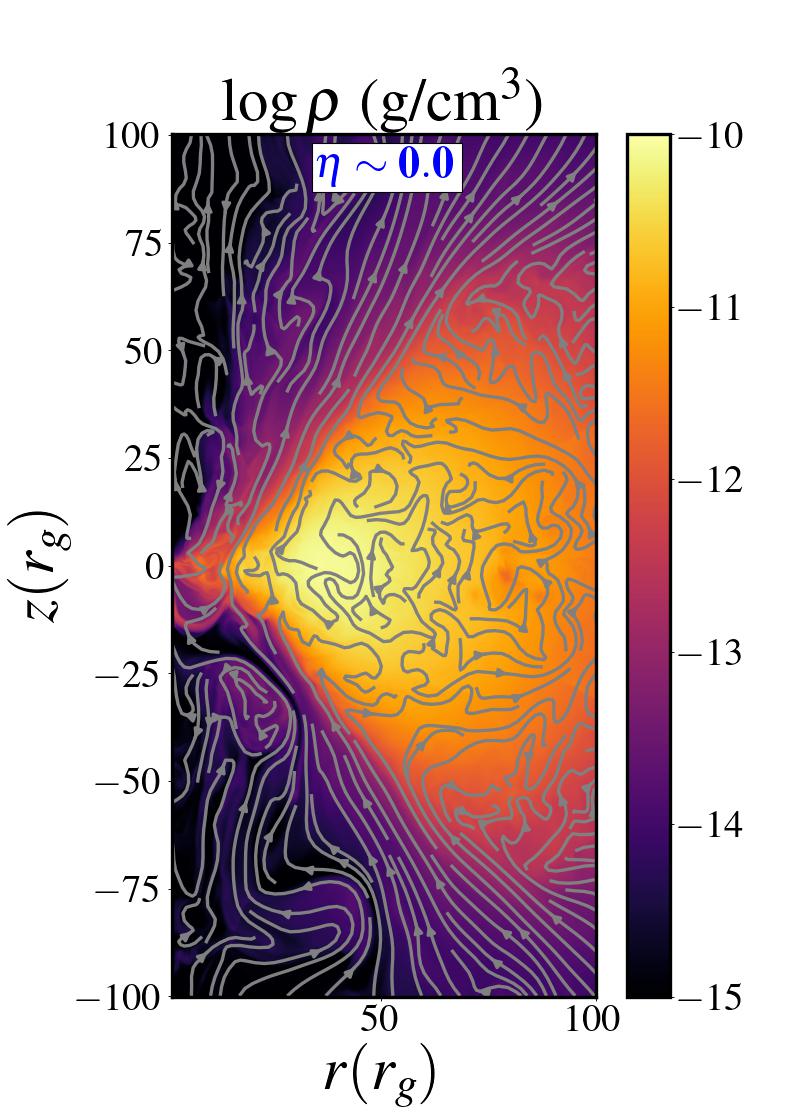} 

        \hskip -2.5mm
        \includegraphics[width=0.17\textwidth]{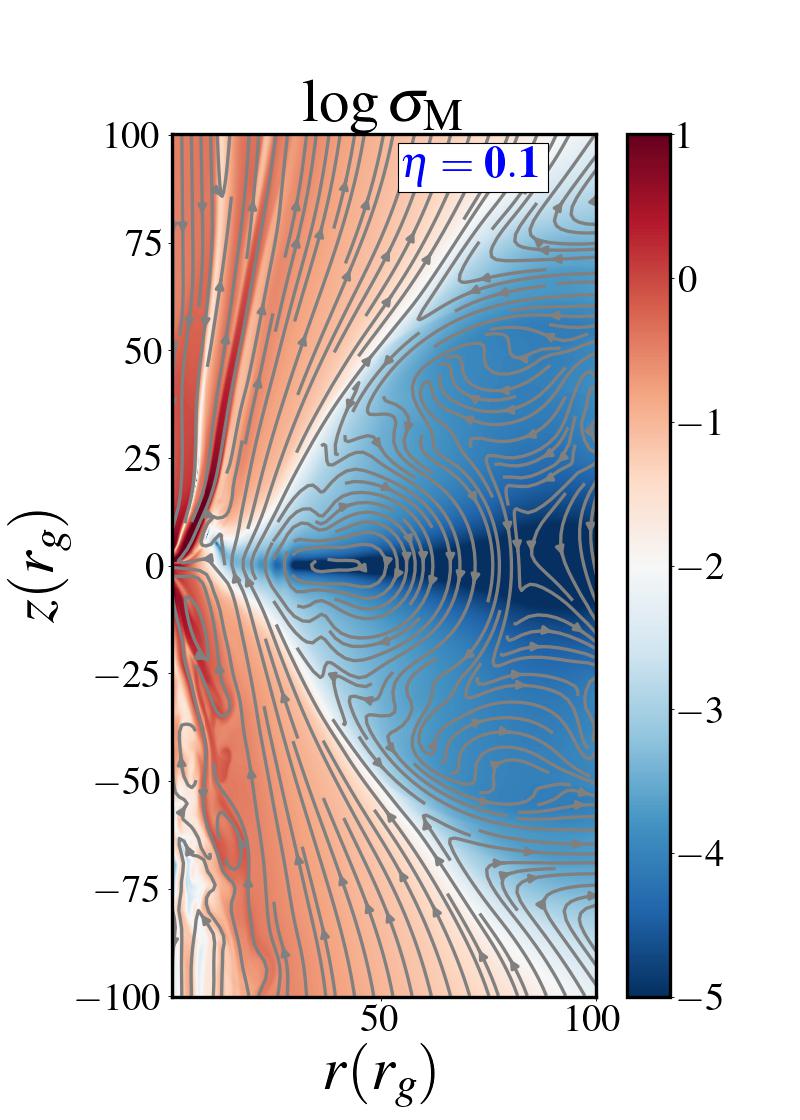} 
        \hskip -2.5mm
        \includegraphics[width=0.17\textwidth]{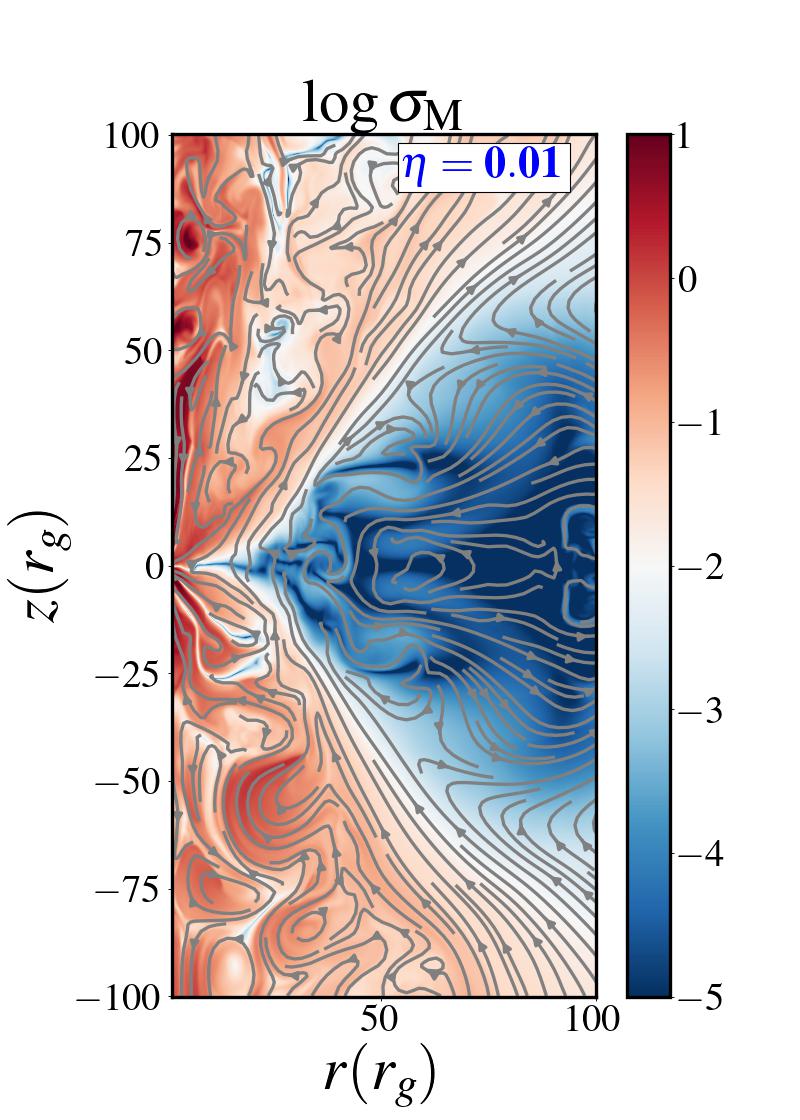} 
        \hskip -2.5mm
	\includegraphics[width=0.17\textwidth]{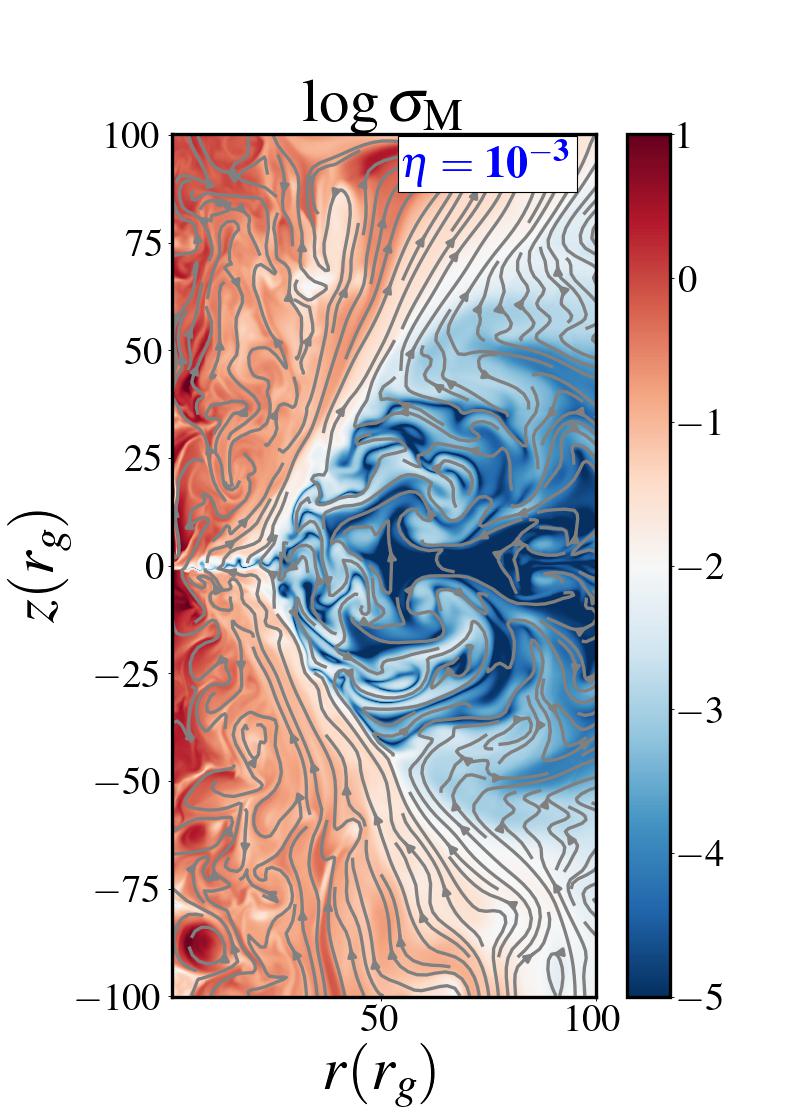} 
        \hskip -2.5mm
        \includegraphics[width=0.17\textwidth]{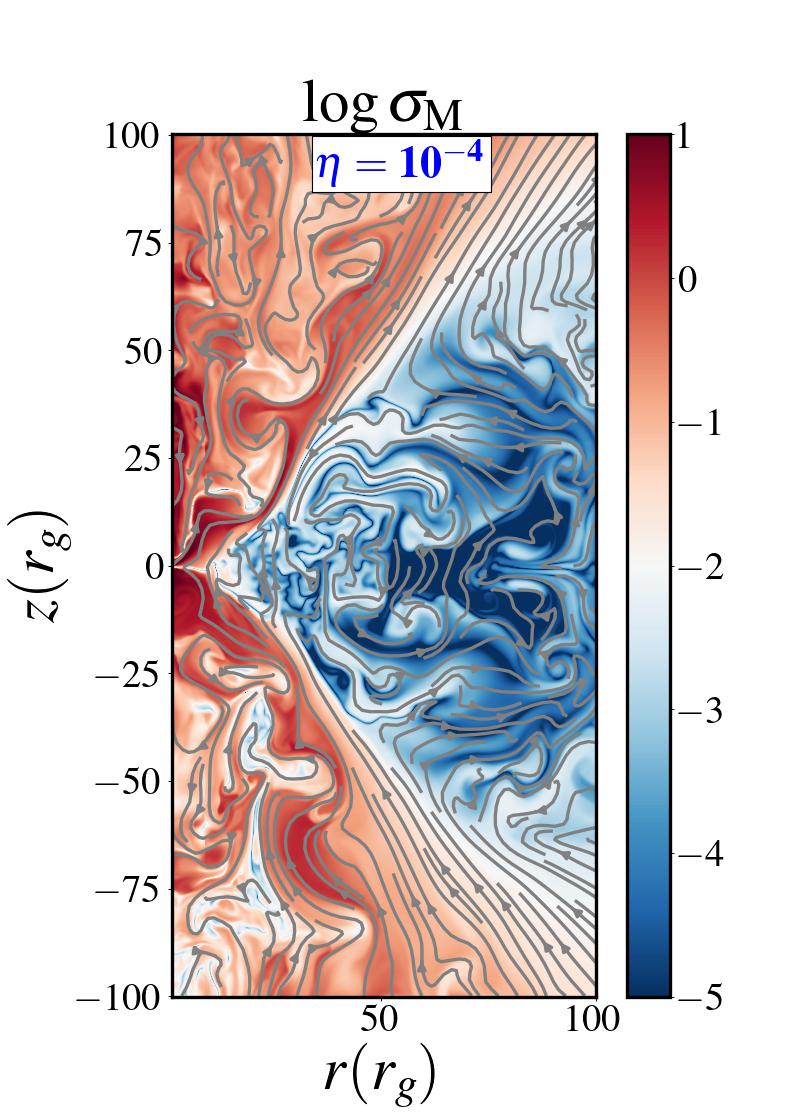} 
        \hskip -2.5mm
        \includegraphics[width=0.17\textwidth]{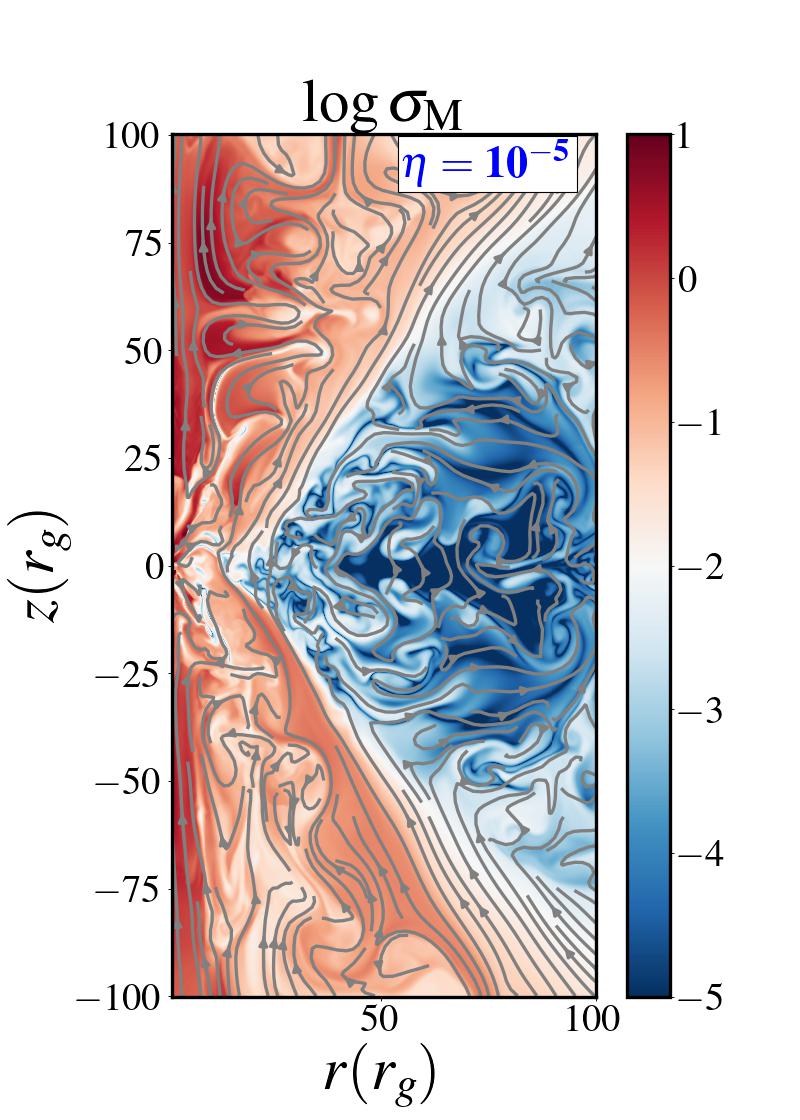} 
        \hskip -2.5mm
        \includegraphics[width=0.17\textwidth]{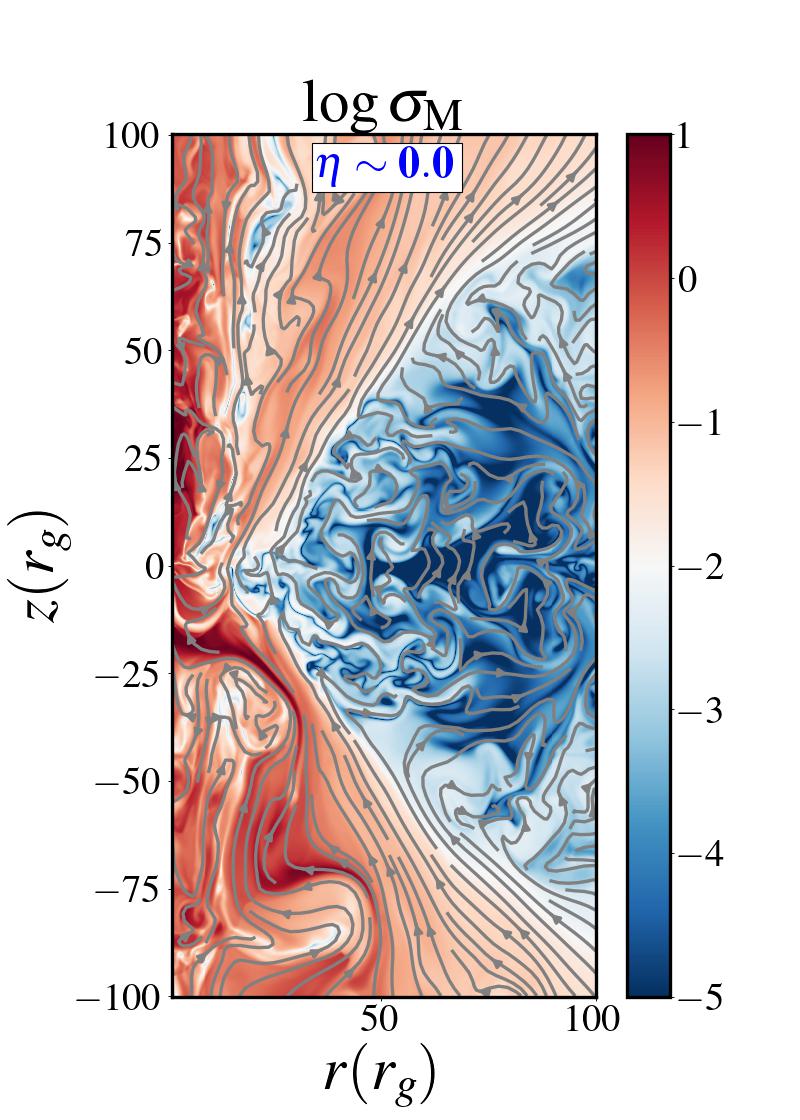} 
        
        \hskip -2.5mm       
        \includegraphics[width=0.17\textwidth]{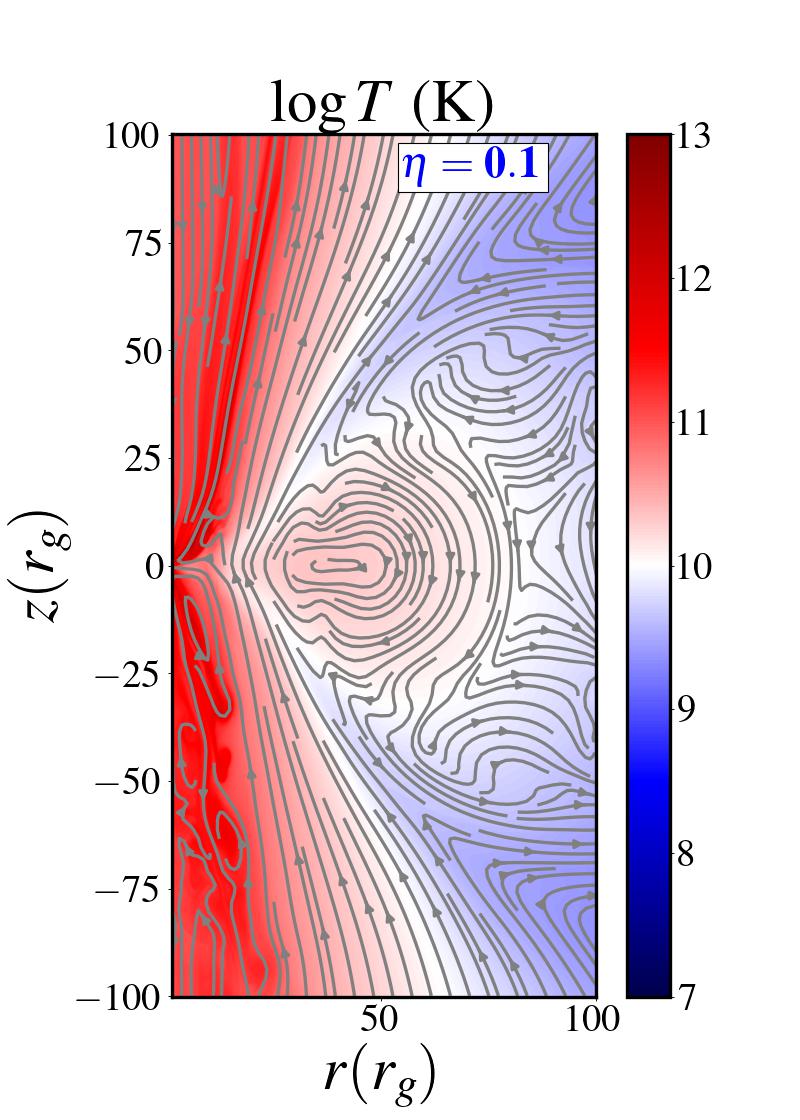} 
        \hskip -2.5mm
        \includegraphics[width=0.17\textwidth]{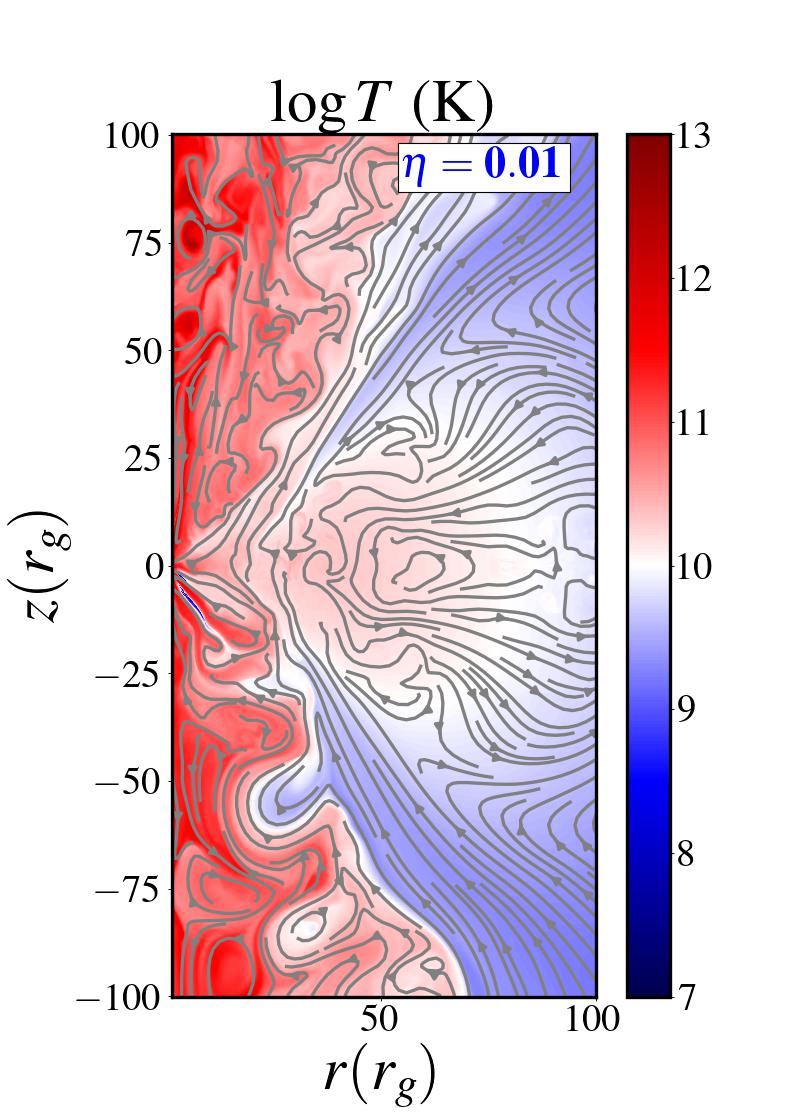} 
        \hskip -2.5mm
	\includegraphics[width=0.17\textwidth]{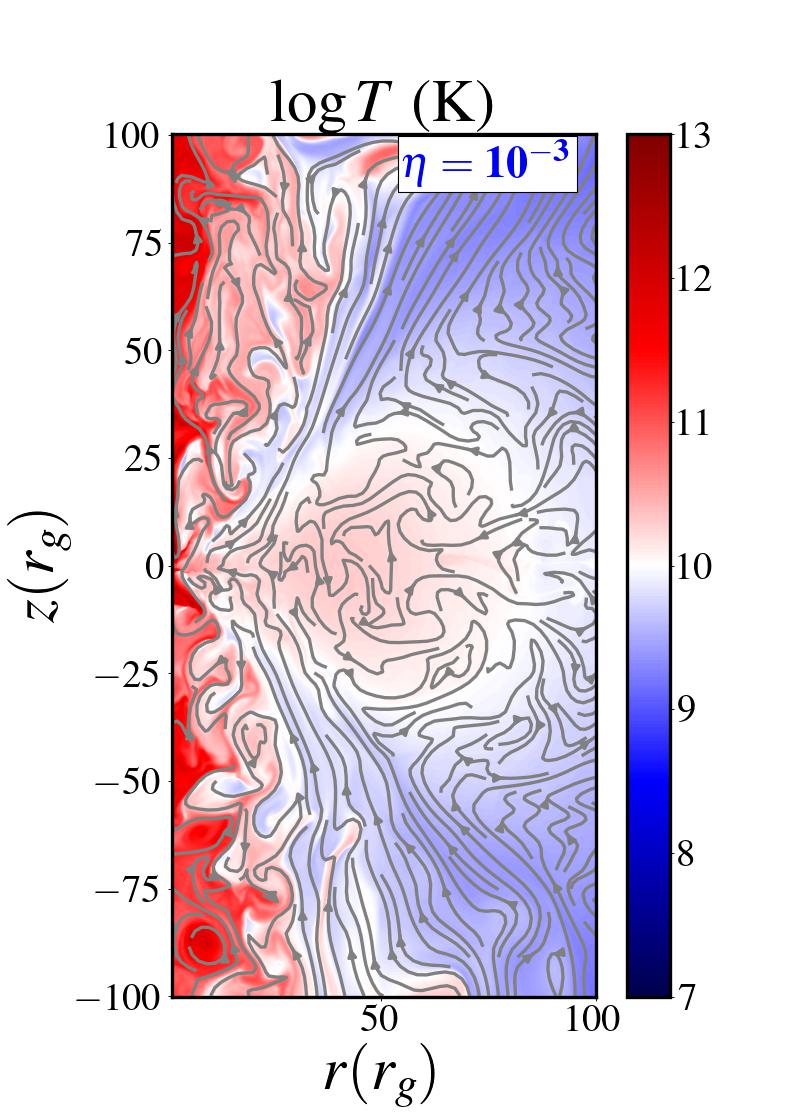} 
        \hskip -2.5mm
        \includegraphics[width=0.17\textwidth]{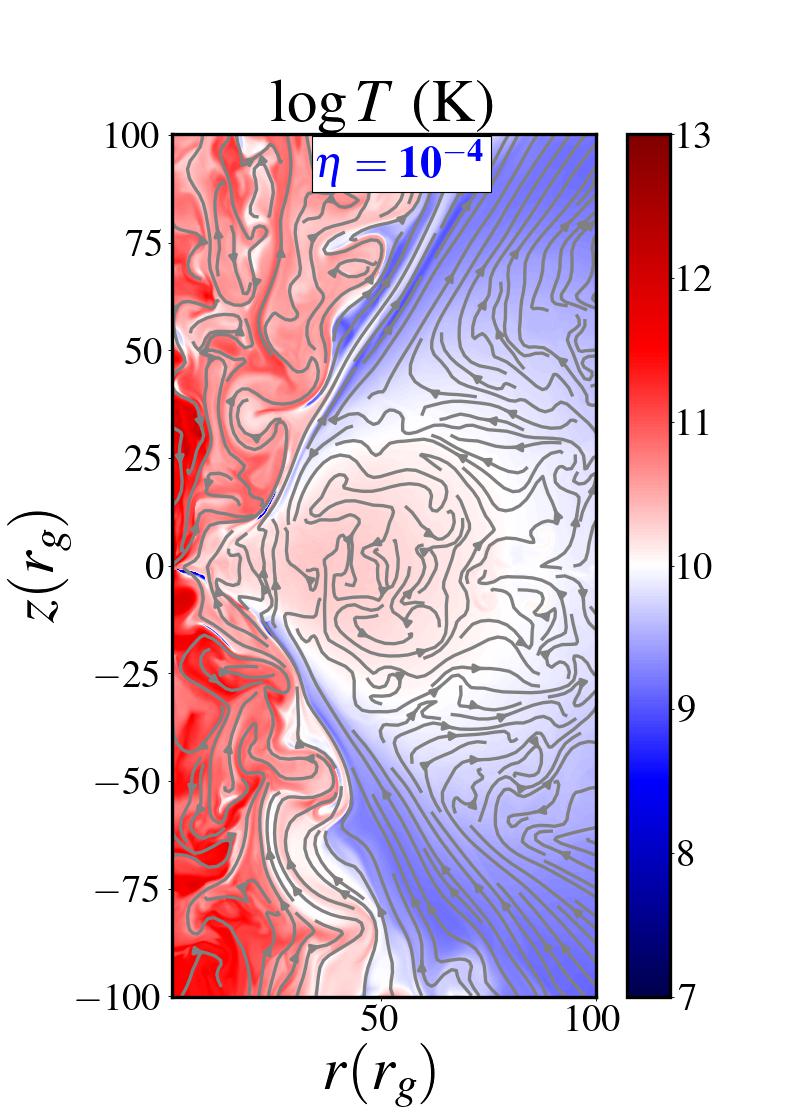} 
        \hskip -2.5mm
        \includegraphics[width=0.17\textwidth]{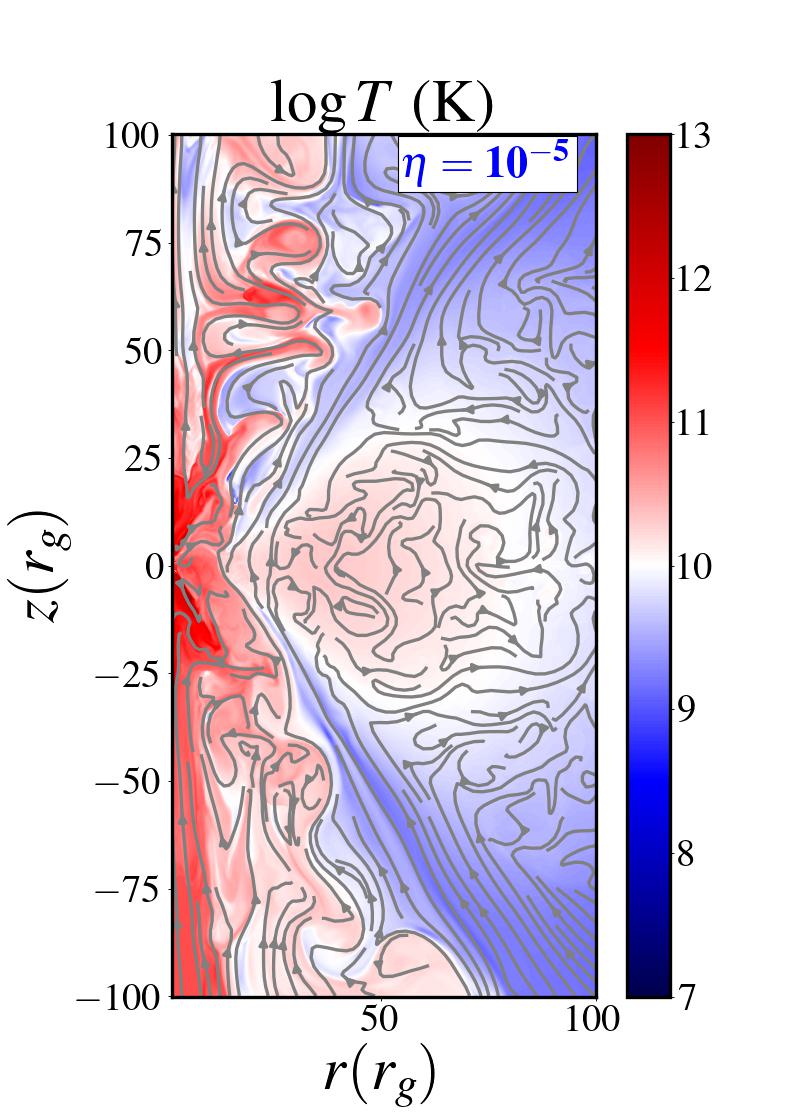} 
        \hskip -2.5mm
        \includegraphics[width=0.17\textwidth]{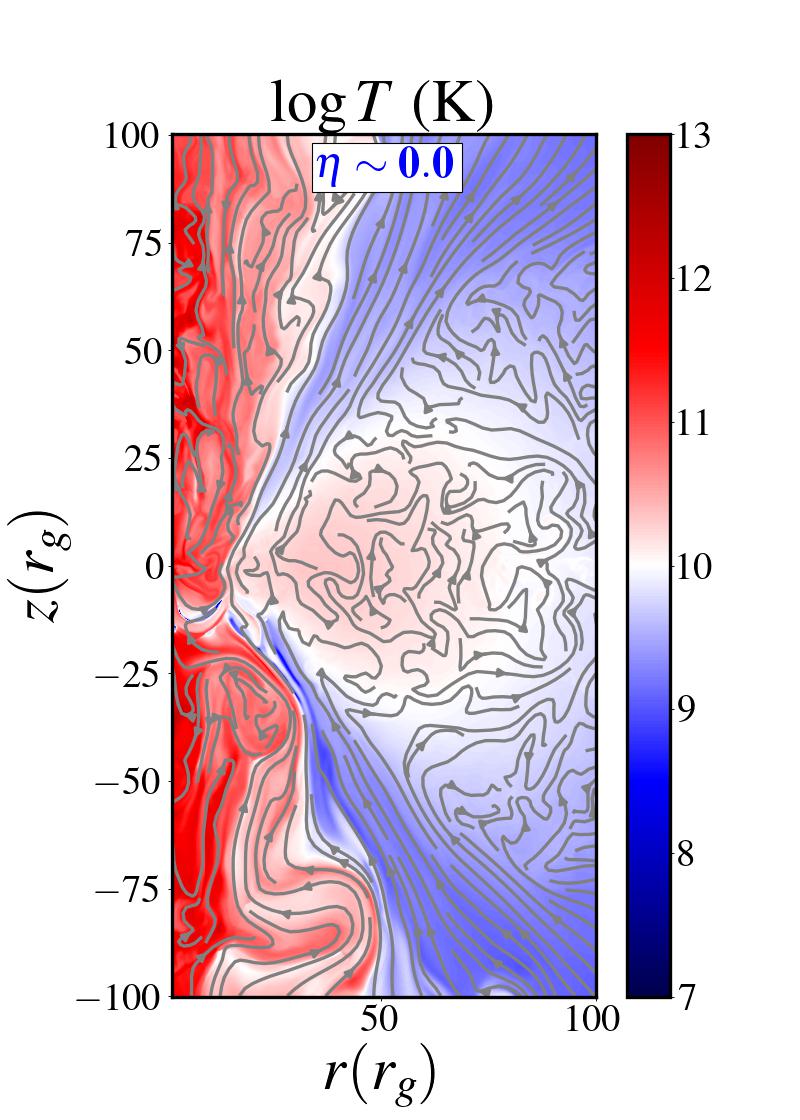} 

        \hskip -2.5mm 
        \includegraphics[width=0.17\textwidth]{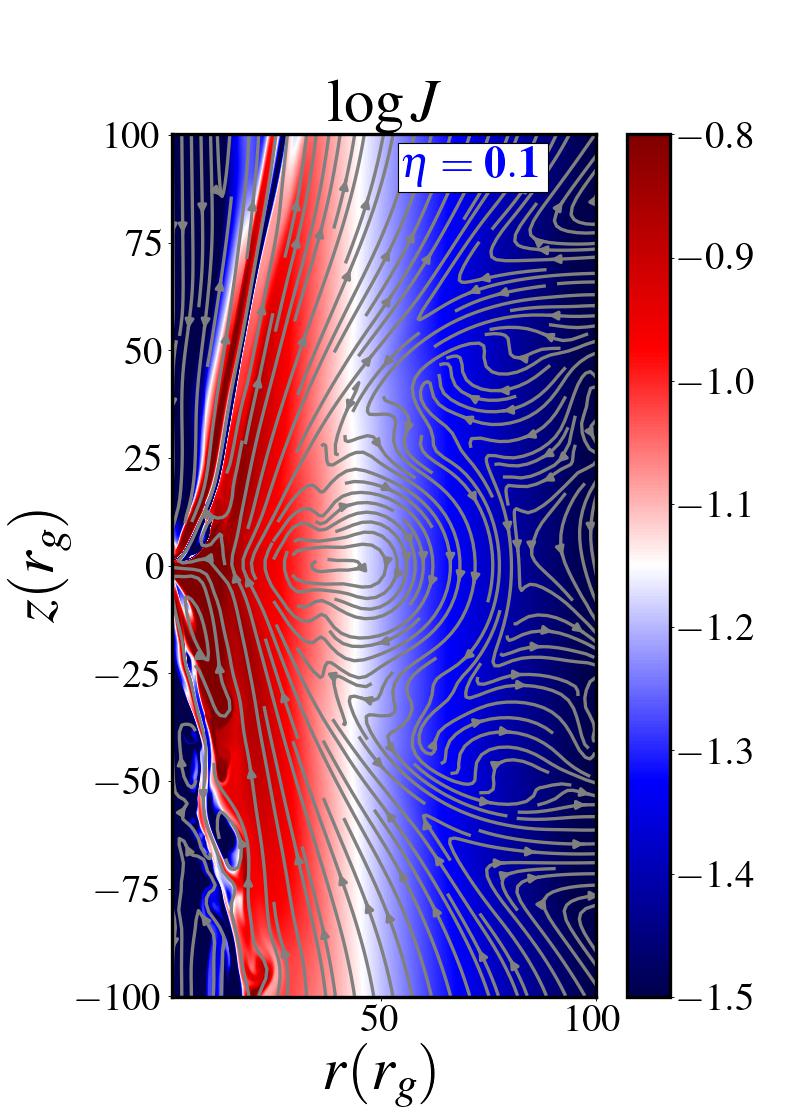} 
        \hskip -2.5mm
        \includegraphics[width=0.17\textwidth]{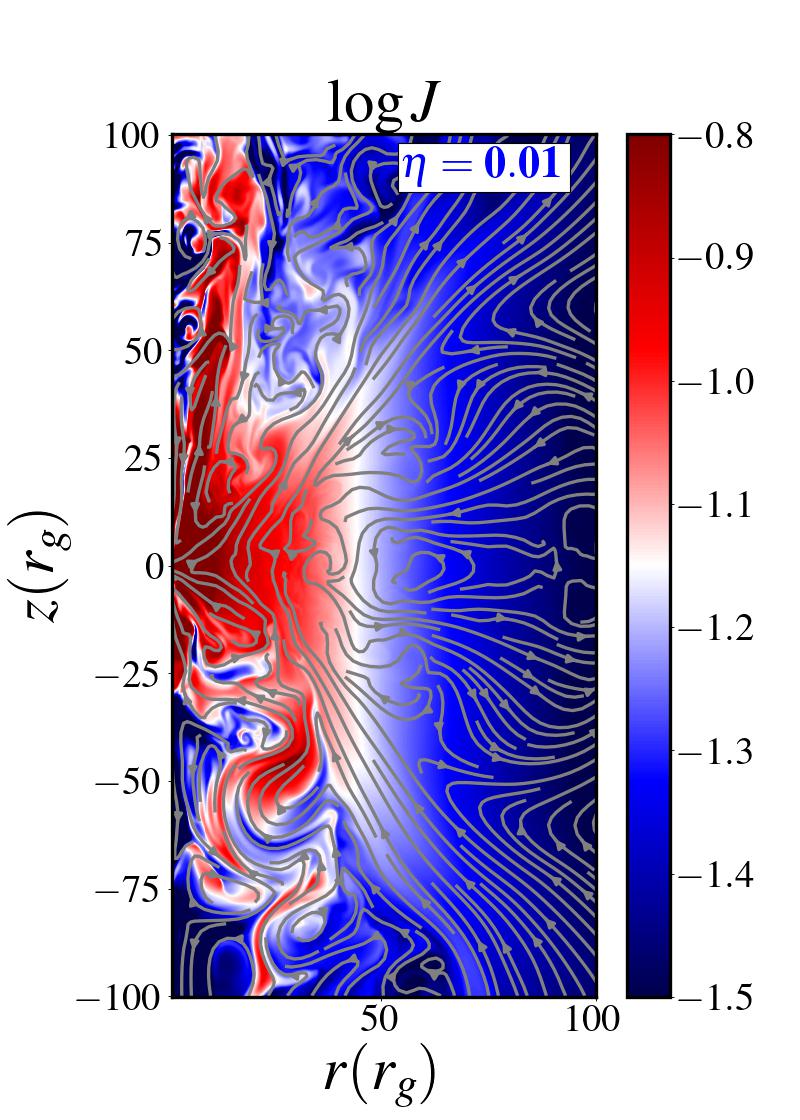} 
        \hskip -2.5mm
	\includegraphics[width=0.17\textwidth]{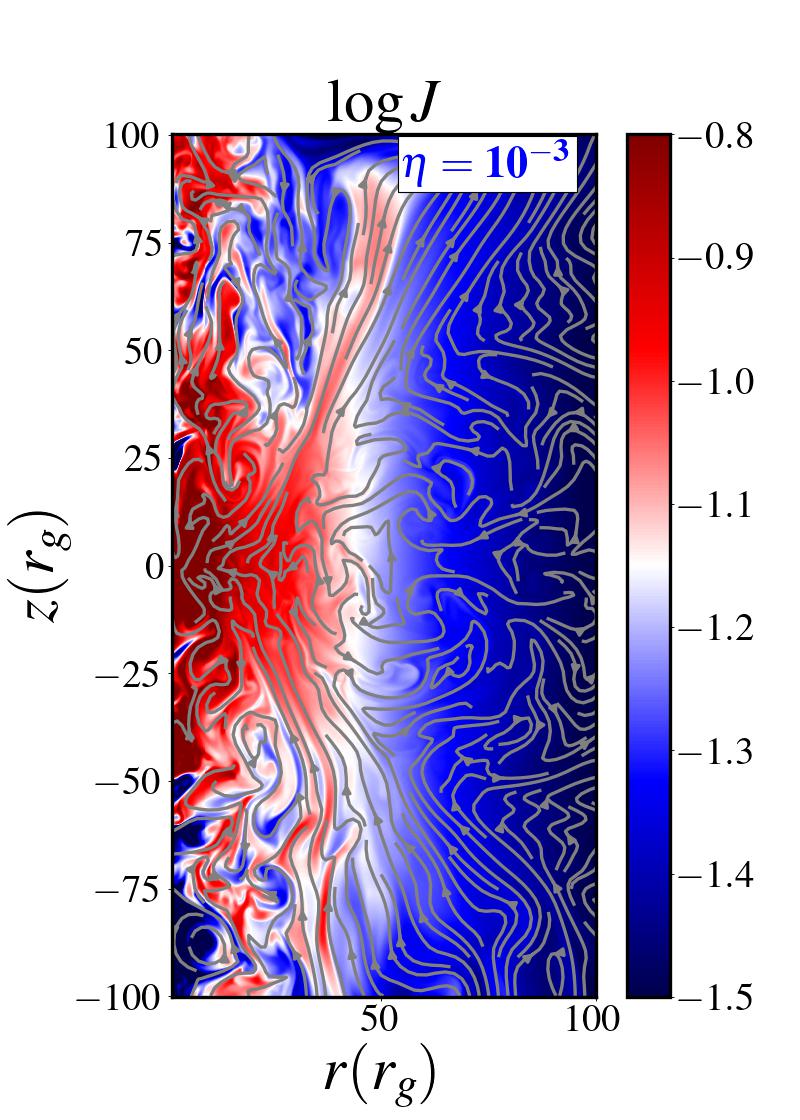} 
        \hskip -2.5mm
        \includegraphics[width=0.17\textwidth]{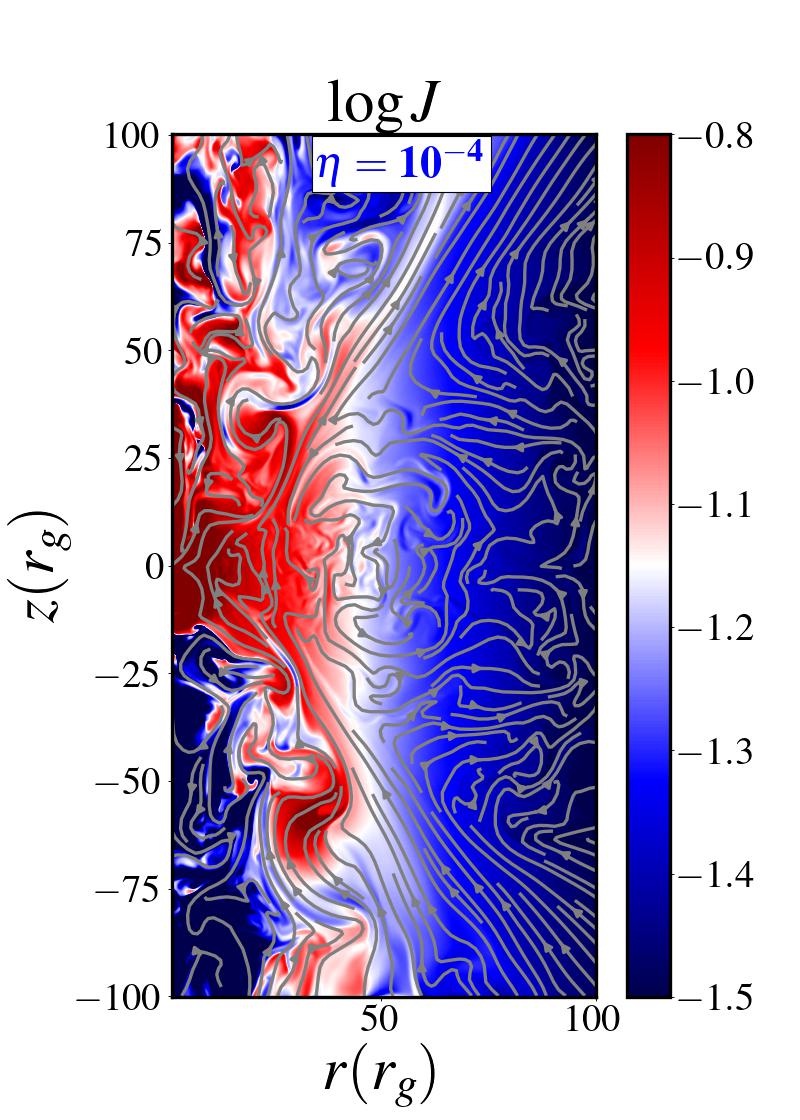} 
        \hskip -2.5mm
        \includegraphics[width=0.17\textwidth]{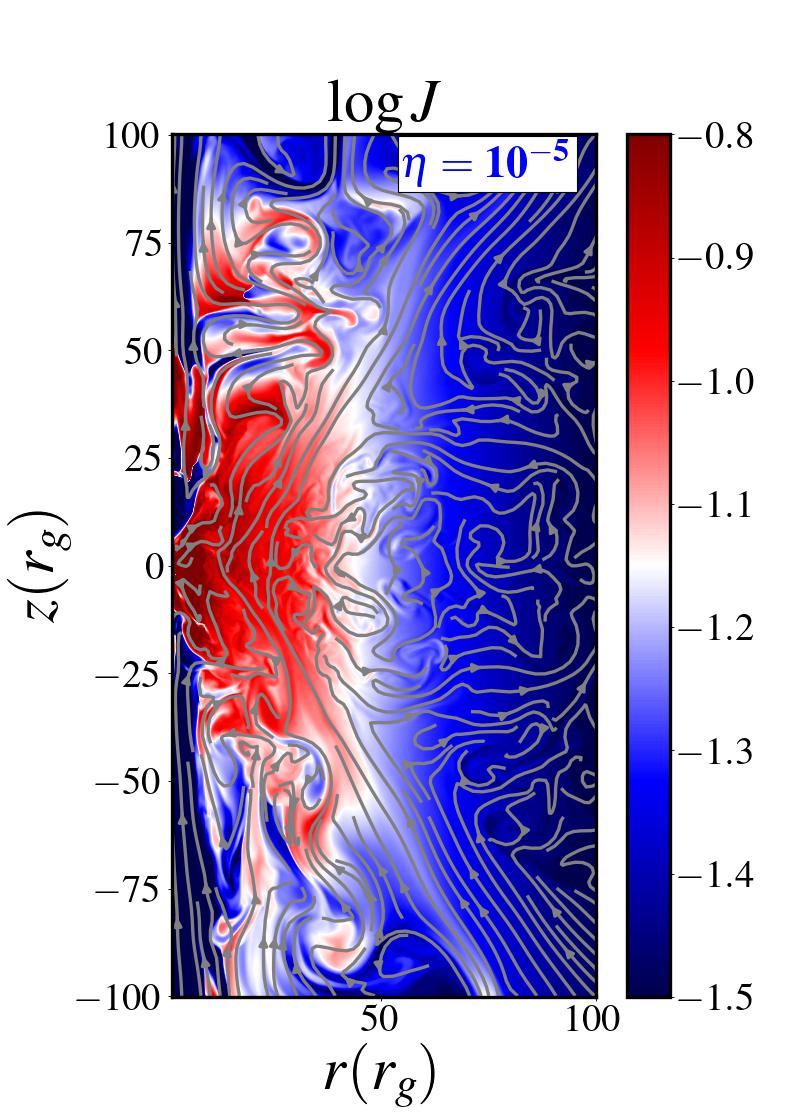} 
        \hskip -2.5mm
        \includegraphics[width=0.17\textwidth]{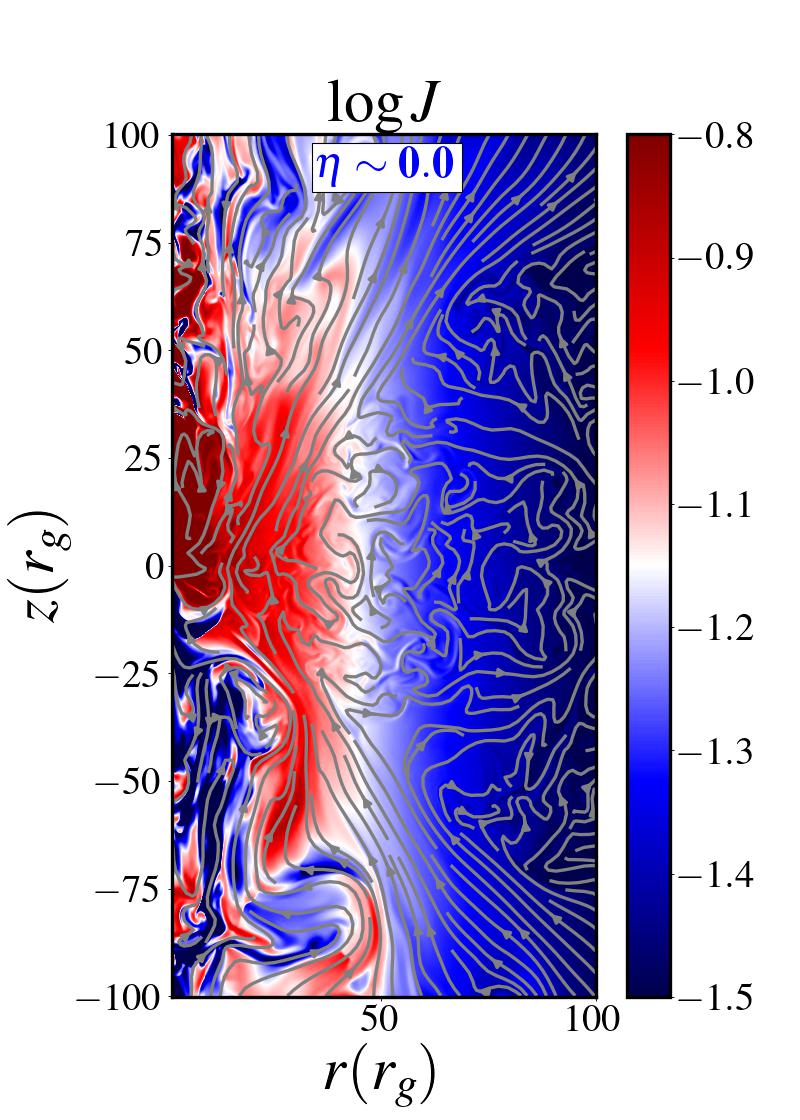} 
	\end{center}
	\caption{2D Model: Distribution of gas density $(\rho)$, magnetization parameter $(\sigma_{\rm M})$, temperature ($T$) and current density ($J$) for 2D model in first, second, third and fourth row, respectively. The grey lines represent the magnetic field lines. Here, we fix the different resistivity as $\eta = 0.1, 0.01, 10^{-3}, 10^{-4}, 10^{-5}$, and $\sim 0$. See the text for details.}
	\label{Figure_5}
\end{figure*}
%%%%%%%%%%%%%%%%%%%%%%%%%%%%%%%%%%%%%%%%%%%%%%%%%%%%

%%%%%%%%%%%%%%%%%%%%%%%%%%%%%%%%%%%%%%%%%%%%%%%%%%%
%%                        Figure 5
%%%%%%%%%%%%%%%%%%%%%%%%%%%%%%%%%%%%%%%%%%%%%%%%%%%
\begin{figure*}
	\begin{center}
        \hskip -3.0mm
        \includegraphics[width=0.32\textwidth]{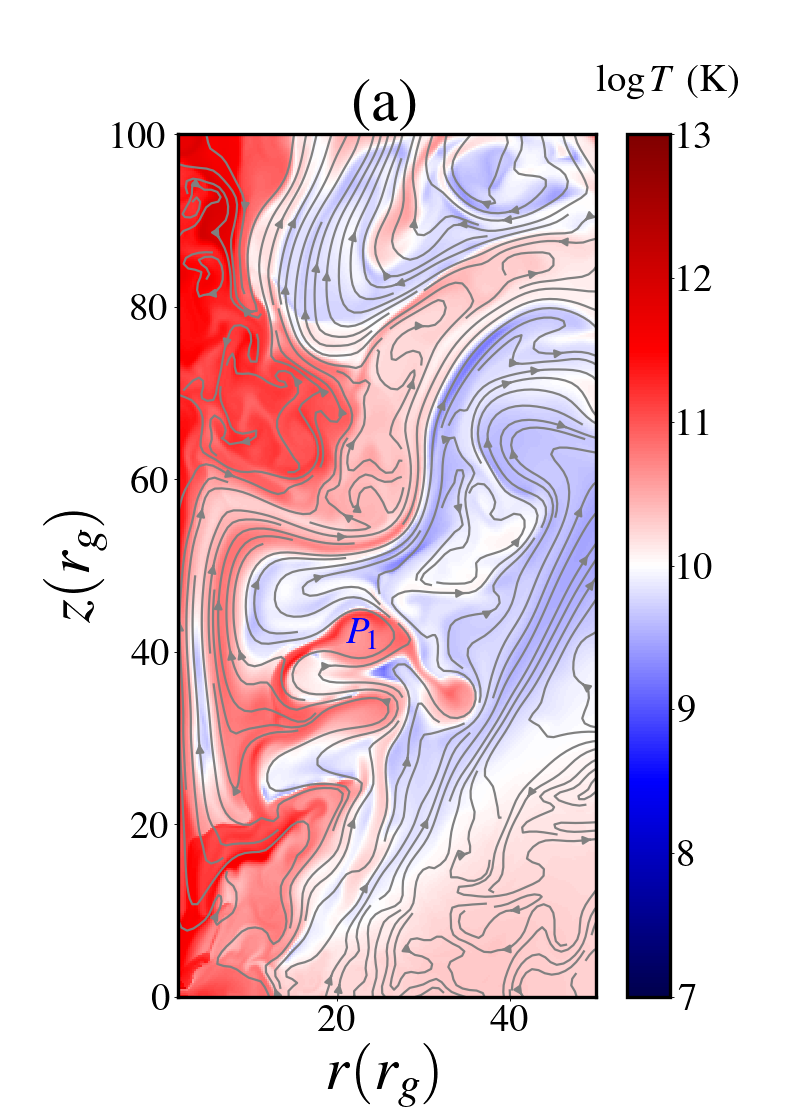} 
        \hskip -3.0mm
        \includegraphics[width=0.32\textwidth]{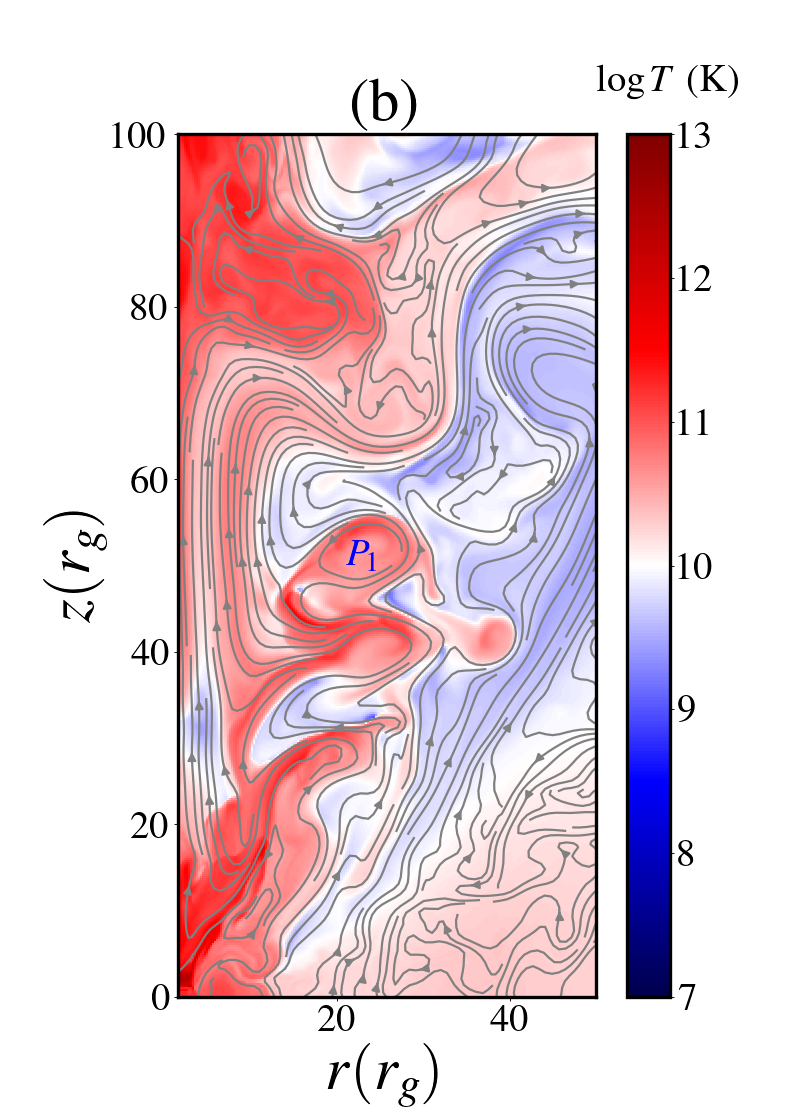} 
        \hskip -3.0mm
	\includegraphics[width=0.32\textwidth]{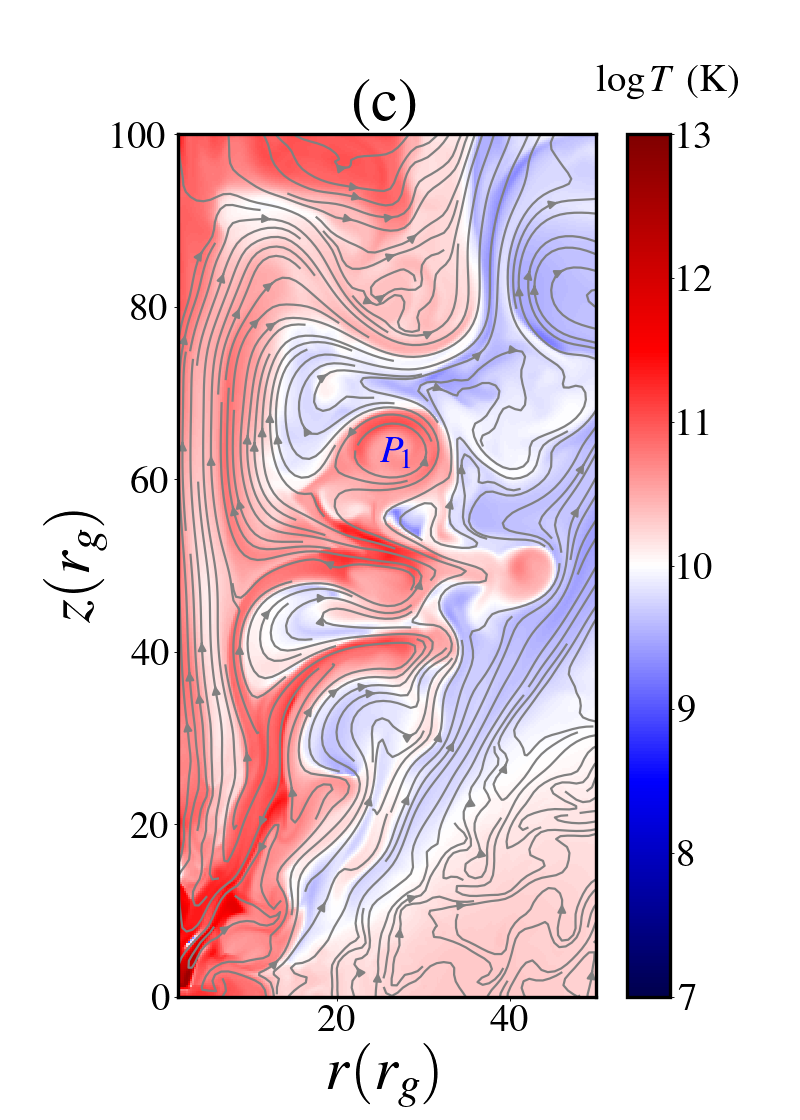} 
        \hskip -3.0mm
        \includegraphics[width=0.32\textwidth]{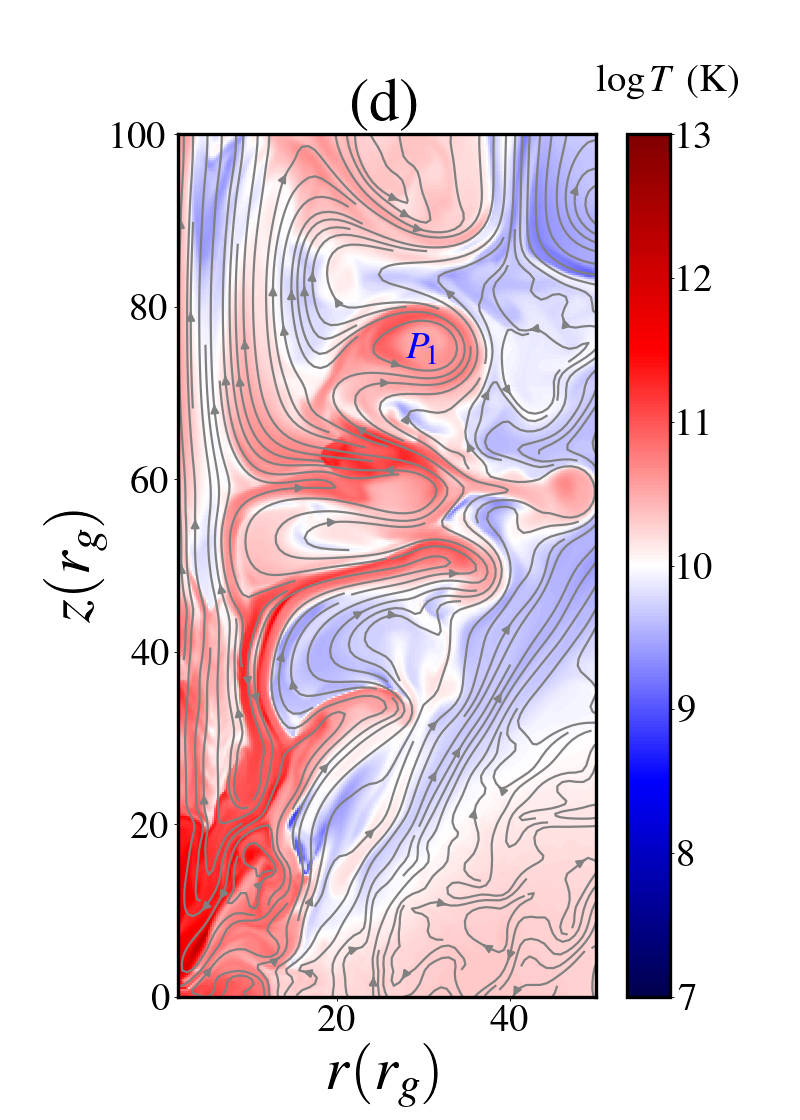} 
        \hskip -3.0mm
        \includegraphics[width=0.32\textwidth]{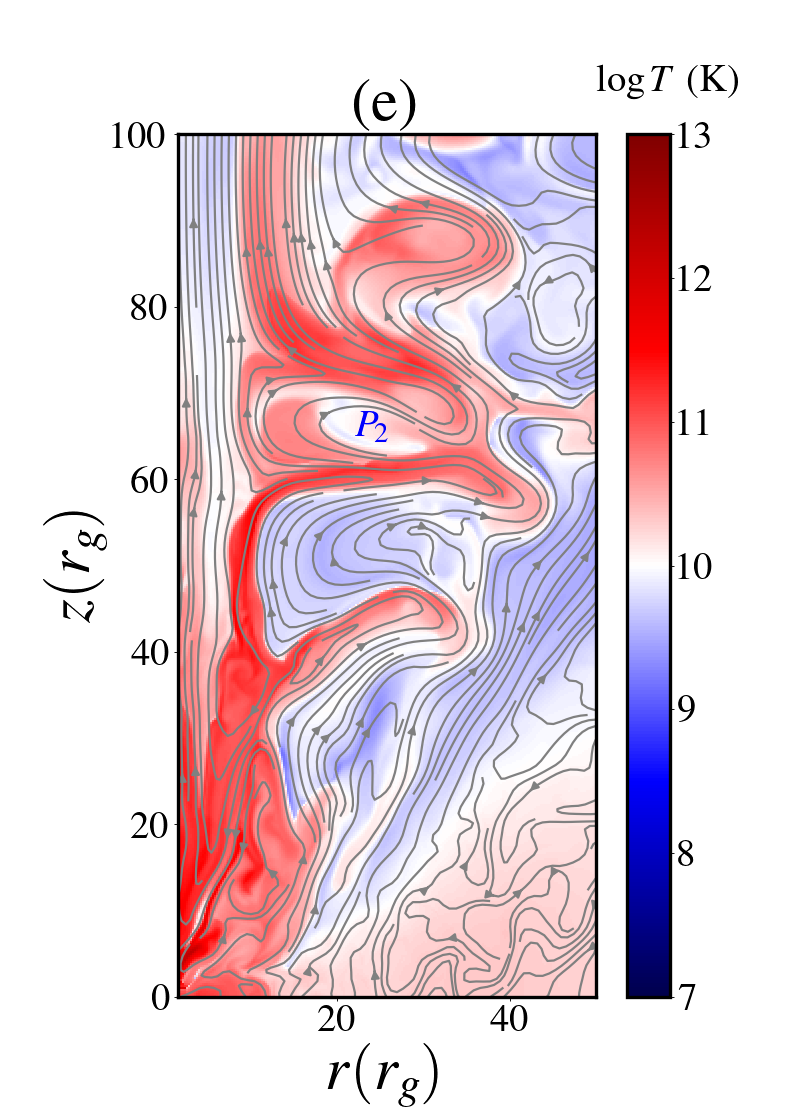} 
        \hskip -3.0mm
        \includegraphics[width=0.32\textwidth]{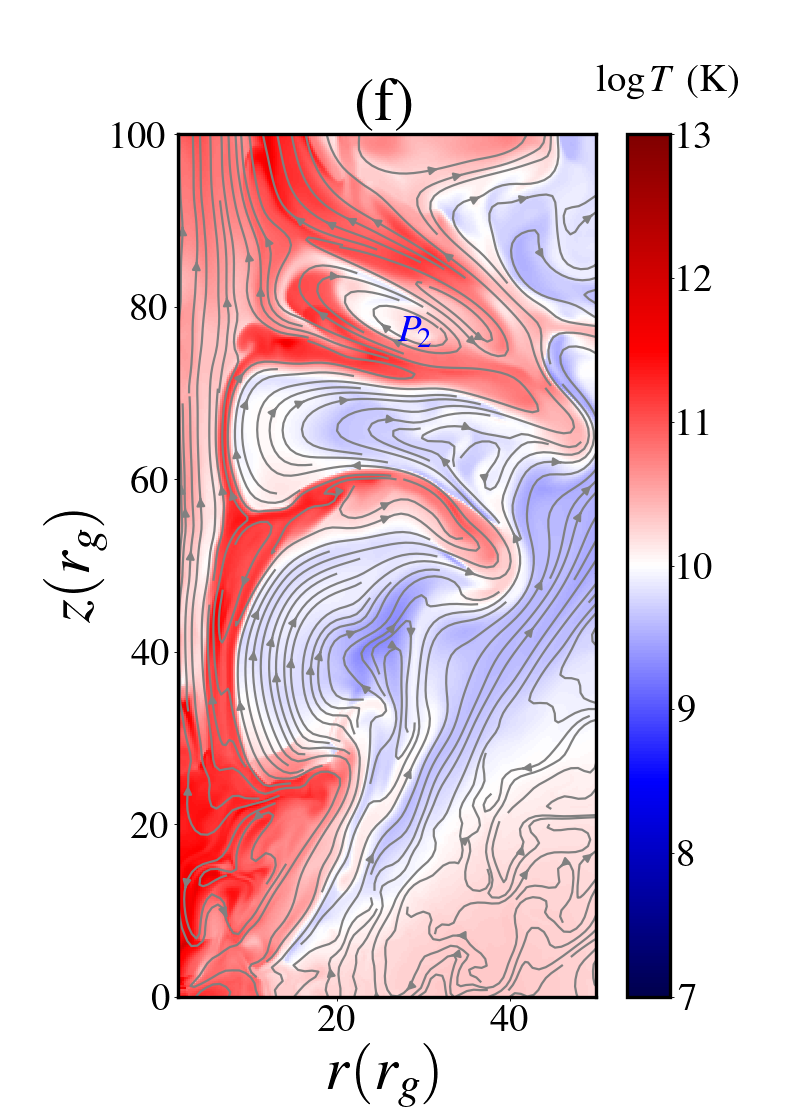} 
        
	\end{center}
	\caption{ The plasmoid formation is shown for various time slices in a 2D model with resistivity of \(\eta = 10^{-5}\). The specific time slices are as follows: (a) \(t = 8200~t_g\), (b) \(t = 8300~t_g\), (c) \(t = 8400~t_g\), (d) \(t = 8500~t_g\), (e) \(t = 8600~t_g\), and (f) \(t = 8700~t_g\). In the diagram, \(P_1\) and \(P_2\) denote the plasmoids. See the text for details.}
	\label{Figure_51}
\end{figure*}
%%%%%%%%%%%%%%%%%%%%%%%%%%%%%%%%%%%%%%%%%%%%%%%%%%%%

%%%%%%%%%%%%%%%%%%%%%%%%%%%%%%%%%%%%%%%%%%%%%%%%%%%
%%                        Figure 6
%%%%%%%%%%%%%%%%%%%%%%%%%%%%%%%%%%%%%%%%%%%%%%%%%%%
\begin{figure}
	\begin{center}
        \includegraphics[width=0.45\textwidth]{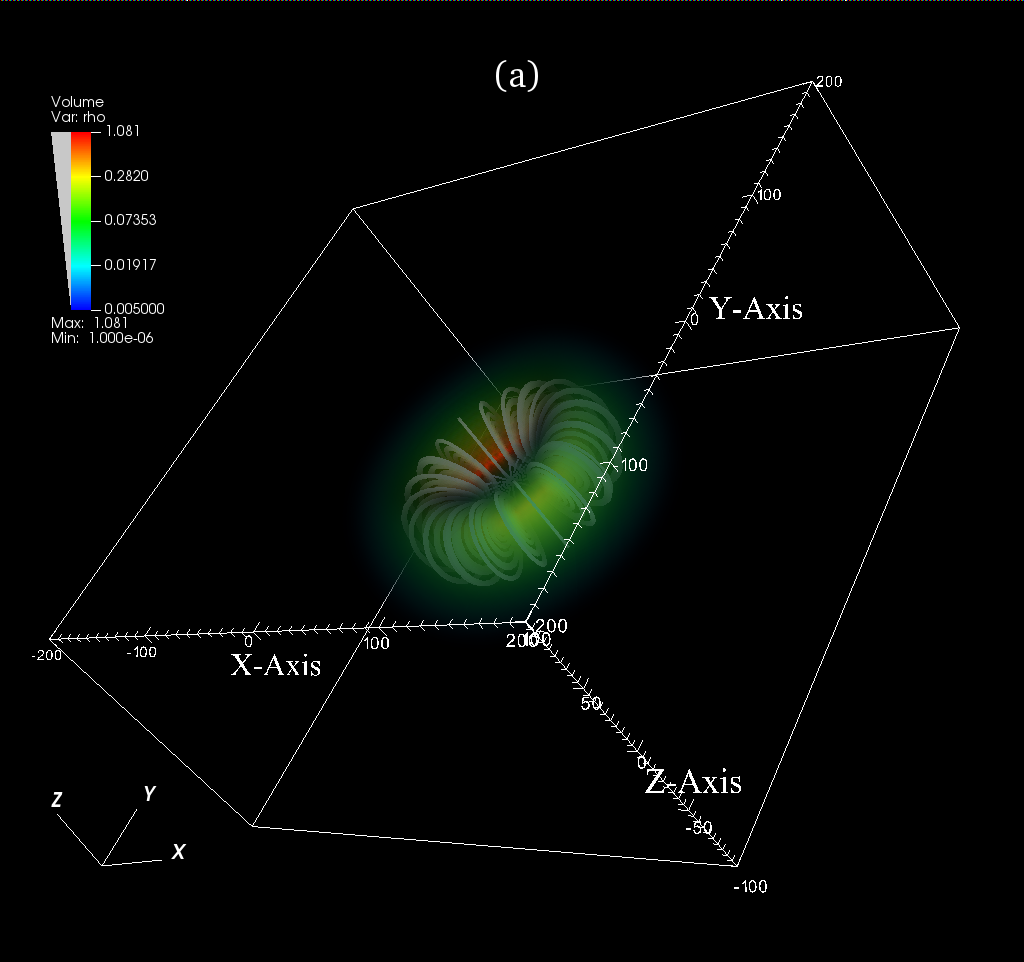} 
          \hskip 5.0mm
	\includegraphics[width=0.45\textwidth]{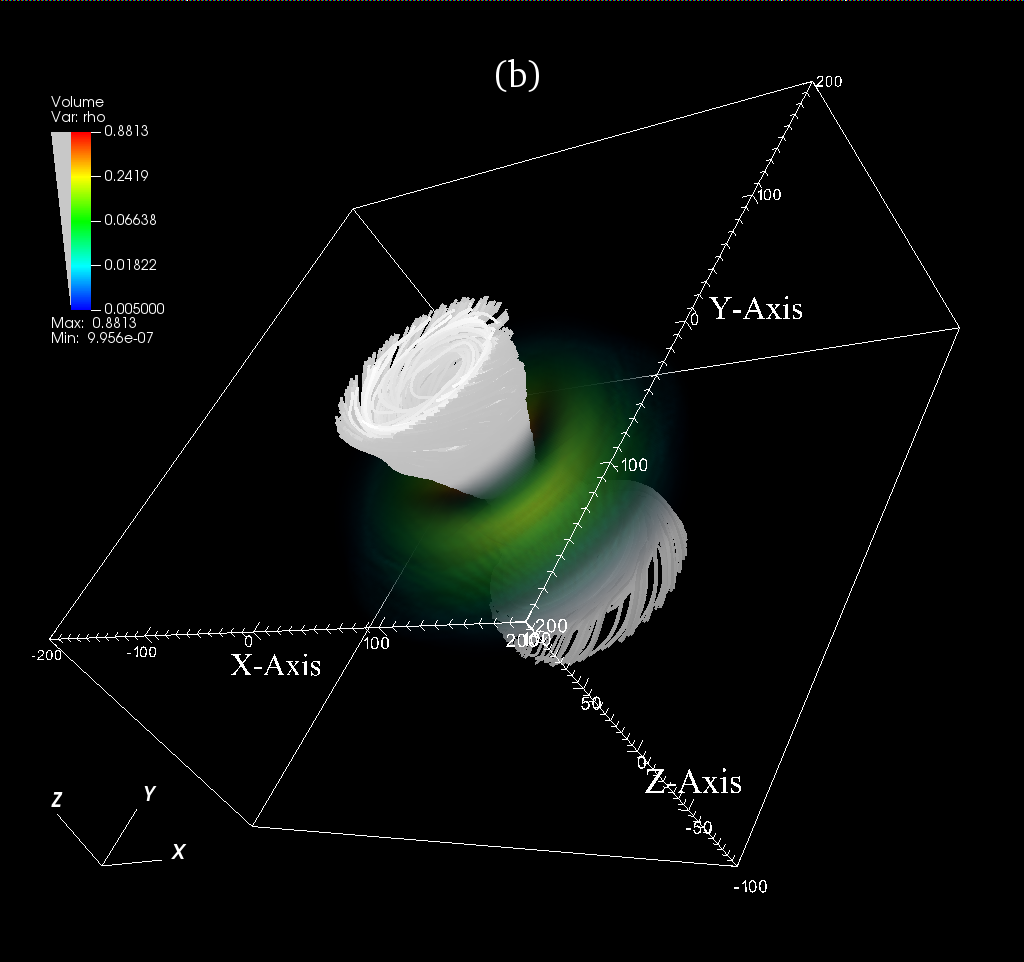} 
	\end{center}
	\caption{Volume rendering of density ($\log \rho$) with the magnetic field lines for ($a$): initial torus at $t = 0t_g$ and ($b$): at the $t = 8500 t_g$.  See the text for details.}
	\label{Figure_6}
\end{figure}
%%%%%%%%%%%%%%%%%%%%%%%%%%%%%%%%%%%%%%%%%%%%%%%%%%%%

%%%%%%%%%%%%%%%%%%%%%%%%%%%%%%%%%%%%%%%%%%%%%%%%%%%
%%                        Figure 7
%%%%%%%%%%%%%%%%%%%%%%%%%%%%%%%%%%%%%%%%%%%%%%%%%%%
\begin{figure*}
	\begin{center}
        \hskip -2.5mm
        \includegraphics[width=0.17\textwidth]{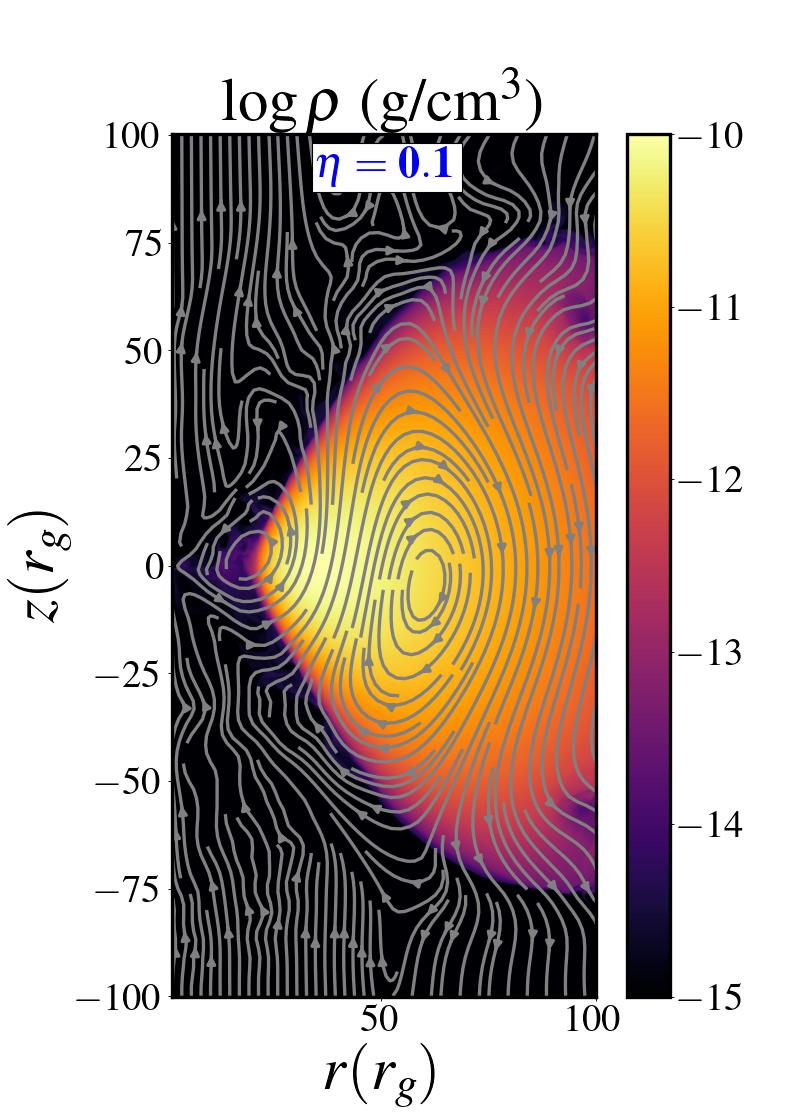} 
        \hskip -2.5mm
        \includegraphics[width=0.17\textwidth]{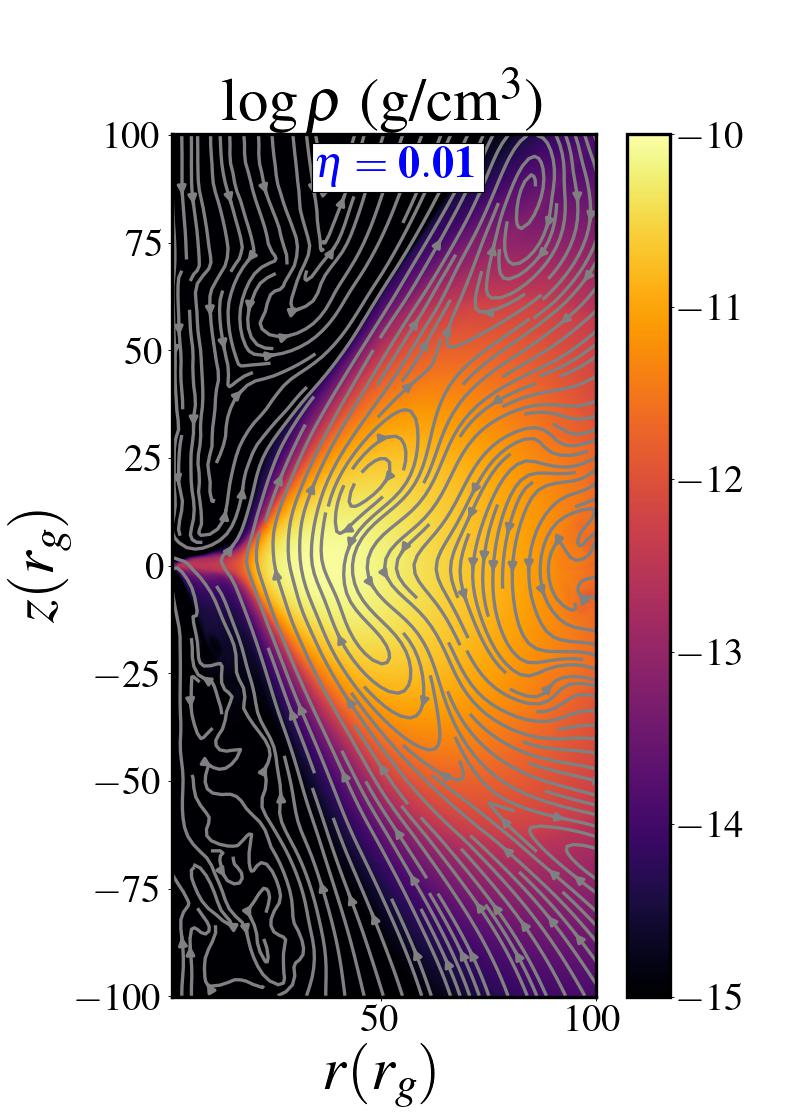} 
        \hskip -2.5mm
	\includegraphics[width=0.17\textwidth]{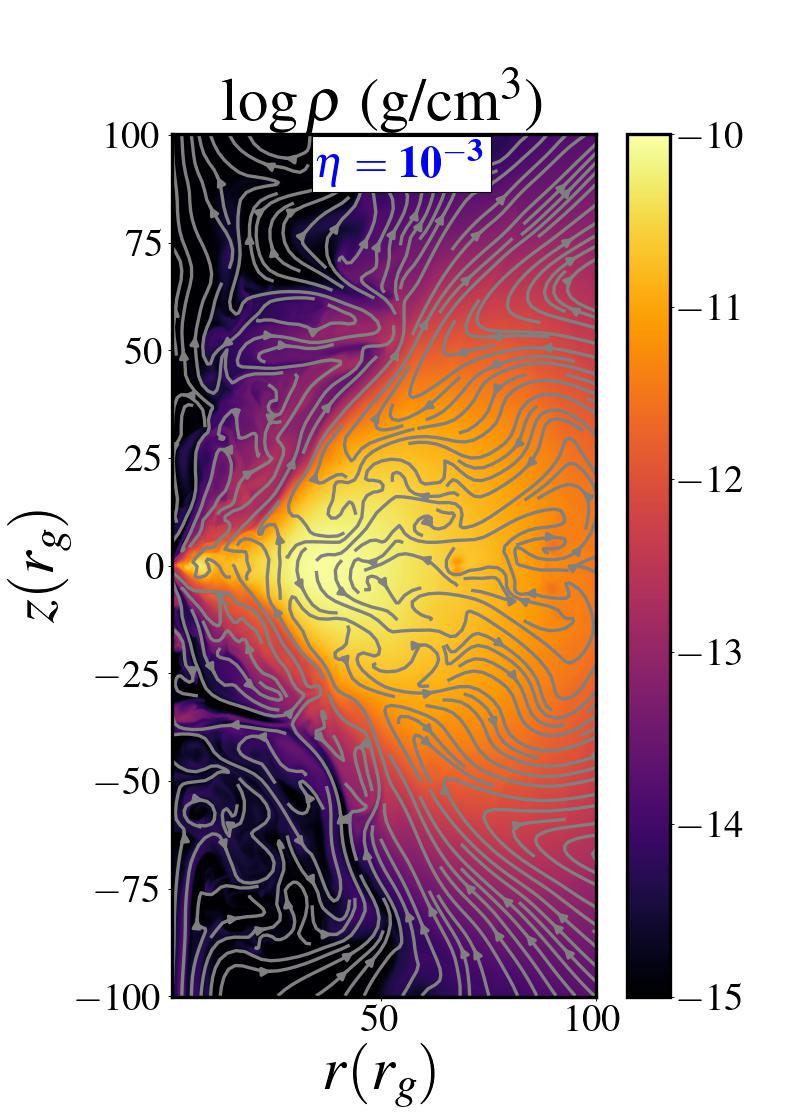} 
        \hskip -2.5mm
        \includegraphics[width=0.17\textwidth]{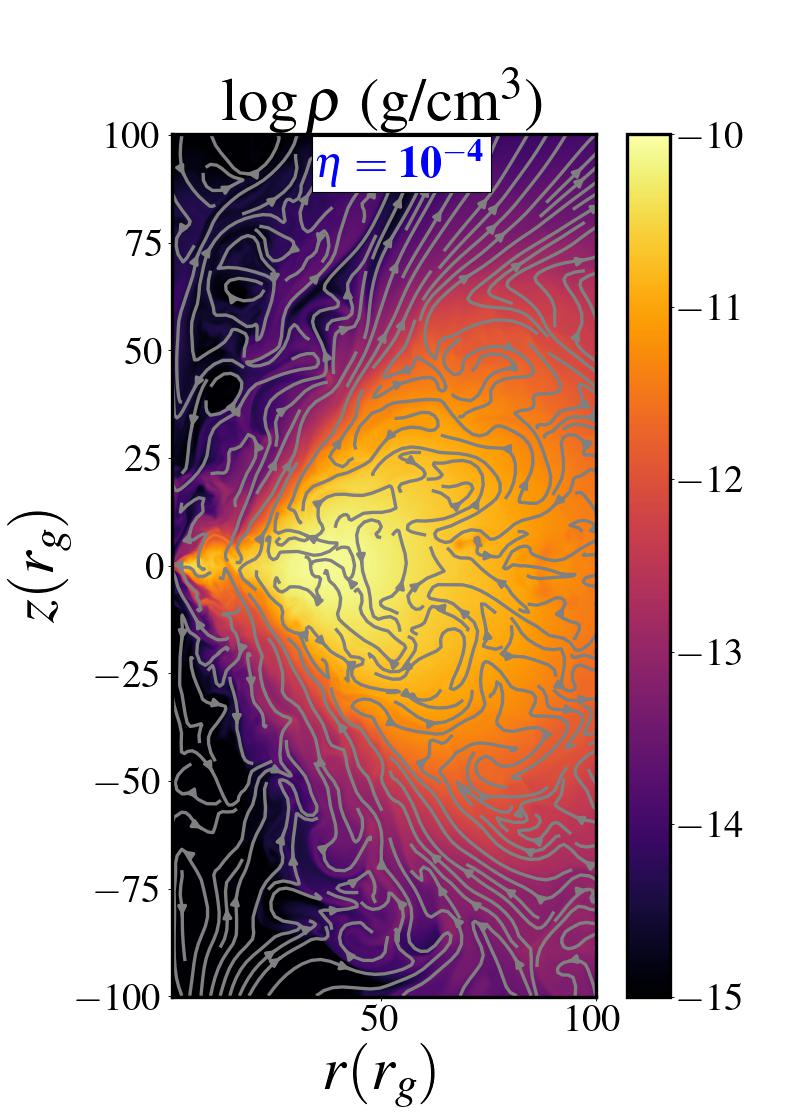} 
        \hskip -2.5mm
        \includegraphics[width=0.17\textwidth]{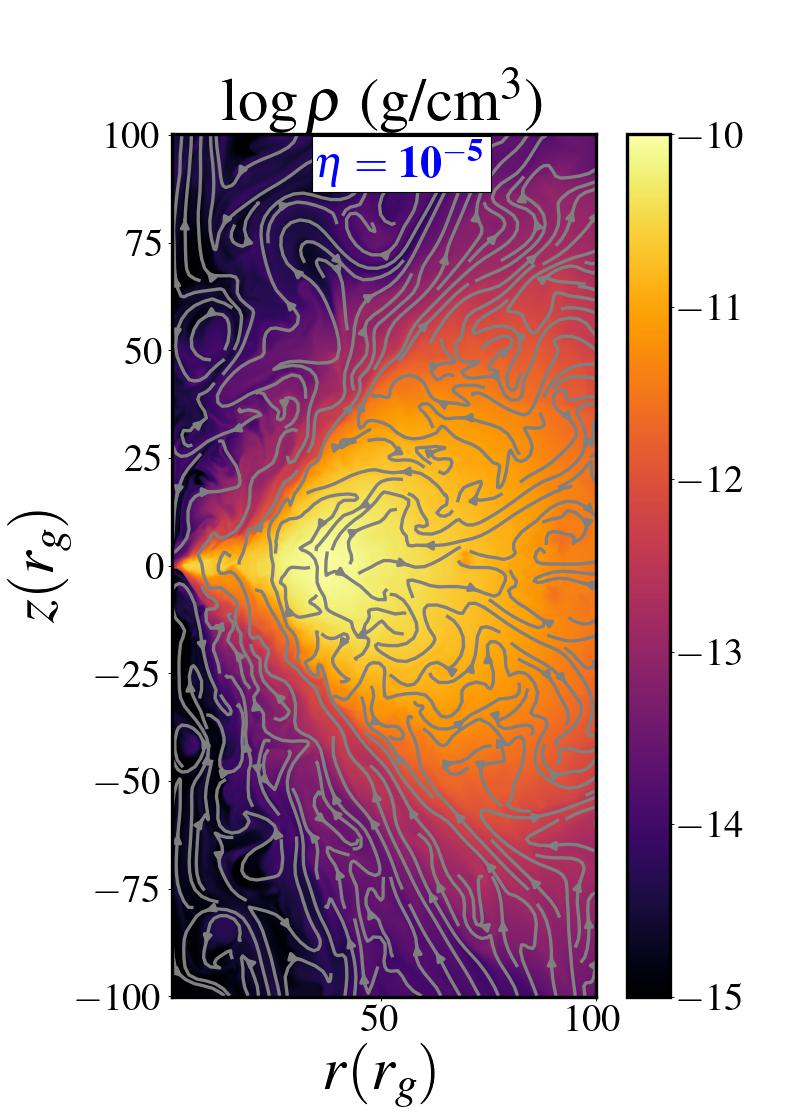} 
        \hskip -2.5mm
        \includegraphics[width=0.17\textwidth]{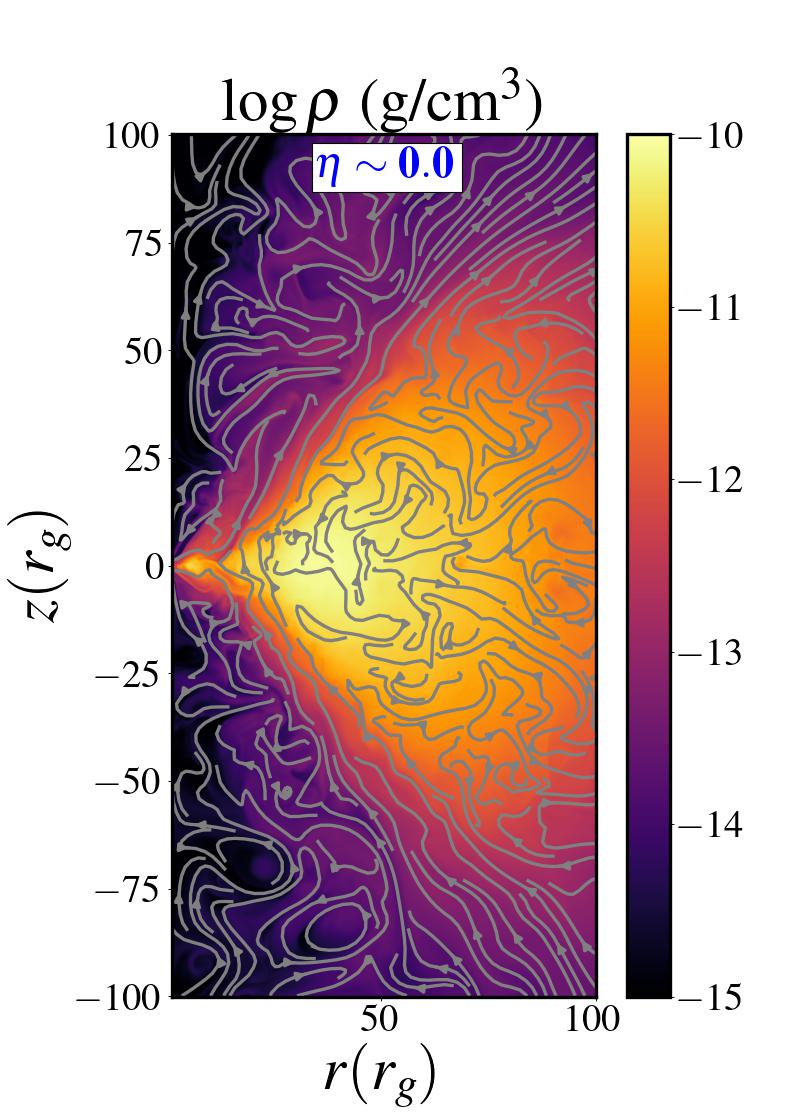} 

        \hskip -2.5mm
        \includegraphics[width=0.17\textwidth]{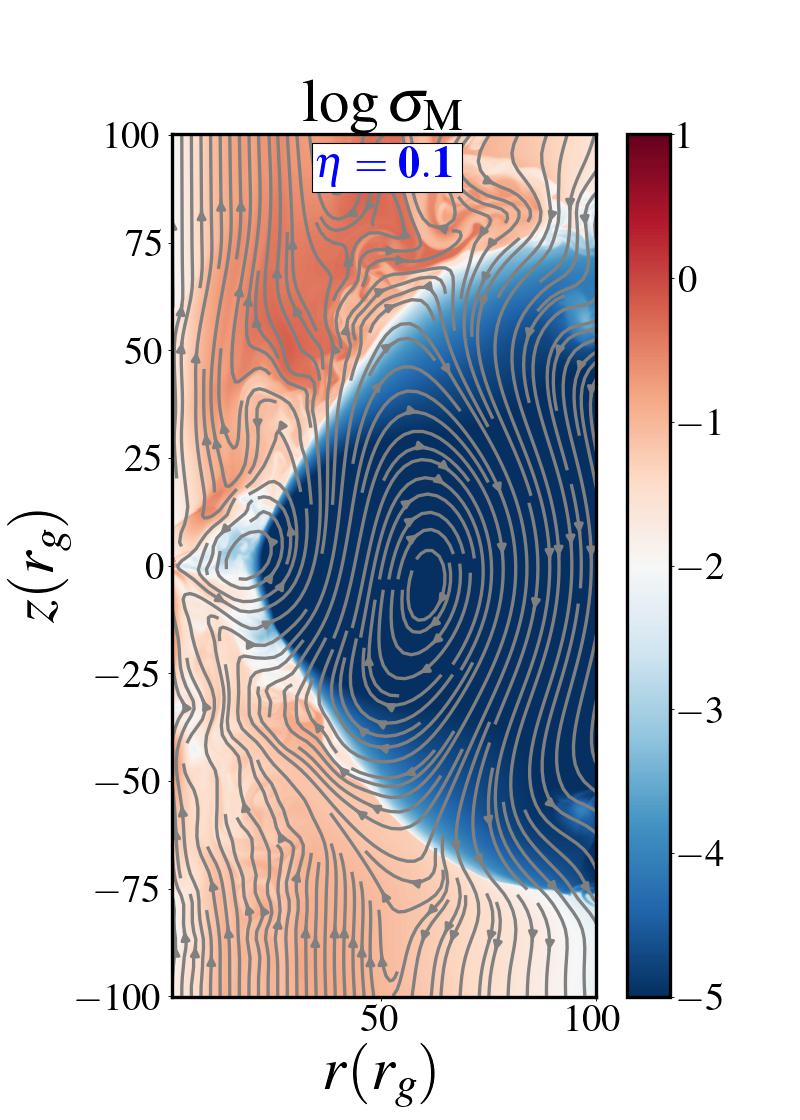} 
        \hskip -2.5mm
        \includegraphics[width=0.17\textwidth]{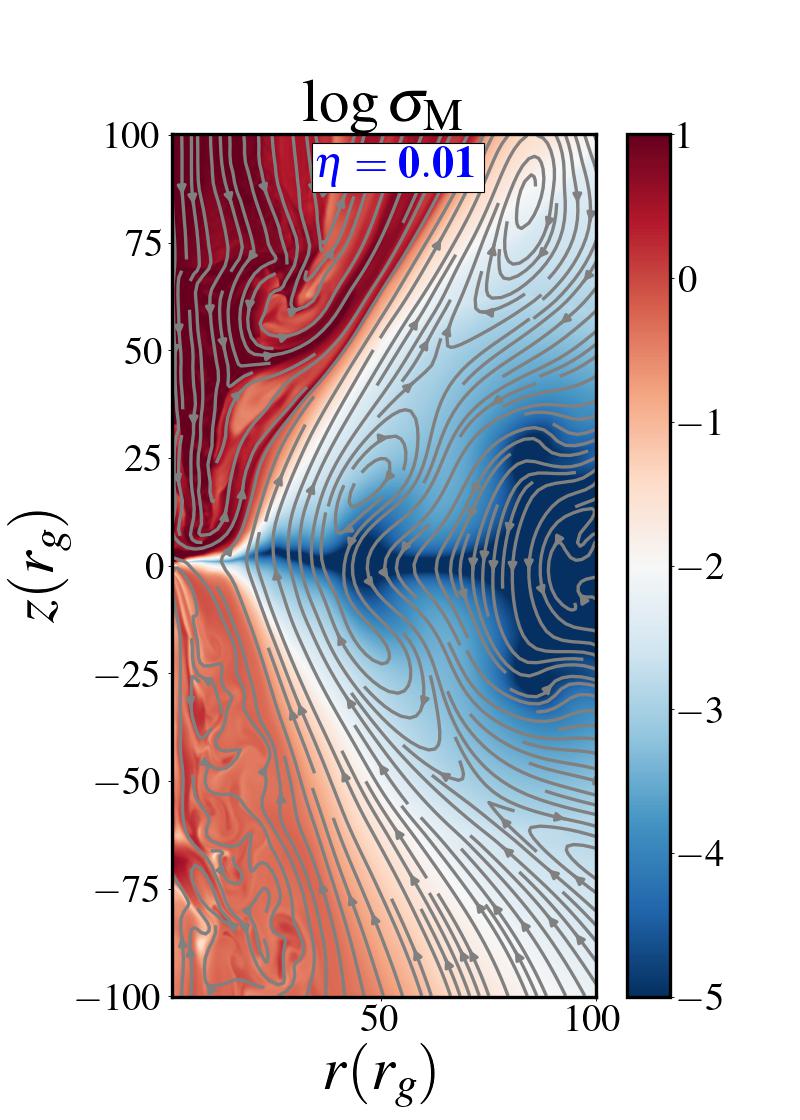} 
        \hskip -2.5mm
	\includegraphics[width=0.17\textwidth]{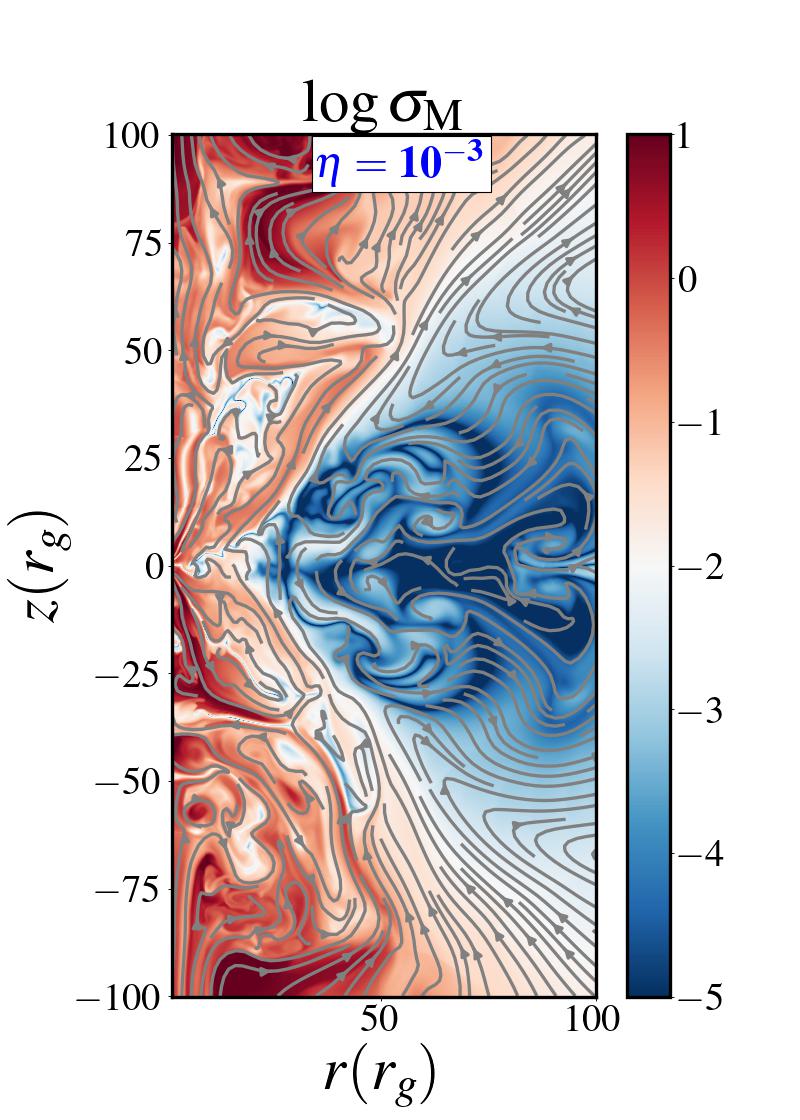} 
        \hskip -2.5mm
        \includegraphics[width=0.17\textwidth]{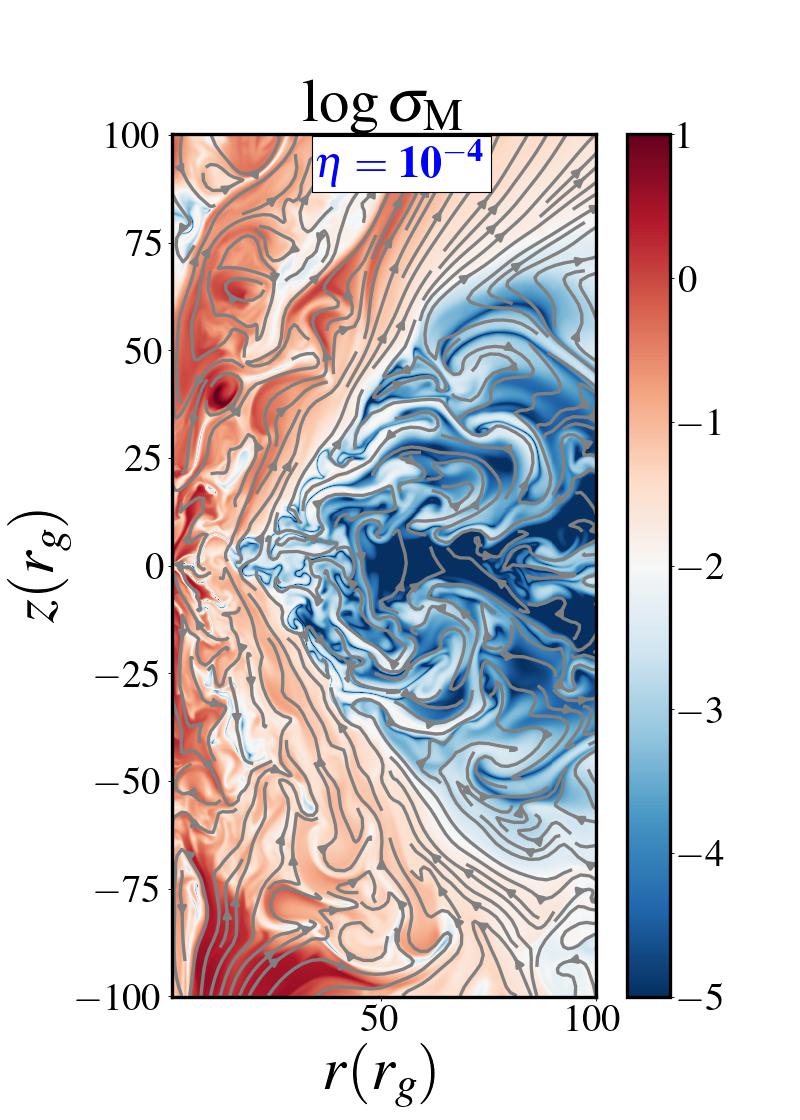} 
        \hskip -2.5mm
        \includegraphics[width=0.17\textwidth]{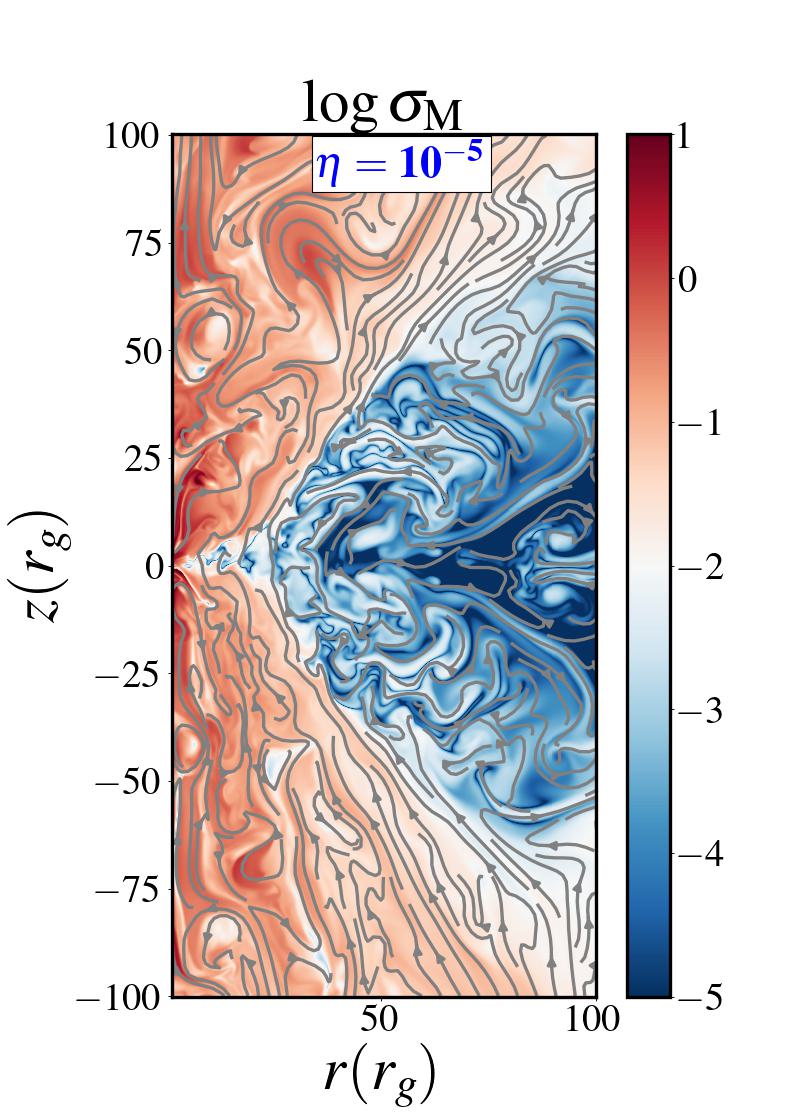} 
        \hskip -2.5mm
        \includegraphics[width=0.17\textwidth]{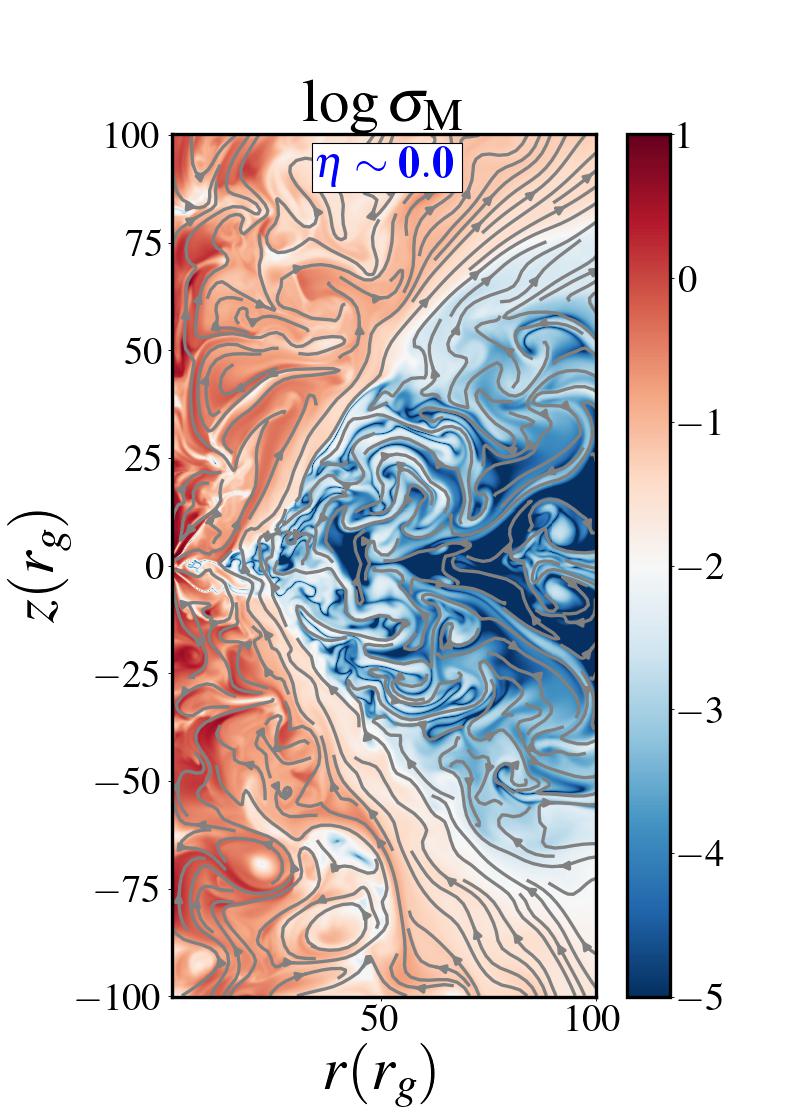} 

        \hskip -2.5mm
        \includegraphics[width=0.17\textwidth]{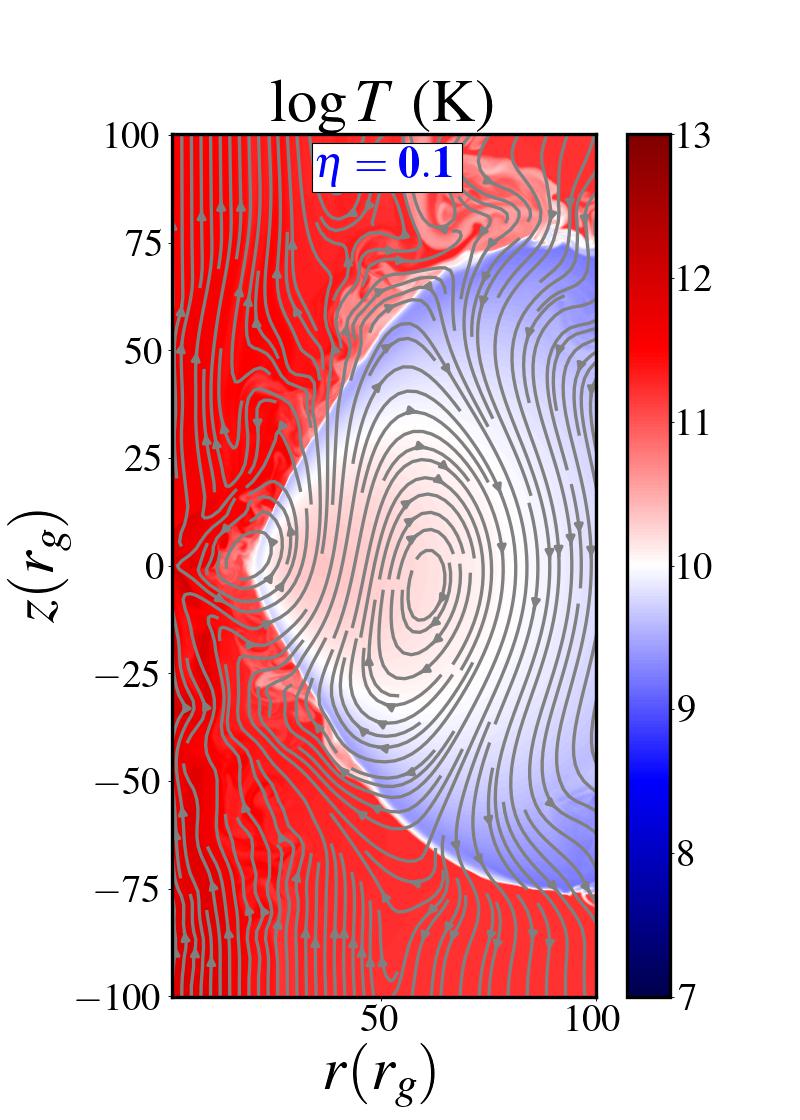} 
        \hskip -2.5mm
        \includegraphics[width=0.17\textwidth]{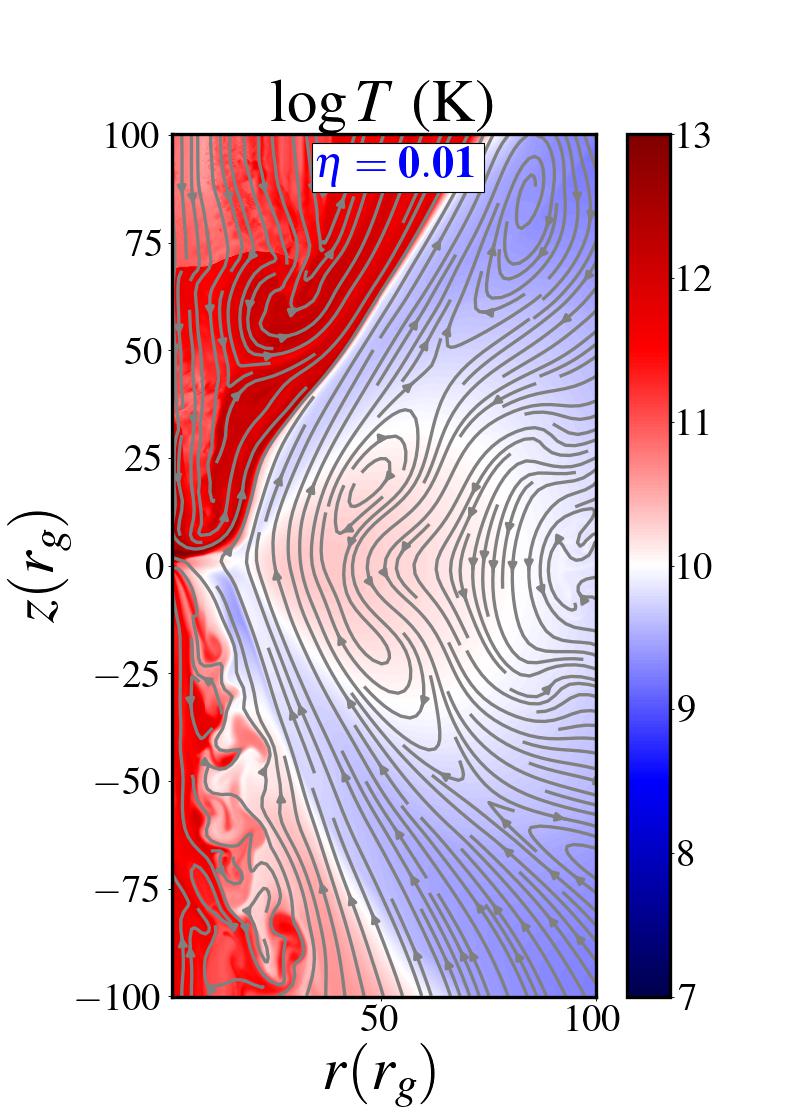} 
        \hskip -2.5mm
	\includegraphics[width=0.17\textwidth]{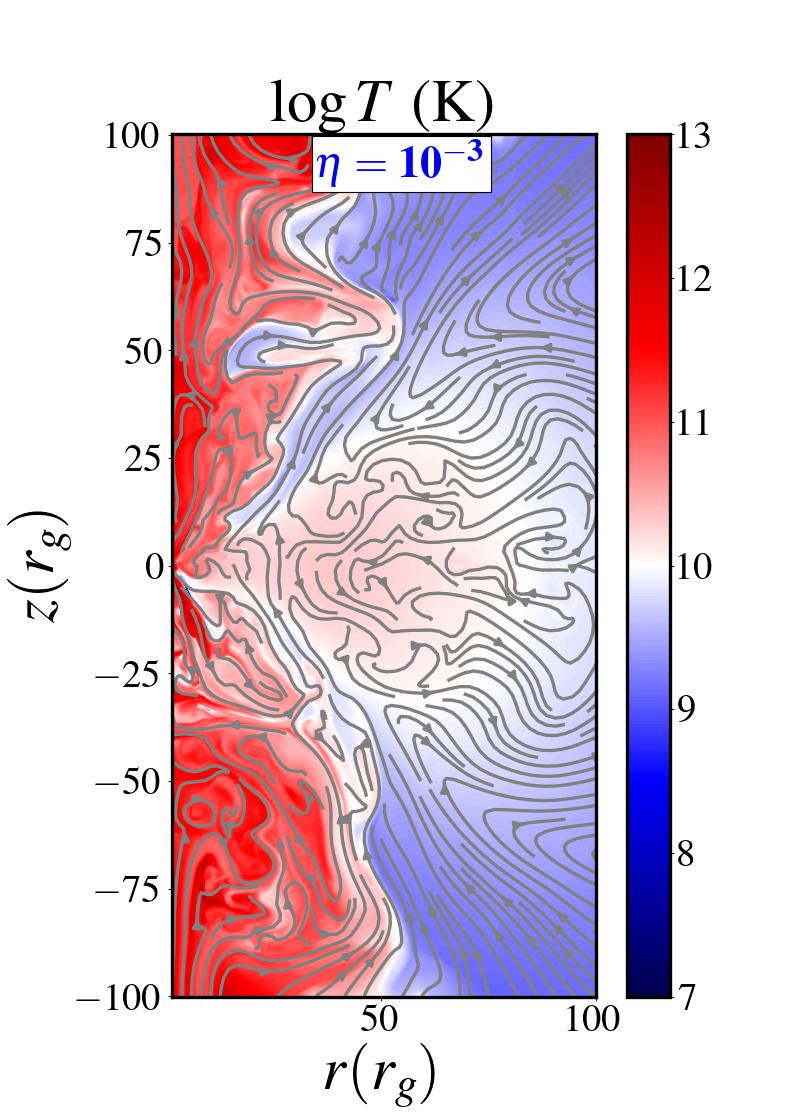} 
        \hskip -2.5mm
        \includegraphics[width=0.17\textwidth]{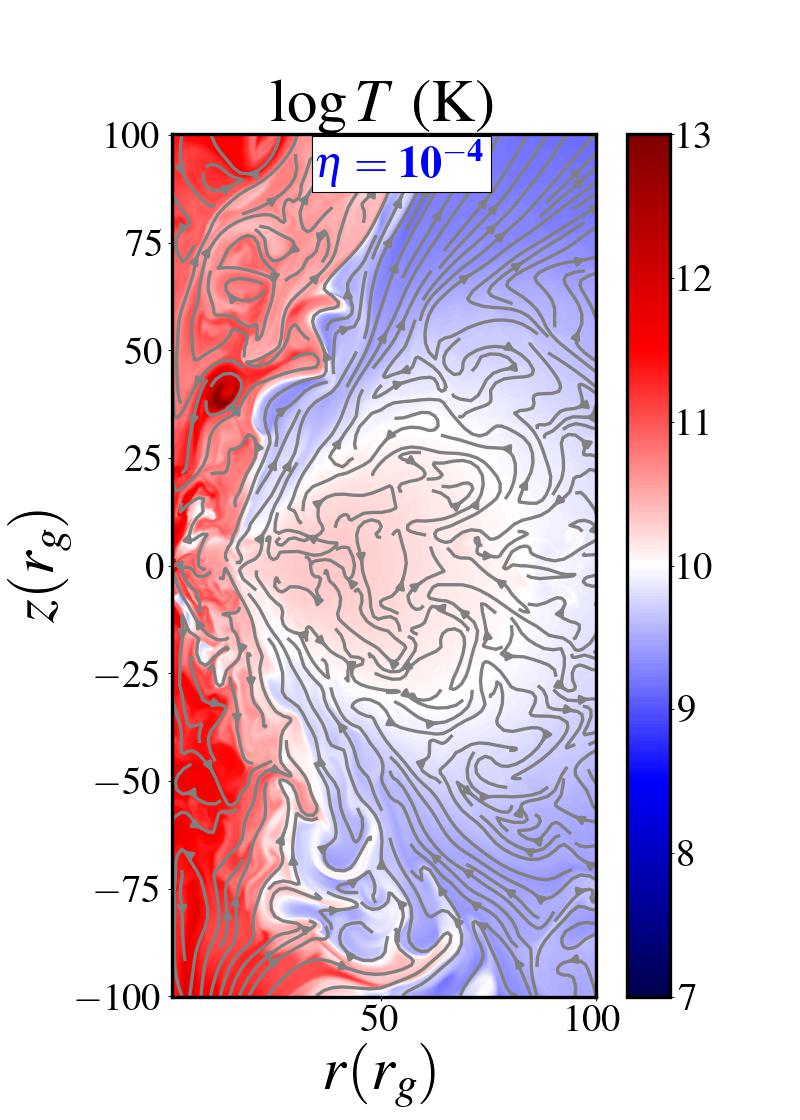} 
        \hskip -2.5mm
        \includegraphics[width=0.17\textwidth]{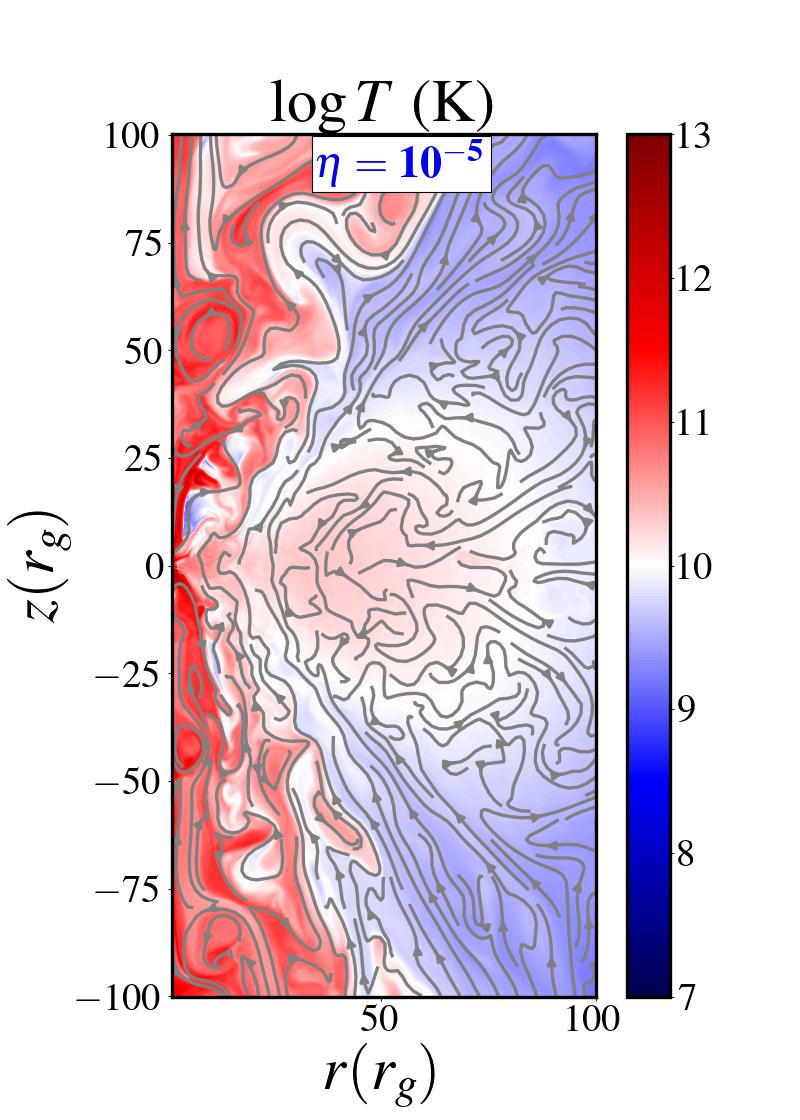} 
        \hskip -2.5mm
        \includegraphics[width=0.17\textwidth]{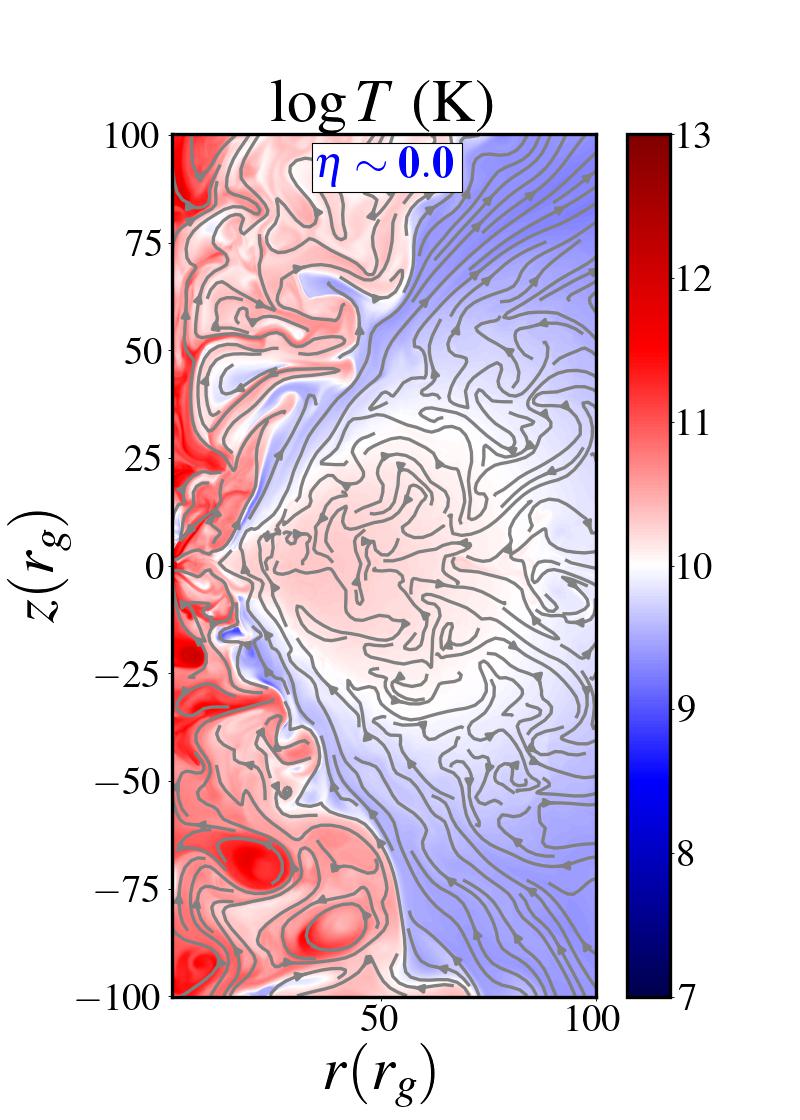} 

        \hskip -2.5mm
        \includegraphics[width=0.17\textwidth]{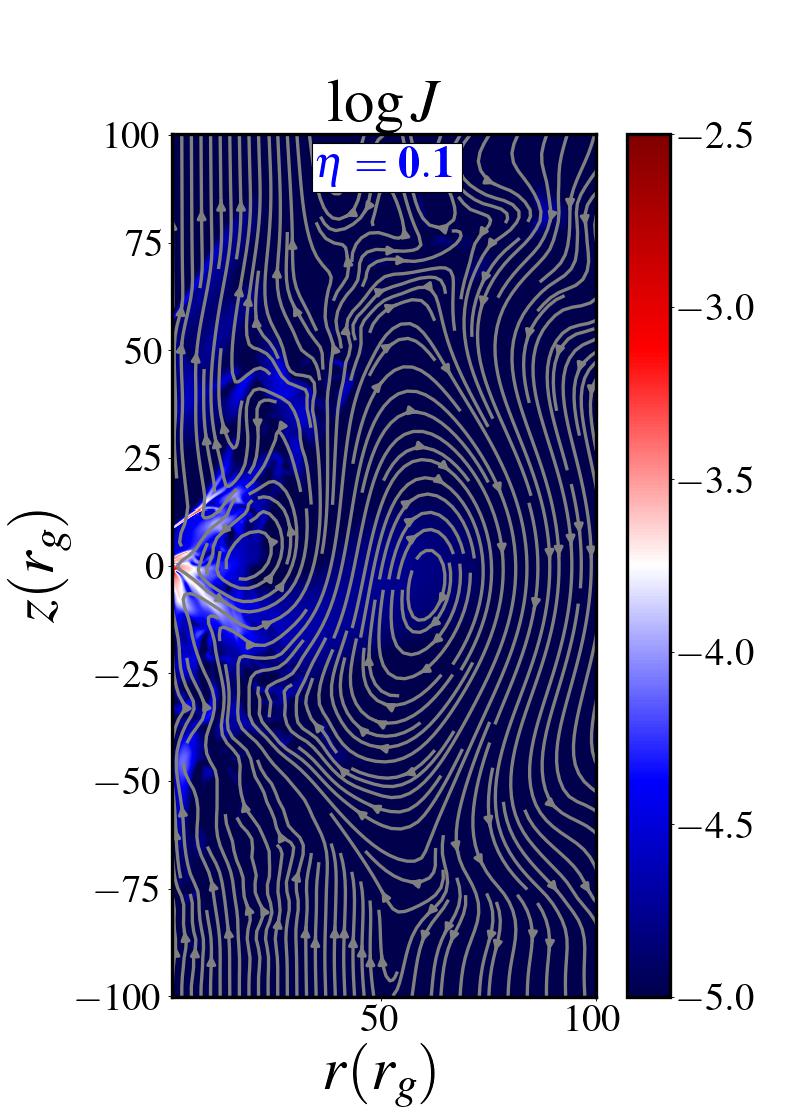} 
        \hskip -2.5mm
        \includegraphics[width=0.17\textwidth]{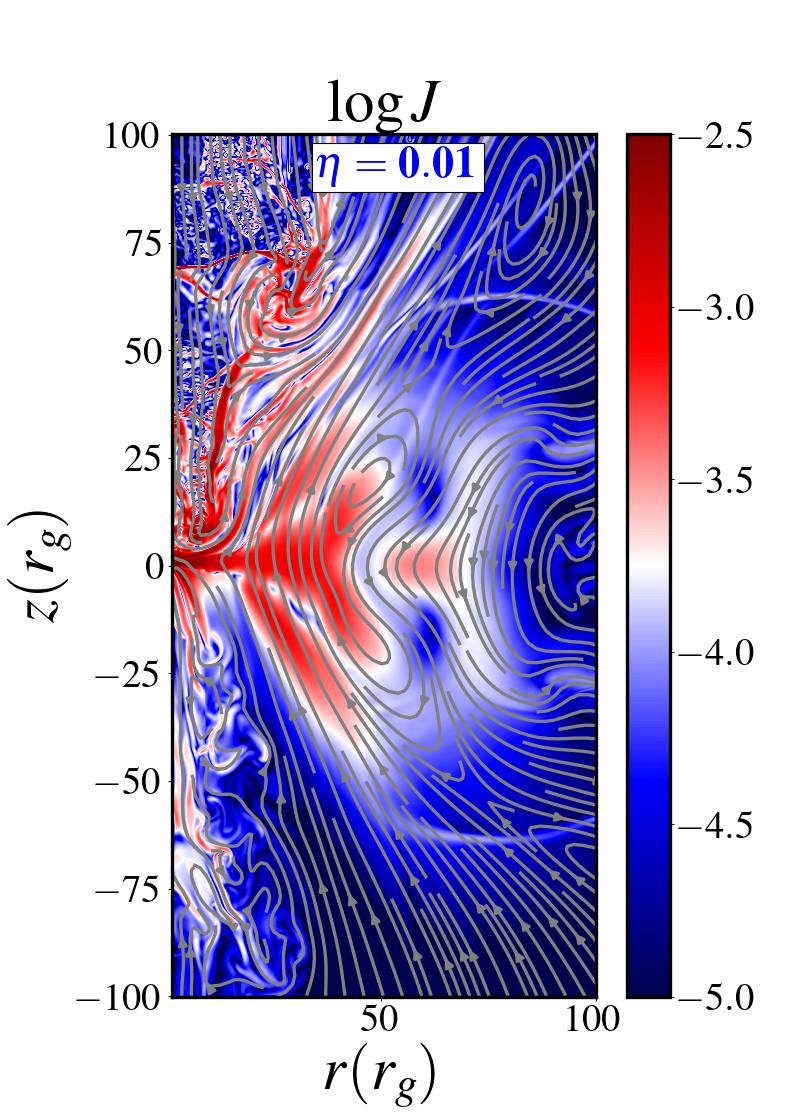} 
        \hskip -2.5mm
	\includegraphics[width=0.17\textwidth]{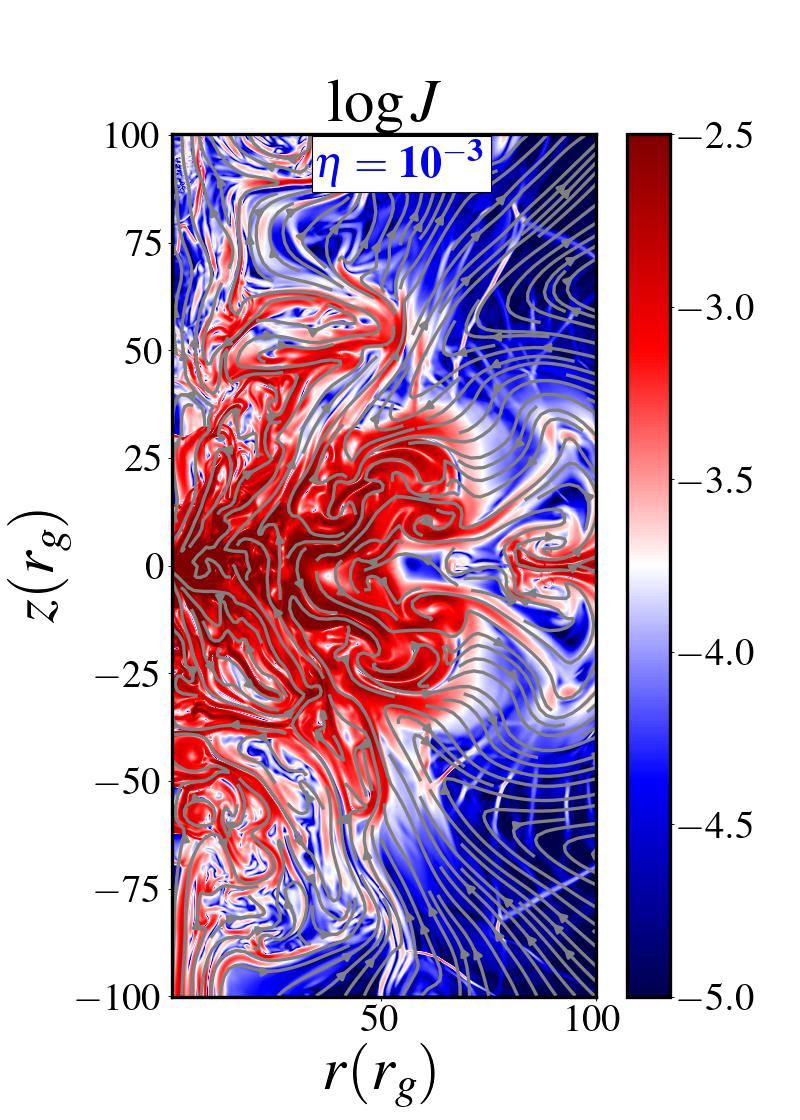} 
        \hskip -2.5mm
        \includegraphics[width=0.17\textwidth]{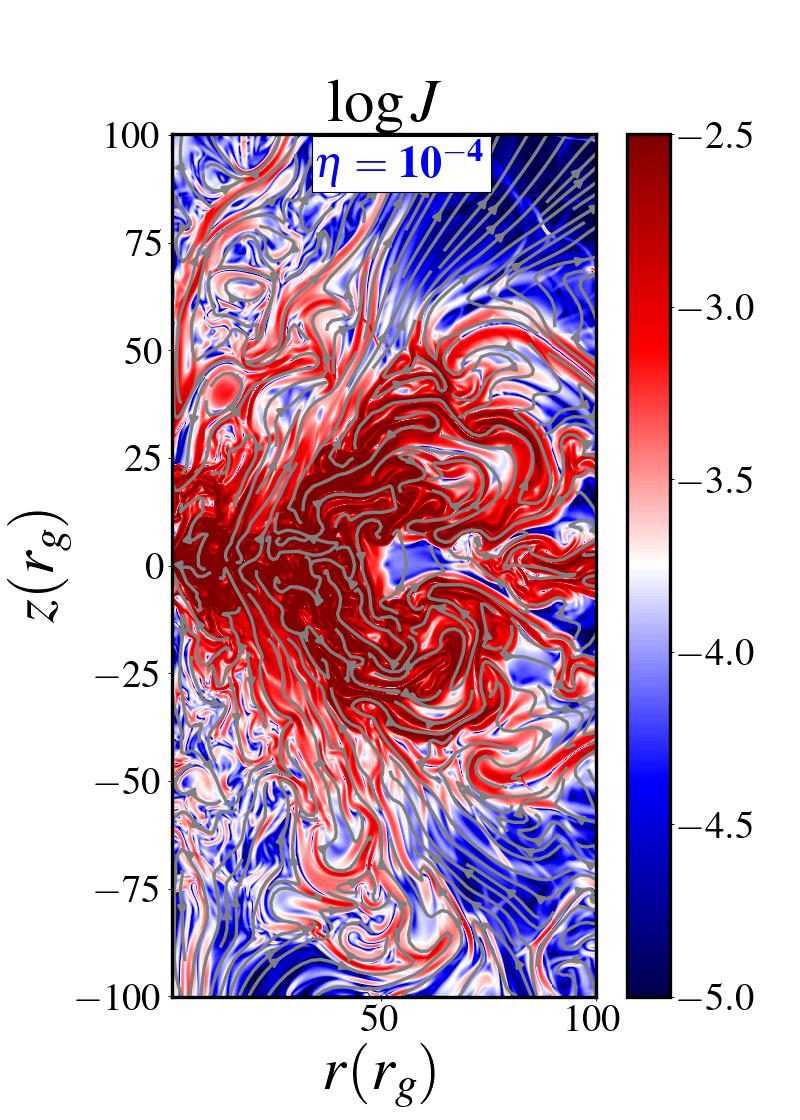} 
        \hskip -2.5mm
        \includegraphics[width=0.17\textwidth]{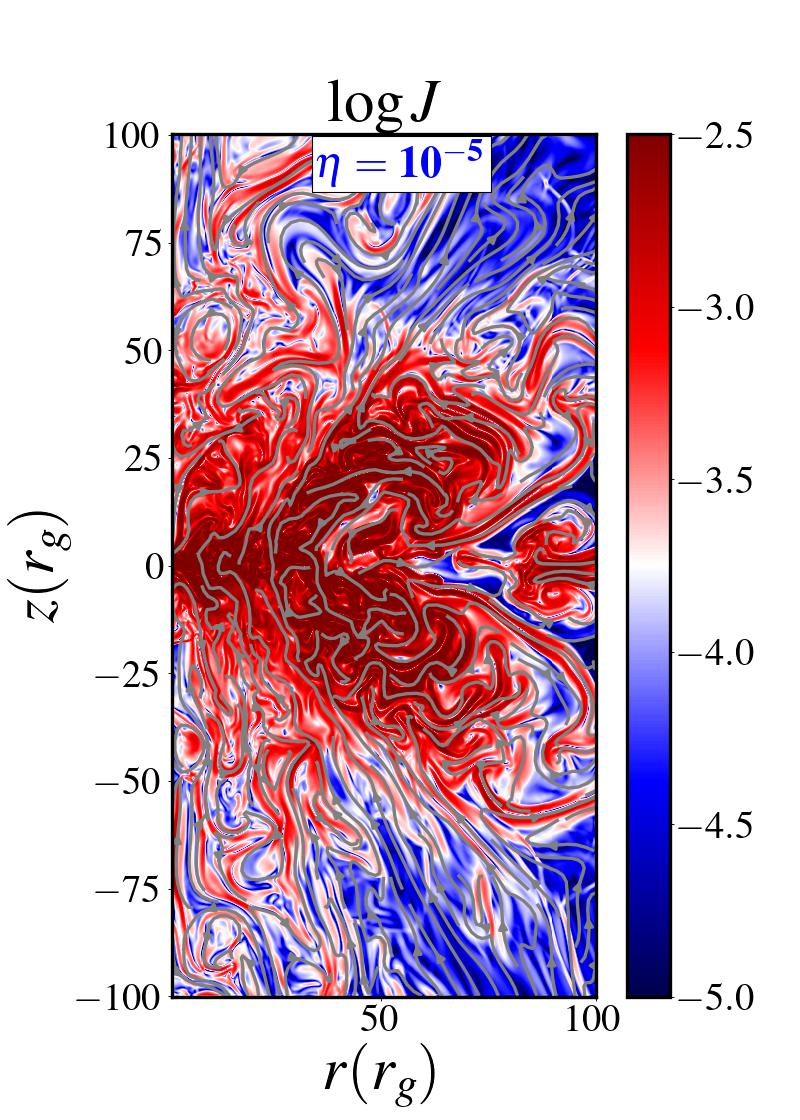} 
        \hskip -2.5mm
        \includegraphics[width=0.17\textwidth]{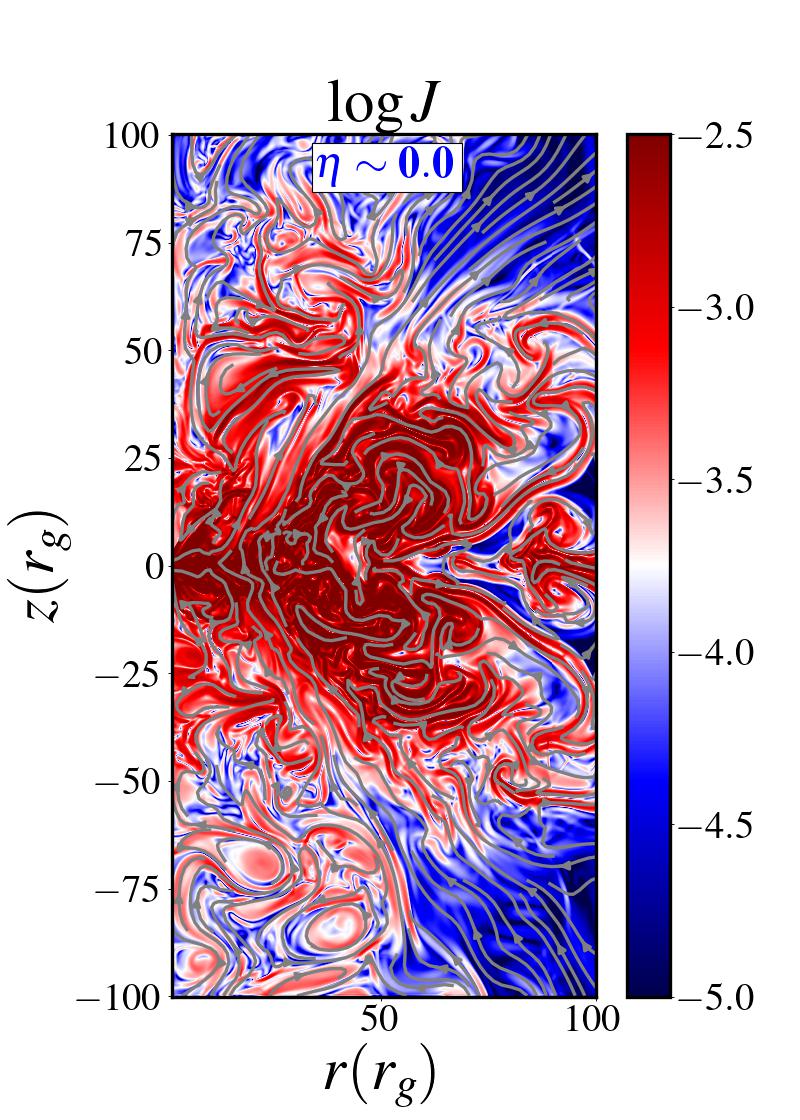} 
	\end{center}
	\caption{3D Model: Distribution of gas density $(\rho)$, magnetization parameter $(\sigma_{\rm M})$, temperature ($T$) and current density ($J$) for 3D model in first, second, third and fourth row, respectively. The grey lines represent the magnetic field lines. Here, we fix the different resistivity as $\eta = 0.1, 0.01, 10^{-3}, 10^{-4}, 10^{-5}$, and $\sim 0$. See the text for details.}
	\label{Figure_7}
\end{figure*}
%%%%%%%%%%%%%%%%%%%%%%%%%%%%%%%%%%%%%%%%%%%%%%%%%%%%

%%%%%%%%%%%%%%%%%%%%%%%%%%%%%%%%%%%%%%%%%%%%%%%%%%%
%%                        Figure 8
%%%%%%%%%%%%%%%%%%%%%%%%%%%%%%%%%%%%%%%%%%%%%%%%%%%
\begin{figure}
	\begin{center}
        \includegraphics[width=0.8\textwidth]{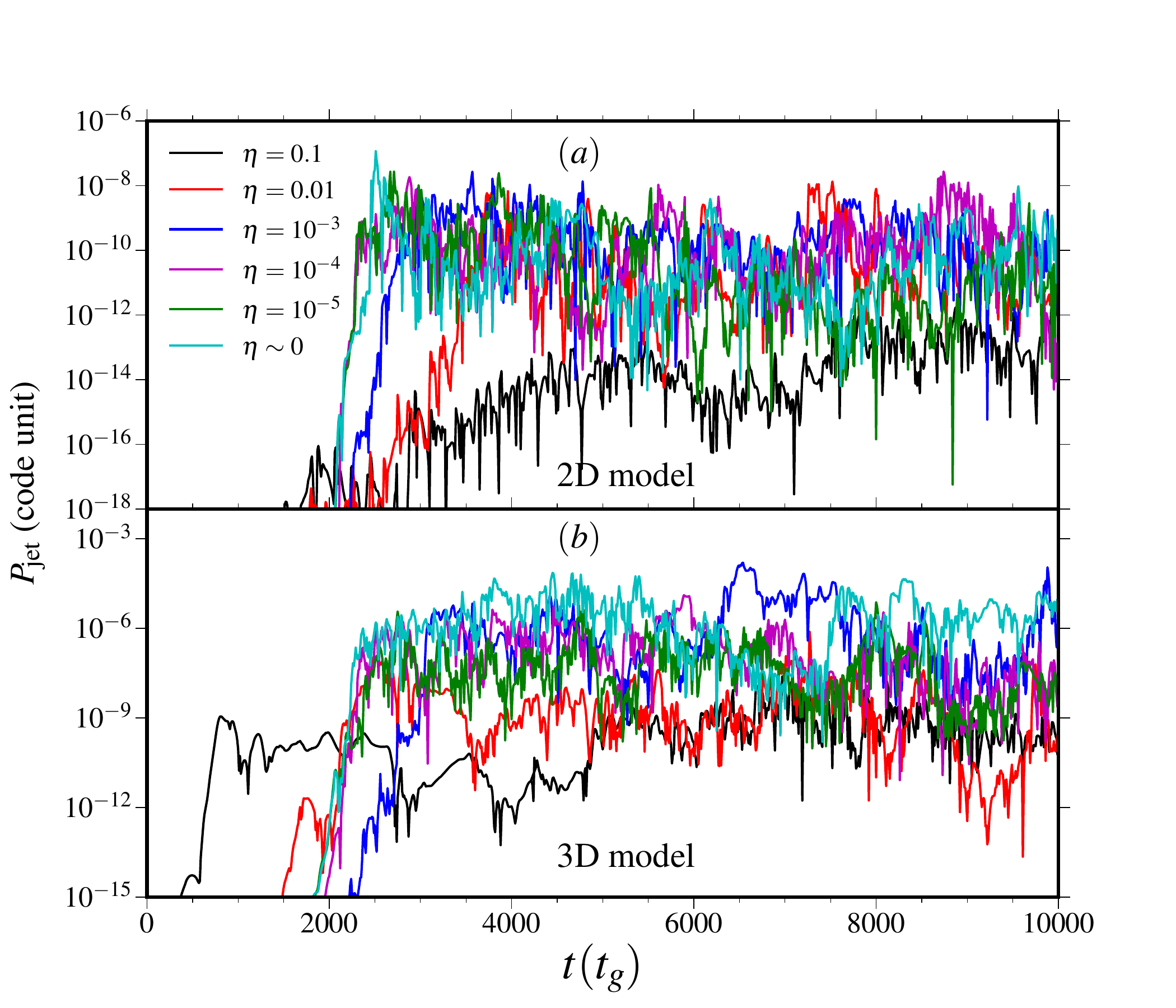} 
	\end{center}
	\caption{Jet power vs resistivity of the flow for 2D and 3D model.  See the text for details.}
	\label{Figure_8}
\end{figure}
%%%%%%%%%%%%%%%%%%%%%%%%%%%%%%%%%%%%%%%%%%%%%%%%%%%%

%============================================================================
\section{Simulation Results}
\label{results}
%============================================================================

We begin by constructing the initial equilibrium torus around a black hole by fixing the constant flow angular momentum (\(\lambda\)) and the spin of the black hole (\(a_k\)). The initial equilibrium torus, illustrated by the distribution of density (\(\rho\)) and temperature (\(T\)) in the (\(r-z\)) plane, is presented in Fig. \ref{Figure_1}. The gray lines with arrows represent the magnetic field lines, with the initial magnetic field embedded within the torus, as shown in Fig.~\ref{Figure_1}. In this study, we conduct a comparative analysis using both 2D and 3D models. We vary the resistivity of the flow, considering values of \(\eta = 0.1, 0.01, 10^{-3}, 10^{-4}, 10^{-5}\), and \(\sim 0\) (ideal MHD), respectively. The comparison of mass accretion rates (\(\dot{M}_{\rm acc}\)) in code unit is presented in Fig.~\ref{Figure_2}a. Interestingly, we observe almost identical mass accretion rates for both the 2D and 3D models at the initial time, up to \( \leq 1000 t_g\). However, the mass accretion rate for the 3D model deviates from that of the 2D model when the MRI becomes more prominent in the flow. As the flow reaches a saturation state and becomes increasingly turbulent, the non-axisymmetric effects in the 3D model become significant, leading to a notable difference in the mass accretion rate when compared to the 2D model \citep{Balbus-Hawley91, Hawley-Krolik01}. Next, we examine the magnetic state of the accretion flow by analyzing the normalized magnetic flux (\(\dot{\phi}_{\rm acc}\)) accumulated at the horizon. We compare the magnetic flux for both the 2D and 3D models under varying resistivity conditions, as shown in Fig.~ \ref{Figure_2}b. We observe that all models belong to a MAD state, regardless of the resistivity. The dashed black lines in the figure represent the threshold value for the MAD state (\(\dot{\phi}_{\rm acc} = 50\)). After time $t\ge 1000t_g$, the magnetic flux shows significant differences between the 2D and 3D models. This variation stems from the fact that the magnetic field structure in the 2D model is considerably simplified due to the axisymmetric assumption, unlike the more complex structure present in the 3D model. Additionally, the value of \(\dot{\phi}_{\rm acc}\) is more than 10 times higher in the 3D model compared to the 2D model. We observe flux eruption events after the magnetic flux has saturated around a specific value for both 2D and 3D cases \citep{Tchekhovskoy-etal11, Chatterjee-Narayan-22}. We also investigate the magnetic state of the accretion flow by examining the spatial average plasma-beta parameter (\(\beta_{\rm ave}\)) across the entire computational domain, as illustrated in Fig.~ \ref{Figure_2}c. The dashed black lines indicate the value at which gas pressure and magnetic pressure are comparable (\(\beta_{\rm ave} = 1\)). We observe that \(\beta_{\rm ave} \lesssim 1\) when the flow transitions into the MAD state in both the 2D and 3D models, as indicated by \(\dot{\phi}_{\rm acc}\) in Fig.~\ref{Figure_2}b. Therefore, \(\beta_{\rm ave}\) serves as a useful indicator for identifying the MAD state. This approach is physically motivated, as the magnetic pressure in the flow for MAD state will either dominate or be comparable to the gas pressure in the accretion flow. Furthermore, we analyze the average magnetic energy within the accretion flow by calculating the spatial average magnetic energy (\(B^2_{\rm ave}\)) over the entire computational domain. We find that magnetic energy increases dramatically, reaching \(\ge 10^{10}\) times of the initial value when the flow enters the MAD state, as shown in Fig.~\ref{Figure_2}d. Moreover, the magnetic energy stabilizes at a value indicative of a sustained MAD state over the duration of the simulation. We note that \(\beta_{\rm ave}\) and \(B^2_{\rm ave}\) undergo significant changes after time $t\geq 1000t_g$ for $\eta = 0.1, 0.01$ and after $t\geq 2000t_g$ for $\eta = 10^{-3}, 10^{-5}, \sim 0$ in both 2D and 3D models, but we do not observe a substantial difference between the 2D and 3D models after reaching the MAD states. In this paper, we compare the simulation results of 2D and 3D models up to \( t \sim 10,000~t_g \). It is well known that MRI cannot sustain itself for long periods in axisymmetric flows, according to the anti-dynamo theory proposed by \citet{Cowling-33}. In these simulation models, we observe that MRI turbulence persists up to \( t \sim 10,000~t_g \) in the 2D configurations.

%{\color{red}In the 2D model of accretion flow, the analysis is typically averaged over the azimuthal direction ($\phi$) based on the assumption of axisymmetry. This simplification restricts the variability of flow variables in the $\phi$-direction and effectively suppresses non-axisymmetric instabilities. In contrast, the 3D model allows for the development of the MRI, which drives turbulence and generates inherently non-axisymmetric structures within the accretion flow \citep{Balbus-Hawley91, Hawley-Krolik01}.}

%In the 2D model, the accretion flow is typically averaged over the azimuthal direction (\(\phi\)) due to the assumption of axisymmetry. This simplification reduces the complexity of the variations in density and velocity around the \(\phi\)-direction. In contrast, the 3D model exhibits a more complex accretion flow due to the presence of {\color{red}MRI} that create a non-axisymmetric structure.

For more comparison between 2D and 3D models, we examine the quality factor of MRI turbulence in our simulations for various resistive flows. The quality factor indicates how effectively the fastest-growing MRI modes are captured in a numerical simulation, helping to ensure that turbulence is not artificially suppressed due to inadequate resolution. To calculate the quality factor, we define the characteristic wavelength of the fastest-growing mode of MRI as \(\lambda_{\text{MRI}} = \frac{2 \pi v_A}{\Omega}\), where \(v_A\) is the Alfvén speed and \(\Omega\) is the angular frequency. Therefore the quality factor for MRI turbulence can be written as \citep{Hawley-etal11, Hawley-etal13, Aktar-etal24a}
\begin{align} \label{quality_factor}
Q_r = \frac{\lambda_{{\rm MRI},r}}{\Delta r}~~~~~~~~~~~~{\rm and}~~~~~~~~~~~~~Q_z = \frac{\lambda_{{\rm MRI},z}}{\Delta z},
\end{align}
where $\lambda_{{\rm MRI},r} = 2 \pi v_{A,r}/\Omega$ and $\lambda_{{\rm MRI},z} = 2 \pi v_{A,z}/\Omega$. Here $v_{A,r}$ and $v_{A,z}$ are the radial and vertical component of Alfv\'en speed, respectively. $\Delta r$ and $\Delta z$ are the grid sizes in the radial and vertical directions, respectively. Here, we calculate \( Q_r \) and \( Q_z \) by considering the vertical average over the range \( -50 r_g < z < 50 r_g \) at a time of \( t = 8500 t_g \). In Fig.~\ref{Figure_3}, we compare the average quality factors \( \langle Q_r \rangle \) and \( \langle Q_z \rangle \) as a function of radial distance for both 2D and 3D models. The black and red curves represent \( \langle Q_r \rangle \) and \( \langle Q_z \rangle \), respectively. Solid curves indicate the 3D models, while dotted curves represent the 2D models. We find that \( Q_r \gtrsim 10 \) and \( Q_z \gtrsim 10 \) for all resistive values, except in the high-density region of the torus for high resistive flows with \( \eta = 0.1 \) and \( \eta = 0.01 \). Notably, \( \langle Q_r \rangle \) and \( \langle Q_z \rangle \) are significantly higher in the 3D models, being approximately \( 10^3 \) times greater than in the 2D models for \( r < 40 r_g \). This indicates that MRI turbulence is much more prominent near the black hole in 3D models compared to the axisymmetric approximation of the 2D models. These results suggest a clear deviation between the 3D and 2D models as the MRI activity becomes more pronounced after the initial time, as illustrated in Fig.~\ref{Figure_2}a,b. Additionally, we observe a dip in the quality factor in the high-density torus region for all resistive flows, since the Alfvén speed is inversely proportional to the density of the flow.  Further, we compare the Maxwell stress ($\alpha_{\rm mag}$) and Reynolds stress ($\alpha_{\rm gas}$) in the left and right panel of Fig.~\ref{Figure_4}, respectively. Black and red curves are for 3D and 2D models, respectively. The normalized Maxwell stress and Reynolds stress are computed as $\alpha_{\rm mag} = \frac{<B_r B_\phi>}{B^2}$ and $\alpha_{\rm gas} = \frac{<\rho v_r \delta v_\phi>}{P_{\rm gas}}$, respectively, where $\delta v_{\phi} = v_\phi - <v_\phi>$ \citep{Hawley-00, Stone-Pringle01, Proga-Begelman-03, Aktar-etal24a}. We observe that average Maxwell stress is close to unity, i.e., $\alpha_{\rm mag}\sim 1$ for low resistive flow ($\eta \lesssim 10^{-4}$) and for high resistive flow ($\eta = 0.1$), the average $\alpha_{\rm mag}\sim 0.1$. On the other hand, we find that average Reynolds stress is $\alpha_{\rm gas} \lesssim 0.01$ for all the resistive models, which is $\le 10-10^2$ times lower than Maxwell stress, i.e., $\alpha_{\rm mag} >> \alpha_{\rm gas}$. It also emphasizes that the outward transport of angular momentum is predominantly driven by Maxwell stress in a MAD model \citep{Aktar-etal24a, Aktar-etal24b}. Here, we calculate $\alpha_{\rm mag}$ and $\alpha_{\rm gas}$ considering vertical average over $-50 r_g < z < 50 r_g$ at time $t=8500 t_g$ similar as MRI quality factor in Fig.~\ref{Figure_3}. In this section, we set $\phi = 0$ to estimate various quantities for 3D models, as illustrated in Fig.~\ref{Figure_3} and Fig.~\ref{Figure_4}. It is important to note that although the MRI is often considered the primary mechanism for driving angular momentum transport in magnetized accretion disks, its role in MAD remains uncertain. In the highly magnetized environments characteristic of MADs, the MRI may be suppressed due to the strong magnetic pressure. Instead, non-axisymmetric Rayleigh-Taylor (RT) instabilities are expected to dominate, facilitating accretion through the intermittent penetration of dense, low-magnetic-pressure blobs into the magnetically supported inner regions \citep{McKinney-etal-12, White-etal-19}. Our 3D simulation models naturally allow for the development of such RT modes. A more detailed analysis of their role and their interaction with any residual MRI activity will be addressed in future work.

We now examine the effect of resistivity on accretion flow by analyzing the spatial distribution of flow variables in the \((r-z)\) plane for the 2D model depicted in Fig.~\ref{Figure_5}. We fix the simulation time at \(t = 8500 t_g\). In the first row, we present the density distribution ($\rho$) for progressively decreasing resistivity values from $\eta = 0.1$ to $\sim 0$. The gray lines with arrows represent magnetic field lines. We find that higher resistivity in the flow leads to significant diffusion and suppression of turbulent structures within the accretion flow. It is observed that below \(\eta \leq 10^{-3}\), the structure and turbulence appear qualitatively similar \citep{Ripperda-etal-2019}. The corresponding magnetization parameter \(\sigma_{\rm M}\) is shown in the second row of Fig.~\ref{Figure_5}. We find that the magnetization parameter is notably high (\(\sigma_{\rm M} \gtrsim 10\)) in the jet region across all resistive cases. We also observe a decrease in MRI turbulence in the higher resistive flow (\(\eta = 0.1, 0.01\)). Generally, higher resistivity leads to increased diffusion, which results in reduced MRI turbulence. Furthermore, we find evidence of plasmoid formation in the jet region (for \(r < 20 r_g\)) under low resistive flow conditions \citep{Ripperda-etal-2020, Ripperda-etal-2022}. This plasmoid formation is particularly visible in the temperature distribution plot presented in the third row. Plasmoids are observed in models with \(\eta = 0.01\), \(10^{-3}\), \(10^{-4}\), and \(10^{-5}\). In low-resistive flow, the magnetic field lines become anti-parallel and form current sheets near the horizon. The fourth row of Fig.~\ref{Figure_5} shows the current density (\(J\)) in the flow. The current density is very high near the horizon for the MAD state, and current sheets form in the jet region. In Fig.~\ref{Figure_51}, we present a zoomed-in version of the temperature distribution across different time slices to identify the plasmoids for a resistivity of $\eta = 10^{-5}$. The time slices considered are as follows: (a) $t = 8200t_g$, (b) $t = 8300t_g$, (c) $t = 8400t_g$, (d) $t = 8500t_g$, (e) $t = 8600t_g$, and (f) $t = 8700t_g$. In Fig.~\ref{Figure_51}a, we observe the formation of a plasmoid at the location $(r, z) \sim (23, 42)r_g$, which we denote as $P_1$. The magnetic field lines appear to form a closed loop around this region. As time progresses, the plasmoid is advected along the jet, as depicted in Figs.~\ref{Figure_51}b, c, and d \citep{Ripperda-etal-2020}. By the time shown in Fig.~\ref{Figure_51}e, plasmoid $P_1$ begins to dissipate, and a new plasmoid, $P_2$, starts to form. Similar to $P_1$, the plasmoid $P_2$ is also advected into the jet, as demonstrated in Fig.~\ref{Figure_51}f. 

Now, we examine the magnetic field structure in the 3D model shown in Fig.~\ref{Figure_6}. For clarity, we present the volume rendering of density ($\rho$) at two different times: $t = 0 t_g$ and $t = 8500 t_g$, as depicted in Fig.~\ref{Figure_6}a and Fig. \ref{Figure_6}b, respectively. In this analysis, we maintain a fixed resistivity of the flow at $\eta = 10^{-5}$. At the initial time, the magnetic field lines are embedded within the torus, as described in section \ref{Mag_config}, and are illustrated in Fig.~\ref{Figure_6}a with grey lines. The spatial distribution is also represented in a 2D slice in the $r-z$ plane, shown in Fig.~\ref{Figure_1}. As the accretion flow enters into the MAD state, we observe twisting magnetic field lines forming near the horizon of the jet, as shown in Fig.~\ref{Figure_6}b. Here, we show the volume rendering plot at time $t = 8500 t_g$.  To analyze the magnetic field structure within the jet region, we illustrate magnetic field lines located within $r <20 r_g$. The twisting magnetic field near the horizon is a key features of the Blandford-Znajek (BZ) mechanism, a pioneering theoretical framework for extracting rotational energy from spinning black holes and launching relativistic jets \citep{Blandford-Znajek77}.

Furthermore, we compare spatial distribution of the 3D model by varying the resistivity and plotting various variables in the \((r-z)\) plane at $\phi = 0$, as shown in Fig.~\ref{Figure_7}, similar to Fig.~\ref{Figure_5}. We observe trends consistent with those in the 2D model: the suppression of MRI turbulence in flows with higher resistivity (\(\eta = 0.1\)), while plasmoid formation occurs in less resistive models (\(\eta = 10^{-3}, 10^{-4}\), and \(10^{-5}\)), which is illustrated in the temperature distribution plot in the third row of Fig.~\ref{Figure_7}. However, the numerical resolution is not sufficient to resolve plasmoids in our global 3D simulation models. We observe that the structure of the disk, jet, and turbulence appears qualitatively similar for low resistive flow ($\leq 10^{-3}$), resembling 2D models.  Additionally, we note a high magnetization parameter \((\sigma_{\rm M} \geq 10)\) near the horizon of the jet for all cases, as depicted in Fig.~\ref{Figure_7}, second row. Current sheets are formed in less resistive flows where \(\eta \leq 10^{-3}\), as shown in the fourth row of Fig.~\ref{Figure_7}. In contrast, the current density in high-resistive flows (\(\eta = 0.1\)) is significantly lower and exhibits a broader distribution over the outer region of the disk in 3D models compared to 2D models. This is illustrated in Fig.~\ref{Figure_5} and Fig.~\ref{Figure_7}, fourth row.

%%%%%%%%%%%%%%%%%%%%%%%%%%%%%%%%%%%%%%%%%%%%%%%%%%
\subsection{Variability vs resistivity}
\label{variability}
%%%%%%%%%%%%%%%%%%%%%%%%%%%%%%%%%%%%%%%%%%%%%%%%%%

To evaluate the variability of the accretion flow, we compute the mean and variance of the mass accretion rate (\(\dot{M}_{\rm acc}\)) and the normalized magnetized flux (\(\dot{\phi}_{\rm acc}\)). The definitions for the mean and variance are as follows:
\begin{align}\label{variabilty_eqn}
\mu = \frac{1}{N} \sum_{i=1}^{N}{x_i}~~~~~~~~~ {\rm and}~~~~~~~~~ s^2 = \frac{1}{N} \sum_{i=1}^{N}{(x_i - \mu)^2},
\end{align}
where \(N\) denotes the number of data points and $x_i$ is the variable under investigation. The variability is subsequently defined as \(\sigma = s/\mu\). We analyze the variability by focusing on both the mass accretion rate and the normalized magnetic flux. We examine this variability by considering the number of data points $N$ over the entire duration for $t \ge 1000 t_g$ \citep{Aktar-etal24b}. The variability measurements for the mass accretion rate \((\sigma_{\dot{M}_{\rm acc}})\) and the magnetic flux \((\sigma_{\dot{\phi}_{\rm acc}})\) are provided in the last column of Table \ref{Table-2}. Our observations reveal no clear relationship between the variability and the resistivity of the accretion flow. \citet{Nathanail-etal-2024} also examined this phenomenon and found that resistivity has minimal effects on the variability of the flow in the MAD state.

%%%%%%%%%%%%%%%%%%%%%%%%%%%%%%%%%%%%%%%%%%%%%%%%%%
\subsection{Jet power vs resistivity}
\label{jet_power}
%%%%%%%%%%%%%%%%%%%%%%%%%%%%%%%%%%%%%%%%%%%%%%%%%%

Jet power from black hole sources is typically expressed in terms of the magnetic flux that threads into the black hole horizon. In this context, \citet{Blandford-Znajek77} describe how the rotational energy of a spinning black hole can be extracted via magnetic fields anchored in the surrounding plasma. The jet power of a black hole can be expressed using the following equation \citep{Tchekhovskoy-etal-10, Tchekhovskoy-etal11}:
\begin{align}\label{jet_eqn}
P_{\rm jet} = \kappa ~ \phi_{\rm acc}^2 ~ \Omega_{\rm H}^2.
\end{align}
In this equation, \( \kappa \) is a constant that depends on the geometry of the magnetic field. For the sake of representation, we use \( \kappa = 0.1 \). The term \( \phi_{\rm acc} \) refers to the magnetic flux threading the black hole's horizon. The horizon frequency of the black hole, denoted as \( \Omega_{\rm H} \), is given by the formula \( \Omega_{\rm H} = \frac{a_k}{2r_{\rm H}} \), where \( r_{\rm H} = 1 + \sqrt{1 - a_k^2} \) represents the black hole's horizon radius. In our study, we keep the spin of the black hole constant across all models, which means that jet power solely depends on the magnetic flux (\(\phi_{\rm acc}\)), particularly influenced by the resistivity of the flow. We present the scaled jet power as we vary the resistivity of the flow for both 2D and 3D models in Fig.~\ref{Figure_8}a and Fig.~\ref{Figure_8}b, respectively. Our observations reveal that jet power is significantly higher in low-resistivity flows compared to high-resistivity flows. Specifically, we find that the average jet power is roughly two order of magnitude lower for high-resistivity flows (\(\eta = 0.1\)) compared to low-resistivity flows (\(\eta \leq 10^{-3}\)) in both the 2D and 3D models. Furthermore, the average jet power stabilizes at a similar value for lower resistivity flows (\(\eta \leq 10^{-3}\)). In higher resistivity flows, the accumulation of magnetic flux is considerably less than in lower resistivity flows, as illustrated in Fig.~\ref{Figure_2}b and Fig.~\ref{Figure_8}. Ultimately, jet power is directly dependent on the accumulation of magnetic flux. In this context, \citet{Vourellis-etal-19} also observe that varying levels of resistivity significantly affect the mass and energy fluxes of disk winds and jets, with lower resistivity resulting in enhanced outflows.

%%%%%%%%%%%%%%%%%%%%%%%%%%%%%%%%%%%%%%%%%%%%%%%%%%
\section{Summary and Discussions}
\label{conclusion}
%%%%%%%%%%%%%%%%%%%%%%%%%%%%%%%%%%%%%%%%%%%%%%%%%%

In this paper, we examine the effect of resistivity on the dynamics of accretion flows around spinning black holes. For our analysis, we employ the resistive MHD (Res-MHD) module in the PLUTO code \citep{Migone-etal07}. The gravitational effects around the black hole are modeled using an effective Kerr potential \citep{Dihingia-etal18b}. Notably, our simulation model achieves higher spatial resolution compared to traditional GRMHD simulations. Additionally, our model is more efficient for 3D simulations given the available computational resources, allowing us to investigate accretion dynamics in multiple dimensions. In this work, we modified the initial atmospheric conditions to resemble those used in GRMHD simulations, which is different from our earlier studies \citep{Aktar-etal24a, Aktar-etal24b}, as discussed in subsection \ref{torus_set_up}. The initial torus size was specifically chosen to ensure consistency with the GRMHD simulation for the MAD state \citep{Wong-etal-21, Fromm-etal-22}. 

We investigate the effect of multi-dimensionality on the mass accretion rate and normalized magnetic flux by comparing the results of 2D and 3D simulations, as shown in Fig.~\ref{Figure_2}a and Fig.~\ref{Figure_2}b, respectively. In this study, we vary the resistivity of the flow, using values of \(\eta = 0.1, 0.01, 10^{-3}, 10^{-4}, 10^{-5}\), and \(\sim 0\). Our findings indicate that the mass accretion rate is similar for both the 2D and 3D models up to an initial time of \(t < 1000 t_g\). However, once the MRI becomes dominant, the 3D models begin to deviate from the 2D model. This divergence occurs because the non-axisymmetric effects in the 3D models are significantly more complex than those in the axisymmetric 2D models. Furthermore, the magnetic fields in the 3D models are much more intricate and varied compared to those in the 2D models. Consequently, we observe notable differences in the normalized magnetic flux between the 2D and 3D models, as illustrated in Fig.~\ref{Figure_2}b. In this context, we compare the MRI quality factor for 2D and 3D models as a function of radial distance, as shown in Fig.~\ref{Figure_3}. We observe that the MRI quality factors are significantly higher for 3D models compared to 2D models in the vicinity of the black hole, regardless of resistive flows. This highlights a notable difference in MRI activity between 3D and 2D models. Additionally, we find that all the models with resistivity belong to the MAD state. Interestingly, we identify an alternative approach to investigating the magnetic state of the flow by examining the average plasma beta parameter (\(\beta_{\rm ave}\)), as depicted in Fig.~\ref{Figure_2}c. Our results show that the MAD state is achieved when \(\beta_{\rm ave} \lesssim 1\) in the accretion flow. Furthermore, we observe that the magnetic energy is significantly high when the flow enters the MAD state, as shown in Fig.~\ref{Figure_2}d.

We also examine the effect of resistivity on the dynamics of accretion flows by comparing the spatial distribution of flow variables in the ($r-z$) plane for both 2D and 3D models, as shown in Fig.~\ref{Figure_5} and Fig.~\ref{Figure_7}. Our observations indicate that MRI turbulence is significantly reduced in high-resistivity flows ($\eta = 0.1$) for both the 2D and 3D models. Higher resistivity leads to increased diffusion, which in turn diminishes MRI turbulence. We find that the turbulence structure and flow dynamics are qualitatively similar for low-resistivity flows ($\eta < 10^{-3}$). A high magnetization jet region, where $\sigma_{\rm M} \sim 10$, is formed in the jet area across all resistive cases in both 2D and 3D models, as illustrated in the second row of Fig.~\ref{Figure_5} and Fig.~\ref{Figure_7}. Interestingly, we observe the formation of plasmoids in the jet region for low-resistivity flows, as seen in the temperature and current density plots. However, to accurately resolve the plasmoids, a much higher resolution is necessary.

Finally, we examine how resistivity affects the variability of the accretion flow by analyzing the time variations of the mass accretion rate and normalized magnetic flux. Our findings show no clear relationship between resistivity and variability, which is consistent with the results reported by \citet{Nathanail-etal-2024} (see Table \ref{Table-2}). Additionally, we compare the scaled jet power in code units by varying resistivity for both 2D and 3D models, as shown in Fig.~\ref{Figure_8}. We observe that the average jet power is significantly higher in low-resistivity flows compared to high-resistivity flows. This indicates that high-resistivity flows are less capable of producing powerful jets.

It is important to address the need for high-resolution simulations to accurately resolve fast, small-scale reconnection dynamics and to determine whether plasmoids can form and grow in accretion flows. Achieving this requires extreme resolution and adaptive mesh refinement (AMR) to effectively resolve thin current sheets and plasmoids \citep{Ripperda-etal-2020, Ripperda-etal-2022}. However, implementing such techniques can be computationally expensive for global accretion simulations. In the present study, we focus on the overall effects of resistivity on flow dynamics using multi-dimensional simulation models. While we intend to investigate plasmoid formation and to provide a more accurate resolution in the future to explain the variability in multiwavelength emissions, particularly as observed for Sgr A*, this specific investigation is beyond the scope of our current work and will be reported elsewhere.

%%%%%%%%%%%%%%%%%%%%%%%%%%%%%%%%%%%%%%%%%%%%%%%%%%%%%%%%%%%%%%%%%%%%%%%

\section*{Acknowledgments}
We sincerely thank the anonymous referee for their valuable suggestions and comments, which have significantly improved the manuscript. This work is supported by the National Science and Technology Council of Taiwan through grant NSTC 112-2811-M-007-038, 112-2112-M-007-040 and 113-2112-M-007-031, and by the Center for Informatics and Computation in Astronomy (CICA) at National Tsing Hua University through a grant from the Ministry of Education of Taiwan. The simulations and data analysis have been carried out on the CICA Cluster at National Tsing Hua University.
%\end{acknowledgments}

%%%%%%%%%%%%%%%%%%%%%%%%%%%%%%%%%%%%%%%%%%%%%%%%%%%%%%%%%%%%%%%%%%%%%%

%\appendix

%\section{Appendix information}


%%%%%%%%%%%%%%%%%%%%%%%%%% References %%%%%%%%%%%%%%%%%%%%%%%%%%%%%%%%%%

\bibliography{refs}{}
\bibliographystyle{aasjournal}



\end{document}